\RequirePackage{fix-cm}
\PassOptionsToPackage{table}{xcolor}
\documentclass[mnsc,nonblindrev]{informs1}
\OneAndAHalfSpacedXI
\usepackage{rotating}
\PassOptionsToPackage{hyphens}{url}
\usepackage{xurl}
\usepackage{pgfplots}
\usepackage[caption=false]{subfig}
\pgfplotsset{compat=1.18}
\usepackage{bbding}
\usepackage[left=1in,top=1in,right=1in,bottom=1in]{geometry}
\usepackage{amssymb,amsmath,amsxtra,amstext,xfrac,mathtools}
\usepackage{graphicx,booktabs,enumerate,paralist,mdwlist}
\usepackage{dsfont,verbatim,latexsym}
\usepackage[titletoc,toc,page]{appendix}
\usepackage{mathrsfs}
\usepackage{bbm}
\usepackage{multirow}
\usepackage[table]{xcolor}
\usepackage[normalem]{ulem}
\newcolumntype{C}[1]{>{\centering\let\newline\\\arraybackslash\hspace{0pt}}m{#1}}
\usepackage{longtable,multirow,threeparttable}
\usepackage{float}
\usepackage{tikz,qtree}
\usetikzlibrary{arrows,positioning,decorations.pathreplacing}
\usepackage{mwe}
\usepackage{makecell}
\usepackage{eurosym}
\usepackage[english]{babel}
\usepackage[utf8]{inputenc}
\usepackage{fancyhdr}
\usepackage{color}
\usepackage{subfig}
\usepackage{array}
\usepackage{graphicx}
\usepackage{tikz}
\usepackage{pgfplots}
\usepackage[T1]{fontenc}
\usepackage{url}
\usepackage{caption}
\usepackage[colorlinks=TRUE,citecolor=MyDarkBlue,urlcolor=MyDarkBlue,linkcolor=MyDarkBlue,hypertexnames=false]{hyperref}
\definecolor{MyDarkBlue}{RGB}{158,0,0}
\usepackage{bookmark}

\usepackage{natbib}
\bibpunct[, ]{(}{)}{,}{a}{}{,}%
\def\bibfont{\small}%
\TheoremsNumberedThrough
\EquationsNumberedThrough

\newcommand{\R}{\mathbb{R}}

\usepackage{tikz}
\usetikzlibrary{patterns}

\hypersetup{pdftitle={Monetizing Intelligence: Advertising and Subscription Design on LLM Platforms},pdfauthor={Yuhang Ji, Fasheng Xu, Tianjun Feng},pdfkeywords={Generative AI, LLM platforms, advertising, subscription design, mechanism design, consumer welfare}}

\begin{document}
\RUNAUTHOR{Ji, Xu, and Feng}
\RUNTITLE{Advertising and Subscription on LLM Platforms}
\TITLE{Monetizing Intelligence:\\ Advertising and Subscription Design on LLM Platforms}
\ARTICLEAUTHORS{%
\AUTHOR{Yuhang Ji$^1$ ~~~ Fasheng Xu$^2$ ~~~ Tianjun Feng$^1$\\[-2mm]}
\AFF{$^1$School of Management, Fudan University}
\AFF{$^2$School of Business, University of Connecticut}
}

\ABSTRACT{%
Large language model (LLM) platforms introducing advertising must decide how much capability ad revenue should fund and what consumers should pay to avoid ads.
We study a monopolistic platform that jointly chooses costly response quality, advertising intensity, and subscription prices when advertising reduces user value but better service raises advertising returns.
Three findings follow.
First, when partly ad-supported and ad-free plans coexist, advertising can finance useful service for consumers who receive none under pure subscription.
Lower effective quality in middle plans weakens the alternatives available to premium users, allowing higher prices for unchanged ad-free service.
Some fully ad-supported users gain utility, while partly ad-supported and ad-free users lose: broader access can coexist with \textit{utility polarization}.
Second, an intermediate advertising cap can maximize consumer surplus by preserving some revenue that finances service while limiting degradation.
Third, cheaper compute can raise quality and reduce ad intensity, yet sufficiently large cost reductions can increase aggregate advertising revenue above its initial level because better service raises the value of the remaining advertised interactions.
An extension shows how advertising-funded upgrades can bring plans to the same maximum capability and advertising exposure, eliminating the basis for different prices.
These results make menu design central to evaluating LLM advertising: the consequences extend beyond users who see ads to the alternatives and prices faced by those who do not.
}%
\HISTORY{Current version: October, 2026}
\maketitle

\vspace{-8mm}
\section{Introduction}

An LLM platform that introduces advertising must decide how much capability to provide on its ad-supported plans and what users must pay to avoid ads.
OpenAI's February 2026 launch description makes this choice concrete: users could avoid ads by upgrading to Plus or Pro, or by accepting fewer daily messages on the Free tier \citep{openai2026ads}.
Advertising exposure, usage capacity, and subscription price are therefore linked choices.
The financial stakes are substantial: on August 31, 2026, OpenAI reported that ChatGPT Ads had reached \$1 billion in annualized revenue run rate \citep{openai2026adsmilestone}.
Microsoft's Copilot-native ad formats and Meta's use of AI interactions to personalize advertising illustrate other ways to monetize conversational activity \citep{microsoft2025copilotads,meta2025recommendations}.
For a platform offering both subscriptions and advertising, the central design problem is how to use ad revenue to fund service while preserving users' incentives to pay for higher tiers.

Advertising can reduce the share of an LLM's supplied capability that becomes useful assistance.
For example, a sponsored product comparison may favor a product that fits the user's needs less well, while concerns about commercial influence may weaken trust in the response.
A recent experimental preprint finds that advertising disclosure attached to a product users explicitly seek reduces trust in the AI by lowering its perceived neutrality \citep{qin2026when}.
Privacy concerns can also lead users to withhold context that the assistant needs to tailor its advice.
Consistent with this concern, a survey finds that some AI-assistant users say they would become less willing to use personalization if advertising were introduced \citep{zappi2026consumers}.
These channels create a trade-off: advertising can help finance greater capability while reducing the value consumers obtain from it.

The same capability that makes an LLM useful to consumers can also make its advertising more valuable.
In a product comparison, for example, a more capable model or greater inference effort can better interpret the user's requirements and identify relevant alternatives, improving the information available for matching ads.
Google reports using both the query and generated-answer content when serving ads in AI Overviews, while Microsoft describes using conversational signals to improve ad relevance \citep{google2026aioverviewads,microsoft2025copilotads}.
These practices motivate our focus on the advertising value of better service.
Yet improving an ad-supported plan also gives premium users a more attractive alternative, potentially reducing what they will pay to avoid ads.
The platform must therefore choose service quality, advertising exposure, and subscription prices together.

Advertising in AI assistants has already drawn legislative scrutiny.
In 2026, U.S. Senator Edward Markey questioned seven major AI companies about advertising safeguards and introduced the \textit{Youth AI Privacy Act}, which would ban chatbot advertising to minors \citep{markey2026youthai}.
We examine the economic trade-off behind restrictions on advertising through a cap on exposure: reducing commercial interference also restricts revenue that can make quality provision profitable.
Because the platform can respond by changing quality and subscription prices, the cap's benefit to consumers depends on how the entire menu adjusts.
Technological progress changes these incentives from the cost side: inference prices have fallen sharply for models reaching comparable benchmark performance \citep{maslej2025aiindex}.
As quality becomes cheaper to supply, platforms can improve service, reduce advertising, or combine the two.
Better assistance can make advertising more valuable, but it also increases the value users lose when ads interfere with the service.
We ask: \textit{How should an LLM platform combine advertising and subscriptions, and how do advertising limits and cheaper compute change its choices and consumers' benefits?}

We study a monopolistic LLM platform that commits to a menu of service plans, each specifying response quality, advertising intensity, and a subscription price.
Consumers privately know how much they value quality and choose their preferred plan or leave the platform.
Quality reflects the capability supplied through model choice, reasoning effort, or usage capacity, and improving it requires computational resources.
The platform earns revenue from advertising exposure and an additional return when better service makes that advertising more valuable.
Advertising also reduces the value consumers obtain from the service.
Our analysis begins with a menu in which advertised and ad-free plans coexist and yields three main findings.

First, advertising can broaden access to useful service while leaving other consumers, including ad-free subscribers, worse off.
Additional advertising revenue can make quality provision profitable for consumers who would receive no useful service under pure subscription, even after accounting for production costs and the potential loss of premium revenue when a better entry plan encourages users to switch down.
We call this \textit{ad-funded quality democratization}.
Yet the benefit does not extend throughout the menu.
When partially advertised and ad-free plans coexist, middle plans provide lower quality than under pure subscription: advertising reduces consumers' willingness to pay for quality, and the added advertising return does not compensate for that loss.
The weaker middle plans make premium subscribers less willing to switch down, allowing the platform to charge more for unchanged ad-free service.
Some fully advertised users consequently gain, while users on partially advertised and ad-free plans lose utility relative to pure subscription, producing \textit{utility polarization}.

Second, a moderate advertising cap can maximize consumers' total surplus, yielding a better outcome than either unrestricted advertising or a complete ban.
Limiting ads can make lower-tier plans more useful.
When these plans become better alternatives to premium service, the platform must leave higher-valuation users more surplus to keep them from switching down.
However, tighter limits also weaken the advertising return that encourages the platform to supply quality, and a complete ban removes this source of revenue.
We identify settings in which consumers do best when advertising is limited enough to improve useful service but still allowed to help finance it.

Third, cheaper compute can support better service with less advertising, while advertising revenue can still rise.
As quality becomes less costly, the platform supplies better service.
For users on partially advertised plans, advertising then interferes with a more valuable service, inducing the platform to reduce exposure and extend ad-free access to more users.
At the same time, better capability can raise the advertising value of users who continue to receive ads.
With sufficiently large efficiency gains, the additional advertising value generated by fully advertised users who already receive useful service can outweigh revenue lost as other users see fewer ads, raising total advertising revenue above its initial level.
This \textit{volume-to-value pivot} shows why lower advertising exposure need not imply a smaller advertising business.

{
An extension shows how higher advertising returns to service quality can first expand the service menu and eventually lead tiers to merge.
Initially, additional advertising revenue makes it profitable to offer useful service to the lowest-valuation consumers, creating a distinct quality tier.
As returns rise further, however, advertising-funded upgrades can bring all groups to the same technological quality ceiling and full advertising exposure.
Because consumers would choose the cheaper of otherwise identical plans, these tiers cannot sustain different prices and thus merge.
}

Our contribution is to explain how the dual role of produced quality shapes an advertising-supported subscription menu.
Classic media models emphasize the trade-off between advertising nuisance and audience size \citep{anderson2005market,peitz2008content}.
Related screening work allows costly quality to influence advertising returns and, in a separate extension, allows advertising nuisance to depend on quality \citep{ichihashi2025mechanism}; other work optimizes subscription fees and ad intensities with endogenous consumption \citep{gutt2026hit}.
Recent LLM-pricing models study compute allocation and access through token budgets or service caps \citep{bergemann2026menupricinglargelanguage,zhong2025token}.
We bring both quality-dependent advertising forces into the platform's joint choice of response quality for each consumer, advertising intensity, and price: better quality raises advertising returns, while advertising reduces the user value of that quality.
The same quality choice therefore changes the value of the service to its intended consumer, the advertising revenue it generates, and the attractiveness of that plan to consumers elsewhere in the menu.
Tracing these effects together connects quality provision at the bottom to the prices and utility of users above, including those who receive no advertising.

\section{Literature Review}\label{sec:literature}

Our analysis connects three streams of work: monopoly screening and nonlinear pricing, the economics of ad-supported media platforms, and the emerging economics of AI platforms. We review each stream in turn and then position the paper against its closest antecedents.

The first stream is monopoly screening and nonlinear pricing.
\citet{mussa1978monopoly} study quality differentiation under private information, while \citet{myerson1981optimal} develops the virtual-valuation approach to optimal mechanism design.
Related work extends nonlinear pricing to multiproduct settings \citep{armstrong1996multiproduct}, random participation \citep{rochet2002nonlinear}, and information goods with negligible marginal cost \citep{sundararajan2004nonlinear}.
Two-sided platform theory explains how prices balance participation across market sides \citep{rochet2003platform}; subsequent work studies second-degree price discrimination by two-sided monopoly platforms \citep{jeon2022second} and screening through content access and advertising exposure \citep{ichihashi2025mechanism}.
\citet{corrao2023nonlinear} study nonlinear pricing when usage generates additional seller revenue and consumers may leave purchased services unused, with advertising among the applications.
\citet{alaei2023subscription} optimize subscription bundles and fees for digital goods consumed by users with heterogeneous usage rates.
Our analysis jointly chooses costly service quality and advertising exposure for each plan, with quality affecting both subscription payments and advertising revenue.

The second stream studies advertising-financed media, beginning with the trade-off between ad nuisance and content provision in broadcasting \citep{anderson2005market,gabszewicz2004programming} and the comparison of advertising and subscription business models \citep{peitz2008content}; \citet{anderson2015advertising} survey this literature.
Subsequent work examines ad targeting \citep{bergemann2011targeting}, two-sided price discrimination by media platforms \citep{lin2020two}, design biases in ad-funded markets \citep{choi2023platform}, attention oligopoly \citep{prat2022attention}, and how revenue models shape content moderation \citep{liu2022implications}.
\citet{sato2019freemium} studies freemium as an optimal menu-pricing problem.
These studies establish how advertising incentives interact with participation, content provision, and pricing.
Our focus is quality produced for each consumer: allocating more capability changes the service delivered, the advertising returns it generates, and the attractiveness of that plan to users elsewhere in the menu.
The closest-paper comparison below identifies how this joint design problem differs from existing quality and advertising extensions.

The third stream studies the economics of AI and LLM platforms.
Empirical work documents differentiation and price dispersion among LLMs \citep{demirer2025emerging}, motivating models with endogenous service quality.
\citet{bergemann2026menupricinglargelanguage} derive token-budget menus from a homogeneous production technology, while \citet{zhong2025token} studies screening through service caps when consumers differ in urgency; neither includes advertising.
On the advertising side, \citet{duetting2024mechanism} design token-level auctions for generated content, while \citet{wang2025sponsored} study platform-controlled ad content and traffic across sequential sponsored-search positions.
\citet{alaei2026dynamic} study ad timing and advertiser selection as conversations reveal user preferences, designing mechanisms for privately informed advertisers.
\citet{zhang2026economic} choose whether to show ads on each query, balancing current advertising returns against future retention and subscription conversion.
Our question concerns consumer screening through response quality, advertising intensity, and subscription prices when quality affects both advertising returns and the user value lost to advertising.

Empirical and computational evidence also documents commercial value in adjacent AI-mediated settings.
Organic LLM referrals exhibit distinct e-commerce performance patterns \citep{kaiser2026chatgpt}, numerical experiments using field data show that LLM-assisted experimentation can improve content allocation \citep{ye2025lola}, and GenAI-personalized video advertisements can increase engagement \citep{kumar2025generative}.
These studies do not directly identify how raising the quality supplied to a consumer changes advertising returns; we therefore treat a positive quality-driven advertising premium as a modeling primitive motivated by the platform evidence summarized in Section~\ref{sec:model}.

Closely related studies examine complementary dimensions of this joint design problem.
\citet{devalve2022menus} characterize subscription-price and advertising menus under monopoly and competition when consumers privately differ in advertising disutility.
Our platform also chooses costly response quality, which affects advertising returns and the value consumers lose to advertising.
\citet{ichihashi2025mechanism} study screening through content access and advertising exposure, with broader access raising both advertising revenue and information rents.
Their extensions allow costly item quality to increase advertising revenue and, separately, allow advertising nuisance to depend on item quality; these channels are used to study quality investment and item--ad matching, respectively.
\citet{gutt2026hit} optimize a continuous menu of ad intensities and subscription fees with endogenous user consumption, while holding content quality fixed.
Their baseline fixes per-impression ad value; an extension lets it decline with ad load, rather than linking it to platform-produced response quality.

We study costly quality supplied to each consumer, jointly chosen with advertising intensity and price.
Higher quality raises advertising returns, while advertising reduces the user value of that quality, so both channels enter the platform's screening problem.
Improving one plan therefore changes its revenue potential and its attractiveness as an alternative to other plans.
The contribution is to trace these effects jointly through quality provision, information rents, and prices across the menu, including on plans that carry no advertising.

\section{Model}\label{sec:model}

We consider a monopolistic Large Language Model (LLM) platform serving a unit mass of consumers who differ in their valuation of response quality.
The platform chooses the quality, advertising intensity, and subscription price of each plan, anticipating which plan each consumer will select.
Higher quality is costly to produce, but it can support both subscription payments and advertising revenue.
Advertising provides another source of revenue while reducing the effective quality consumers receive.

\subsection{Consumers}

Each consumer has a privately known type $\theta\in[0,1]$, drawn from a cumulative distribution function $F$ with a continuous and strictly positive density $f$ on $[0,1]$.
Type $\theta$ is the consumer's marginal valuation of effective response quality.
Improving a lower-quality plan makes it more attractive to consumers who would otherwise pay for a higher-quality plan.
To keep these consumers on the higher-quality plan, the platform may need to lower its price.
Providing better service to one group can therefore reduce subscription revenue from another.
Virtual value adjusts a consumer's valuation for this indirect revenue cost and can be negative even when the consumer values the service.
We impose the standard regularity condition that virtual value increases with type:\footnote{The continuous-type model is a tractable baseline. Section~\ref{sec:tier_design} studies a discrete-type formulation and shows that advertising can non-monotonically change the number of positive-quality tiers.}

\begin{assumption}\label{ass:regularity}
The virtual value $v(\theta)=\theta-\frac{1-F(\theta)}{f(\theta)}$ is strictly increasing and continuous.
\end{assumption}

Under this assumption, virtual value crosses zero at a unique type $\theta_0\in(0,1)$.
In the corresponding pure-subscription benchmark, the platform supplies no positive quality below this threshold: the screening cost outweighs the subscription return.
Advertising can change that decision.
If improving service also raises advertising revenue, the platform may profitably supply positive quality to some of these consumers and may also collect subscription payments.

\subsection{Platform's Instruments}\label{sec:instruments}

The platform offers a menu of plans, $\{q(\theta),\alpha(\theta),T(\theta)\}_{\theta\in[0,1]}$.
Each plan specifies response quality, advertising intensity, and a subscription price.
Here, $q(\theta)\ge0$ is a scalar index of the response capability supplied to type $\theta$.
The platform can raise this index through model choice, inference effort, or usage capacity; these operational choices are summarized by the cost function introduced below.
The subscription price is $T(\theta)\ge0$.
Advertising intensity is $\alpha(\theta)\in[0,1]$.
At $\alpha=0$, the consumer experiences no advertising-induced degradation; under the on-interface interpretation, the plan is ad-free.
Positive values of $\alpha$ represent greater advertising intensity, up to the architectural limit $\alpha=1$.
Effective quality $g(q,\alpha)$ measures the service quality the consumer experiences after advertising-induced degradation.

OpenAI's original February 2026 launch description illustrates the joint design of these instruments: users could avoid ads by upgrading to a paid plan or by opting out of ads in the Free tier in exchange for fewer daily free messages \citep{openai2026ads}.
This menu trades advertising exposure against either subscription payments or usage capacity, consistent with our interpretation of quality as including usage allowances.

Consumers benefit from supplied capability $q$ only when it produces assistance they can use and rely on.
Advertising can reduce the share of this capability that becomes useful assistance through two channels.
First, advertising can weaken the usefulness or credibility of a response.
Consider a consumer asking an AI assistant to compare software products.
If sponsorship steers the recommendation toward a product that fits the consumer's needs less well, the same capability delivers less useful advice.
Such \textit{output interference} can arise through the selection, ranking, or framing of recommendations.
{
Even when ads do not influence response content, as OpenAI states for ChatGPT (\citealp{openai2026ads}), concerns about commercial influence may make the consumer less willing to rely on it.
}
Evidence on perceived credibility comes from \citet{qin2026when}, who find that advertising disclosure attached to a product users explicitly seek can reduce trust in the AI by lowering perceived neutrality.

Second, \textit{privacy friction} can limit the information available to the assistant.
For example, a business user seeking advice on improving a workflow may omit internal procedures or operational constraints out of concern that conversational data will be used for advertising.
The assistant then has less context with which to tailor its advice, leaving some benefits of its capability unrealized.
Context-based advertising makes this concern relevant: OpenAI uses conversational context to match ads, while Meta uses interactions with its AI features to personalize advertising across its apps \citep{openai2026ads,openai2026adsmilestone,meta2025recommendations}.
Consistent with the possibility of reduced engagement, 27\% of respondents in a survey of 1,000 U.S. AI-assistant users say they would become less willing to use personalization after advertising is introduced \citep{zappi2026consumers}.

We assume that advertising-induced losses in effective quality are weakly larger at higher supplied quality, reflecting complementarity between capability, useful context, and consumer reliance.
Losing a given fraction of useful assistance, for example, entails a larger absolute loss when the service has greater potential value.

\subsection{Quality Production}\label{sec:quality_production}

The platform incurs computational costs when it generates responses for a consumer.
Raising supplied quality $q$ can require a more capable model, more inference tokens, or a larger usage allowance on the same model.
We summarize the cost of providing this service by $c(q)$ and impose the following conditions.

\begin{assumption}[Cost Function]\label{ass:cost}
The cost function $c: \mathbb{R}_+ \rightarrow \mathbb{R}_+$ is twice continuously differentiable with $c(0)=0$, $c'(0)=0$, $c'(q)>0$ for $q>0$, $c''(q)>0$, and $\lim_{q \rightarrow \infty} c'(q)=\infty$.
\end{assumption}

Strict convexity means that each additional unit of quality costs more to provide.
This assumption summarizes two ways the platform can improve service.
First, routing users to more capable models can entail higher inference costs.\footnote{\citet{demirer2025emerging} document a nonlinear relationship between model intelligence and API prices, with incremental intelligence improvements associated with larger price increases at the upper end. This pricing evidence motivates costly quality upgrading but does not identify the curvature of the platform's production costs.}
Second, it can allocate more reasoning effort within a model.
If additional inference yields progressively smaller quality gains, achieving a further quality improvement requires more computation.
These considerations motivate convexity; they do not imply that every operational route to higher quality has the same cost curvature.
We use continuous quality as a tractable approximation to model choice, inference effort, and usage capacity, consistent with standard vertical differentiation models \citep{mussa1978monopoly}.
Section~\ref{sec:tier_design} examines the core economic forces in a discrete-type formulation.

Quality is bounded by a technological ceiling, $q\le\bar q$, representing limits on the capability the platform can supply.
We set $c'(\bar q)=1$, so the ceiling equals the efficient quality for the highest-type consumer in an ad-free benchmark.
This identity is a maintained link between production costs and feasible capability.
Under the main specification and its regularity restrictions, the ceiling does not constrain quality for a positive mass of consumers.
Section~\ref{subsec:beta} allows the ceiling to expand as compute costs fall, while Section~\ref{sec:tier_design} examines tier pooling when it binds.

\subsection{Consumer Utility}

A type-$\theta$ consumer assigned quality $q$, advertising intensity $\alpha$, and transfer $T$ obtains utility
\begin{equation}\label{eq:utility}
U = \theta g(q,\alpha) - T,
\end{equation}
where $g(q,\alpha)$ is the \textit{effective quality} the consumer experiences after advertising-induced degradation.
The single private type $\theta$ determines both the value of this quality and the utility loss from reducing it.
For a given plan with a positive advertising-induced quality loss, consumers who value quality more also lose more utility from that loss.
The baseline has no separate dimension of private ad aversion.
Output interference and privacy friction motivate the following restrictions on effective quality.

\begin{assumption}[Effective Quality]\label{ass:effective_quality}
The function $g: \R_+ \times [0,1] \to \R_+$ satisfies:
\begin{enumerate}[$(a)$]
\item $g_q(q,\alpha)>0$: Higher quality improves the user experience.
\item $g_\alpha(q,\alpha)<0$ for $q>0$: Advertising degrades perceived quality.
\item $g_{q\alpha}(q,\alpha)\leq0$: Advertising weakly reduces the marginal value of quality.
\item $g(0,\alpha)=0$ for all $\alpha$, and $g(q,0)=q$.
\end{enumerate}
\end{assumption}

Condition (c) says that advertising weakly reduces the marginal value of supplied quality.
Consequently, the effective-quality loss $g(q,0)-g(q,\alpha)$ is non-decreasing in $q$ for a fixed advertising intensity.
The condition allows advertising to destroy more potential value when greater capability depends on trust or useful context.
Because consumers differ in their valuation of effective quality, changing a plan's ad load also changes its attractiveness relative to other plans.
Advertising therefore enters both the platform's revenue calculation and the incentive constraints governing plan choice.

\subsection{Advertising Revenue}

The platform earns advertising revenue $R(q,\alpha)$ from a consumer receiving supplied quality $q$ and advertising intensity $\alpha$.
We allow better service to make advertising more valuable by supporting richer interactions and more informative signals of commercial intent.
This \textit{quality-driven ad premium} gives the platform a return to quality in addition to subscription payments.

\begin{assumption}\label{ass:ad_revenue}
The advertising revenue function $R: \mathbb{R}_+ \times [0,1] \rightarrow \mathbb{R}_+$ is twice continuously differentiable with $R(q,0)=0$, $R_\alpha(q,\alpha)>0$ for $\alpha<1$, $R_{\alpha\alpha}(q,\alpha)<0$, and $R_q(q,\alpha)>0$ for $\alpha>0$.
\end{assumption}

The condition $R_\alpha>0$ for $\alpha<1$ gives the platform a direct return to additional advertising, while $R_{\alpha\alpha}<0$ imposes diminishing marginal returns.
The condition $R_q>0$ for $\alpha>0$ means that better supplied quality increases the revenue from a given positive advertising intensity.
For example, better matching can increase the probability that an ad is clicked; at a fixed payment per click, this raises expected revenue without increasing advertising exposure.
Google reports that both the user query and the content of an AI Overview enter ad serving, while Microsoft attributes higher Copilot ad-relevance metrics to the richer signals contained in entire conversations \citep{google2026aioverviewads,microsoft2025copilotads}.
OpenAI similarly reports using conversational context to match advertisements and states that CPC and outcome-optimized bidding account for most campaigns \citep{openai2026adsmilestone}.
{
These practices motivate our assumption that better service raises advertising returns, although they do not directly establish this causal relationship.
}

With quality-linked advertising revenue, improving a plan changes both consumers' willingness to pay and the advertising value of their interactions.
Advertising-induced degradation (Assumption~\ref{ass:effective_quality}) and convex quality costs (Assumption~\ref{ass:cost}) therefore enter the same design problem.
Sections~\ref{sec:optimal_mechanism}--\ref{sec:tier_design} examine how the platform balances these forces across the menu.

\subsection{Mechanism Design Problem}

The platform chooses measurable quality allocations $q(\theta)\in[0,\bar q]$, advertising intensities $\alpha(\theta)\in[0,1]$, and subscription payments $T(\theta)\ge0$ to maximize expected profit from subscription revenue and advertising revenue net of quality-production costs:
\begin{equation}\label{eq:platform_problem}
\max_{q(\cdot), \alpha(\cdot), T(\cdot)}
\int_0^1 \left[ T(\theta) + R(q(\theta), \alpha(\theta)) - c(q(\theta)) \right] dF(\theta)
\end{equation}
subject to incentive compatibility (IC), so that each consumer weakly prefers her intended plan to every other plan, and individual rationality (IR), so that she weakly prefers it to the zero-utility outside option:
\begin{align}
\text{(IC)} \quad
& \theta g(q(\theta), \alpha(\theta)) - T(\theta)
\geq
\theta g(q(\theta'), \alpha(\theta')) - T(\theta'),
\quad \forall \theta,\theta' \in [0,1],
\label{eq:IC}\\
\text{(IR)} \quad
& \theta g(q(\theta), \alpha(\theta)) - T(\theta) \geq 0,
\quad \forall \theta \in [0,1].
\label{eq:IR}
\end{align}

\paragraph{Timing, Information, and Solution Concept.}
The platform publicly posts and commits to its menu.
Each consumer privately knows her type and selects a plan or the zero-utility outside option, participating when indifferent.
The distribution $F$ and functions $c$, $R$, and $g$ are common knowledge.
The platform then delivers the chosen allocation and collects payment, earning subscription and advertising revenue net of production cost.
By the revelation principle, we represent the problem as a direct mechanism satisfying \eqref{eq:IC} and \eqref{eq:IR}.
Full commitment is maintained; Appendix~\ref{sec:discussion_assumptions} discusses its role.

Aggregate consumer surplus is $CS:=\int_0^1[\theta g(q(\theta),\alpha(\theta))-T(\theta)]\,dF(\theta)$.
We define social welfare, $SW$, as consumer surplus plus the platform's expected profit in \eqref{eq:platform_problem}.
This measure includes the platform's advertising revenue; it does not separately account for advertiser surplus.

\subsection{Main Specification}\label{sec:Main Specification}

For an explicit characterization of the optimal menu, we use a uniform type distribution, $\theta\sim U[0,1]$.
Then $f(\theta)=1$, virtual value is $v(\theta)=2\theta-1$, and the virtual-zero type is $\theta_0=1/2$.
Computational cost is quadratic, $c(q)=\beta q^2/2$, where $\beta\in(0,1]$ is the compute-cost parameter.
Lower $\beta$ means that any given quality costs less to provide.
The maintained ceiling condition $c'(\bar q)=1$ then gives $\bar q=1/\beta$.
The baseline normalizes $\beta=1$ and hence $\bar q=1$; Section~\ref{subsec:beta} studies lower compute costs together with the implied expansion of the quality ceiling.

We capture the interaction between quality and advertising with a multiplicative function:
\begin{equation}\label{eq:multiplicative}
g(q, \alpha) = q(1 - \delta \alpha),
\end{equation}
where $\delta\in(0,1)$ measures the proportional quality loss per unit of advertising intensity.
Consumers experience full supplied quality $q$ when $\alpha=0$ and retain $q(1-\delta)$ when $\alpha=1$.
Thus, a fixed positive ad load removes the same fraction of supplied quality at every $q$, but a larger absolute amount when supplied quality is higher.
This reduced-form loss represents the output-interference and privacy-friction channels described in Section~\ref{sec:instruments}, without modeling either channel separately.

Finally, the advertising revenue is specified as:
\begin{equation}\label{eq:revenue}
R(q, \alpha) = \lambda q \alpha + \bar{r}\left(\alpha - \frac{\alpha^2}{2}\right),
\end{equation}
Here, $\lambda>0$ measures the advertising return to supplied quality: at a fixed positive ad load, higher $q$ raises the quality-linked component $\lambda q\alpha$.
The parameter $\bar r>0$ governs the quality-independent component, which has diminishing returns to advertising intensity and remains available at $q=0$.
At $q=0$, baseline revenue comes from impressions generated by consumers who participate in a minimal-capability ad-supported layer.
Consumers assigned $q=T=0$ obtain zero utility and participate under the indifference convention stated above.
This specification satisfies Assumption~\ref{ass:ad_revenue}.

A larger quality-driven premium $\lambda$ can make positive quality profitable for consumers who receive none under pure subscription.
Greater degradation $\delta$ lowers the intended consumer's willingness to pay but also makes the plan less attractive to higher-valuation consumers.
This can reduce the utility the platform must leave higher types to discourage them from choosing plans intended for lower types.
This utility, which arises because consumers' valuations are private, is their \textit{information rent}.
Baseline ad profitability $\bar r$ adds a quality-independent return, including at the zero-quality baseline.
Section~\ref{sec:optimal_mechanism} combines these effects to determine quality and advertising across the menu.

To ensure an endogenous ad-free premium tier, we impose the following parameter restrictions.

\begin{assumption}[Market Regularity]\label{ass:market_regularity}
$\delta \in (0, 1)$, $\lambda \in (0, \delta)$, and $\bar{r} \in (0, \delta - \lambda)$.
\end{assumption}

{
The bounds on $\lambda$ and $\bar r$ in Assumption~\ref{ass:market_regularity} keep advertising profitability low enough for an ad-free premium segment (e.g., ChatGPT Plus/Pro) to coexist with ad-supported segments.
Appendix~\ref{app:extended_parameters} characterizes the remaining parameter regions and their implications, with even the highest type receiving advertising when $\lambda+\bar r>\delta$.
Under the baseline normalization $\beta=1$, these restrictions also ensure that the quality ceiling does not bind for a positive mass of consumers.
}

Appendix~\ref{sec:discussion_assumptions} discusses the roles of the functional forms, one-dimensional private information, full commitment, and monopoly.
It also explains the consumer- and platform-side extensions, the limits of the reduced-form competition diagnostic, and the applicability of the framework beyond LLM platforms.

\section{Optimal Mechanism}\label{sec:optimal_mechanism}

The platform must decide how much quality advertising can profitably support and which consumers should be offered ad-free service.
At a given positive quality, more advertising brings revenue but weakens the plan's appeal to consumers who value quality, including premium users who could switch down.
Upgrading the plan instead raises its advertising return while making it a stronger alternative to premium service.
The return to either decision therefore includes its effect on subscription revenue elsewhere in the menu.
We use virtual-surplus maximization to combine these effects and derive the segmented menu in Theorem~\ref{thm:optimal_mechanism}.

\subsection{Reduction to Virtual Surplus}

Let $U(\theta):=\theta g(q(\theta),\alpha(\theta))-T(\theta)$ denote type $\theta$'s equilibrium utility.
Incentive compatibility links the utility offered to each type to the effective quality available to lower types.
This link determines transfers up to the lowest type's utility and allows us to express the platform's problem in terms of quality and advertising alone.

\begin{lemma}[Virtual Surplus Representation]\label{lem:envelope}\label{lem:virtual_surplus}
The platform's problem has the following representation:
\begin{enumerate}[$(a)$]
    \item \textit{Incentive compatibility.} In any incentive-compatible mechanism, $U(\theta)$ is absolutely continuous with $U'(\theta) = g(q(\theta), \alpha(\theta))$ almost everywhere, and effective quality $g(q(\theta), \alpha(\theta))$ is non-decreasing in $\theta$.
    \item \textit{Profit representation.} Individual rationality and nonnegative payments imply $U(0)=0$. The platform's problem is therefore equivalent to maximizing expected virtual surplus:
\begin{equation}\label{eq:vs_objective}
\max_{q(\cdot), \alpha(\cdot)} \int_0^1 \mathcal{V}(\theta; q(\theta), \alpha(\theta)) dF(\theta)
\end{equation}
subject to the original allocation bounds and the requirement that $g(q(\theta),\alpha(\theta))$ be non-decreasing in $\theta$, where
\begin{equation}\label{eq:pointwise_vs}
\mathcal{V}(\theta; q, \alpha) := \underbrace{v(\theta)g(q, \alpha)}_{\text{Virtual Quality Value}}
+ \underbrace{R(q, \alpha)}_{\text{Ad Revenue}} - \underbrace{c(q)}_{\text{Quality Cost}},
\end{equation}
and $v(\theta)=\theta-\frac{1-F(\theta)}{f(\theta)}$ is the virtual value.
\end{enumerate}
\end{lemma}

Part (a) expresses each subscription payment as the consumer's gross value of service less the utility required for incentive compatibility:
\begin{equation}\label{eq:transfer}
T(\theta) = \theta \cdot g(q(\theta), \alpha(\theta)) - U(0) -
\int_0^\theta g(q(t), \alpha(t)) dt.
\end{equation}
The lowest type has $U(0)=-T(0)$, so individual rationality and $T(0)\ge0$ force $U(0)=T(0)=0$.
The integral is therefore the consumer's information rent, which increases with the effective quality available to lower types.
Because $U'(\theta)=g(q(\theta),\alpha(\theta))\ge0$, higher types also satisfy individual rationality.
Substituting \eqref{eq:transfer} into the objective and integrating by parts yields part (b).

\subsection{Virtual Advertising Profit}

To isolate the advertising trade-off, decompose the pointwise virtual surplus into a pure-subscription baseline and an advertising-driven component:
\begin{equation}
\mathcal{V}(\theta; q, \alpha) = \underbrace{v(\theta)g(q, 0) - c(q)}_{\text{Pure Subscription Surplus}} + \underbrace{ R(q, \alpha) - v(\theta)(g(q, 0)-g(q, \alpha))}_{\text{Virtual Advertising Profit}},
\end{equation}
where $g(q,0)-g(q,\alpha)\ge0$ captures the effective-quality loss from advertising.
\begin{definition}[Virtual Advertising Profit]\label{def:virtual_ad_profit}
For type $\theta$, define
\begin{equation}\label{eq:virtual_ad_profit}
\rho_\theta(\alpha; q) := R(q, \alpha) - v(\theta)(g(q, 0) - g(q, \alpha)).
\end{equation}
\end{definition}

Virtual advertising profit measures the change in the platform's virtual surplus from adding advertising at fixed quality.
Expanding $v(\theta)=\theta-\frac{1-F(\theta)}{f(\theta)}$ separates three effects.
The platform earns advertising revenue $R(q,\alpha)$, but the effective-quality loss lowers type $\theta$'s willingness to pay by $\theta\bigl(g(q,0)-g(q,\alpha)\bigr)$.
At the same time, making this allocation less attractive to higher types reduces the information rents the platform must leave them, yielding the offsetting term $\frac{1-F(\theta)}{f(\theta)}\bigl(g(q,0)-g(q,\alpha)\bigr)$.
The last two effects combine into $-v(\theta)\bigl(g(q,0)-g(q,\alpha)\bigr)$.
For non-positive virtual values, the rent savings weakly outweigh the loss in willingness to pay, so degradation does not reduce virtual surplus.
For positive virtual values, willingness-to-pay losses dominate, creating a net cost that is weakly increasing in quality under Assumption~\ref{ass:effective_quality}(c).
This difference helps explain why advertising can fund quality for low types while inducing the platform to limit advertising on higher tiers.

\subsection{Pointwise Optimization}

We now use the baseline specification in Section~\ref{sec:Main Specification} and maintain Assumption~\ref{ass:market_regularity} throughout the subsequent analysis, unless otherwise stated.
In the pure-subscription benchmark ($\alpha\equiv0$), maximizing virtual surplus yields $q^{sub}(\theta)=\max\{0,2\theta-1\}$: types $\theta\le1/2$ receive zero quality, while higher types receive positive quality.
This allocation provides the benchmark for identifying the quality changes induced by advertising.
We first solve the hybrid problem type by type while retaining the allocation bounds and temporarily ignoring the requirement that effective quality be non-decreasing in type.\footnote{Under our specification, effective quality $g(q^*,\alpha^*)$ is non-decreasing (and strictly increasing wherever $q^*(\theta)>0$), as verified in the proof of Theorem~\ref{thm:optimal_mechanism}; non-monotonic alternatives would require standard ironing \citep{myerson1981optimal}.}

Virtual advertising profit is
$\rho_\theta(\alpha; q) = \lambda q\alpha + \bar{r}(\alpha - \frac{\alpha^2}{2}) - (2\theta - 1)q\delta\alpha$.
Maximizing the full pointwise virtual surplus in \eqref{eq:pointwise_vs}, including subscription value and production cost, gives the quality choice conditional on advertising intensity:
\begin{equation}\label{eq:foc_q}
    q^*(\theta, \alpha) = \max\{0, (2\theta - 1)(1 - \delta\alpha) + \lambda\alpha\}.
\end{equation}
The term $\lambda\alpha$ adds an advertising return to supplied quality, while $-(2\theta-1)\delta\alpha$ captures the change in its virtual value caused by degradation.
For negative virtual values, the latter term is positive because degradation reduces the information rents associated with providing quality to lower types.

For fixed quality, the platform adds advertising as long as marginal ad revenue exceeds the net degradation cost:
\begin{equation}\label{eq:foc_alpha}
\frac{\partial \rho_\theta}{\partial \alpha} = \underbrace{\lambda q +
\bar{r}(1 - \alpha)}_{\text{Marginal Ad Revenue}} -
\underbrace{(2\theta - 1)q\delta}_{\text{Net Degradation Cost}}.
\end{equation}

Consider first consumers with non-positive virtual type, $\theta\le1/2$.
For these consumers, the rent savings from ad-induced degradation weakly outweigh the loss in willingness to pay.
Together with direct advertising revenue, this makes the maximal ad load optimal.
Positive quality requires an additional incentive: the quality-driven ad premium $\lambda$ must be large enough to offset the negative marginal virtual value of serving these consumers.

\begin{proposition}[Advertising-Funded Quality]\label{prop:negative_types}
For every $\theta\in[0,1/2]$, the optimal advertising intensity is $\alpha^*(\theta)=1$, and the optimal quality is $q^*(\theta)=\max\{0,(2\theta-1)(1-\delta)+\lambda\}$.
In particular, every type $\theta<1/2$ satisfying $(2\theta-1)(1-\delta)+\lambda>0$ receives strictly positive quality despite having a negative virtual value.
\end{proposition}

Proposition~\ref{prop:negative_types} separates the incentive to show advertising from the incentive to supply positive quality.
Baseline ad profitability ($\bar r>0$) supports the maximal ad load for $\theta\le1/2$, even when quality is zero.
Supplying positive quality to negative-virtual-value consumers requires the quality-driven premium: because $R_q=\lambda\alpha>0$, improving their service also increases advertising revenue.
In the counterfactual with $\lambda=0$, advertising revenue is independent of quality, the positive-quality threshold returns to $\underline{\theta}=1/2$, and $q^*(\theta)=\max\{0,(2\theta-1)(1-\delta)\}=0$ for all $\theta<1/2$.
{
Conversely, in the boundary counterfactual $\bar r=0$, a positive $\lambda$ can still support positive quality for some consumers with negative virtual values.
}
Appendix~\ref{app:ext_concave_ad_premium} replaces $\lambda q\alpha$ with $\lambda h(q)\alpha$ and shows that advertising can still induce $q^*(\theta)>0$ for some consumers with negative virtual values.

{
Consumers with positive virtual type ($\theta>1/2$) all receive positive quality.
For fixed $q$, $\partial\mathcal V/\partial\alpha=\bar r(1-\alpha)-q\Delta(\theta)$, where $\Delta(\theta):=(2\theta-1)\delta-\lambda$.
Here, $q\Delta(\theta)$ measures the marginal cost of advertising to the platform, after accounting for the quality-driven ad return.
The platform retains the full advertising load when $\Delta(\theta)\le0$.
As $\theta$ rises, $\Delta(\theta)$ increases.
When $\Delta(\theta)>0$, the platform reduces advertising, with the following interior solutions whenever the optimal ad load remains positive:\footnote{See Appendix~\ref{app:proof_pointwise} for the derivations of \eqref{eq:foc_q} and \eqref{eq:interior_allocations}.}
}
\begin{equation}\label{eq:interior_allocations}
    \alpha^{int}(\theta) = \frac{\bar{r}-(2\theta - 1)\Delta(\theta)}{\bar{r}-\Delta(\theta)^2},
    \quad q^{int}(\theta) = \frac{\bar{r}((2\theta - 1)-\Delta(\theta))}{\bar{r}-\Delta(\theta)^2}.
\end{equation}

{
For the highest types, the marginal gain from adding advertising at their optimal ad-free quality $q=2\theta-1$ is non-positive, so the platform sets $\alpha^*=0$.
} Theorem~\ref{thm:optimal_mechanism} combines these choices with Proposition~\ref{prop:negative_types} to characterize the optimal menu.

\subsection{Characterization of the Optimal Mechanism}

The pointwise choices form a menu with three distinct boundaries: where positive quality begins, where the platform starts reducing the ad load, and where it removes advertising.

\begin{definition}[Critical Thresholds]\label{def:thresholds}
Define the following thresholds:
\begin{enumerate}[(1)]
    \item \textit{Positive-quality threshold}: $\underline{\theta}:=\frac{1}{2}-\frac{\lambda}{2(1-\delta)}$, above which consumer types receive positive quality.
    \item \textit{Partial-ad threshold}: $\theta_1:=\frac{1}{2}+\frac{\lambda}{2\delta}$, above which the platform reduces the ad load below one.
    \item \textit{Ad-free threshold}: $\theta_2\in(\theta_1,1)$ uniquely solving $(2\theta_2-1)[(2\theta_2-1)\delta-\lambda]=\bar r$, equivalently $\theta_2=\frac{2\delta+\lambda+\sqrt{\lambda^2+4\delta\bar r}}{4\delta}$, at and above which the platform eliminates advertising.
\end{enumerate}
\end{definition}

Using these thresholds, the optimal allocations are summarized as follows.

\begin{theorem}[Optimal Mechanism]\label{thm:optimal_mechanism}
The optimal mechanism partitions the market into up to four segments:
\begin{enumerate}[$(a)$]
    \item \textbf{Zero-quality baseline} ($\theta\in[0,\underline{\theta}]$, empty when $\underline{\theta}<0$): consumers receive $q^*(\theta)=0$, $\alpha^*(\theta)=1$, and $T^*(\theta)=0$.
    \item \textbf{Positive quality with full advertising (low types)} ($\theta\in[0,1]$ with $\underline{\theta}<\theta\le\theta_1$): consumers receive $q^*(\theta)=(2\theta-1)(1-\delta)+\lambda>0$ and $\alpha^*(\theta)=1$.
    Payments satisfy $T^*(\theta)>0$ for $\theta>0$, with $T^*(0)=0$ when $\underline{\theta}<0$.
    \item \textbf{Partial advertising (medium types)} ($\theta\in(\theta_1,\theta_2)$): consumers receive $q^*(\theta)=q^{int}(\theta)$ and $\alpha^*(\theta)=\alpha^{int}(\theta)\in(0,1)$, as defined in \eqref{eq:interior_allocations}.
    \item \textbf{Ad-free service (high types)} ($\theta\ge\theta_2$): consumers receive the pure-subscription monopoly quality $q^*(\theta)=2\theta-1$, face no advertising ($\alpha^*(\theta)=0$), and pay the highest subscription prices on the menu.
\end{enumerate}
Payments are recovered from \eqref{eq:transfer} with $U(0)=0$.
Effective quality $g(q^*(\theta),\alpha^*(\theta))$ is non-decreasing in type and strictly increasing wherever $q^*(\theta)>0$, so the pointwise allocations are implementable without ironing.
\end{theorem}

Relative to pure subscription, the optimal menu raises supplied quality on positive-quality, fully advertised plans below $\theta_1$.
Here, the quality-driven advertising return exceeds the virtual cost of degradation, so improving service is more profitable than under pure subscription.
This return can support positive quality for some consumers who would receive zero quality under pure subscription; fully advertised positive-quality plans can also carry subscription charges.
For middle types, some advertising remains profitable, but the degradation it causes lowers the net return to improving service, so the platform supplies less quality than under pure subscription.
The advertising return and virtual degradation cost balance at $\theta_1$, while supplied quality also matches the benchmark for ad-free consumers and consumers who continue to receive zero quality.
Evaluating who benefits additionally requires tracing subscription payments and information rents across the menu.

OpenAI's Go launch announcement paired higher usage allowances and longer memory than Free with planned advertising in Free and Go and ad-free Plus and Pro tiers \citep{openai2026chatgptgo,openai2026ads}.
These choices illustrate the separate roles of supplied capability and advertising exposure; they do not establish the predicted quality distortions relative to pure subscription.

\begin{remark}[Zero-Quality Baseline]\label{rem:zero_quality_baseline}
The allocation $(q^*,\alpha^*,T^*)=(0,1,0)$ represents the limiting case of a minimal-capability, ad-supported layer.
Zero quality means zero effective service quality; consumers obtain zero utility and participate under the maintained indifference convention, allowing the platform to earn baseline advertising revenue $R(0,1)=\bar r/2$.
Within the regular parameter region, this segment has positive mass when $\lambda<1-\delta$.
At $\lambda=1-\delta$, only the lowest type receives zero quality; when $\lambda>1-\delta$, the quality-driven advertising return supports positive quality for every type.
Appendix~\ref{app:supp_tables} tabulates the menu within this region, and Appendix~\ref{app:extended_parameters} analyzes the broader parameter regions.
\end{remark}

\section{How Advertising Reshapes Quality, Prices, and Welfare}\label{sec:comparative_statics}

Evaluating advertising requires tracking both the service consumers receive and what they pay after the platform redesigns its menu.
We first ask whether greater harm from advertising necessarily reduces supplied capability, and then whether more profitable impressions lower subscription prices (Sections~\ref{subsec:statics_delta} and~\ref{subsec:statics_rbar}).
These are parameter changes within the hybrid model.
Section~\ref{subsec:statics_lambda} instead compares the optimized hybrid menu with pure subscription to determine which users benefit, separating individual utility from aggregate consumer surplus and social welfare.

\subsection{Quality Responses to Ad Degradation ($\delta$)}\label{subsec:statics_delta}

\begin{figure}[htbp]
    \centering
    \subfloat[Quality $q^*(\theta)$]{
        \includegraphics[height=0.31\textwidth]{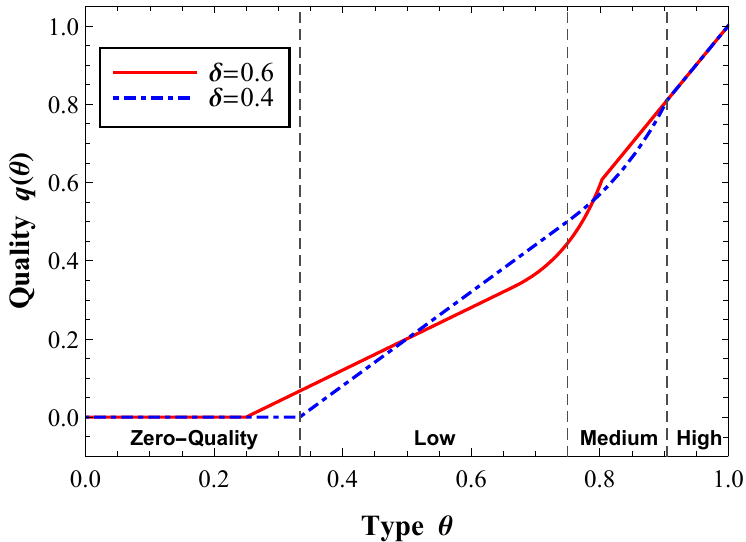}}\quad
    \subfloat[Ad Intensity $\alpha^*(\theta)$]{
        \includegraphics[height=0.31\textwidth]{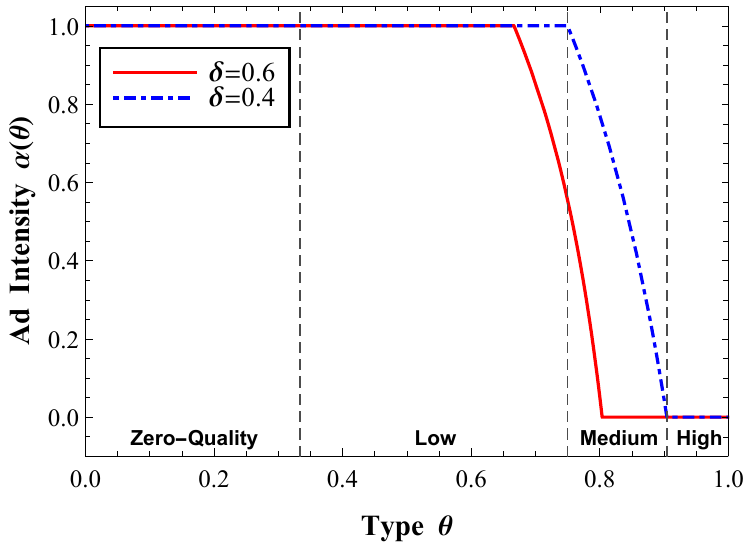}}

    \par\vspace{0.2cm}
    \begin{minipage}{1\textwidth}
        \centering \scriptsize \textit{Notes:} The vertical dashed lines mark, from left to right, $\theta=\underline{\theta}$, $\theta_1$, and $\theta_2$ under $\delta=0.4$; the segment labels also refer to this menu. The curves compare finite changes in $\delta$, so a given consumer may change segments.
    \end{minipage}
    \caption{The Impact of Ad Degradation ($\delta$) on Platform Allocations ($\bar{r} = 0.1$, $\lambda = 0.2$)}
    \label{fig:delta_impact}

\end{figure}

We first increase ad degradation $\delta$, holding the advertising-revenue parameters fixed.
On a positive-quality plan carrying advertising, greater degradation reduces effective service at a fixed supplied quality and ad load.
The platform can respond by changing both quality and advertising, so this direct loss does not determine the direction of its quality choice.
Proposition~\ref{prop:delta_quality} identifies the resulting local responses and the expansion of positive-quality provision when a zero-quality segment exists.
Figure~\ref{fig:delta_impact} illustrates how quality and advertising move together at the stated parameter values; its curves compare finite changes, whereas the proposition distinguishes local responses within the relevant regions.

\begin{proposition}[Quality Distortion]\label{prop:delta_quality}
A marginal increase in $\delta$, holding $\lambda$ and $\bar r$ fixed and remaining in the regular region, changes supplied quality as follows:
\begin{enumerate}[$(a)$]
    \item \textbf{Entry into positive quality:} The threshold $\underline{\theta}$ decreases. When it lies inside the type support, consumers between the new and old cutoffs gain positive quality, while those below the new cutoff remain at zero.
    \item \textbf{Positive-quality low types ($\alpha^*=1$):} For feasible types $\theta\in[0,1]$ with $\underline{\theta}<\theta\le\theta_1$, quality increases if $\theta<1/2$, is unchanged if $\theta=1/2$, and decreases if $\theta>1/2$.
    \item \textbf{Medium types ($\alpha^{int}\in(0,1)$):} Locally within this segment, quality decreases under a heavy ad load ($\alpha^{int}>1/2$), is unchanged to first order at $\alpha^{int}=1/2$, and increases under a light ad load ($\alpha^{int}<1/2$).
    \item \textbf{High types ($\alpha^*=0$):} Quality remains unchanged at $2\theta-1$ for consumers who remain ad-free.
\end{enumerate}
All increases and decreases in parts (b) and (c) are strict. At moving segment boundaries, the comparison concerns an increase in $\delta$, rather than asserting two-sided differentiability.
\end{proposition}

For fully advertised users below the median valuation, greater degradation makes their plans less attractive to higher types who might switch down.
The resulting rent savings lower the screening cost of supplying quality at the bottom.
Because the quality-driven premium $\lambda$ rewards that quality, the platform raises supplied quality for these users whenever it is already positive and expands positive-quality provision when the cutoff lies inside the type support.
This response requires $\lambda>0$: with quality-independent advertising ($\lambda=0$), below-median users continue to receive zero quality, as in Proposition~\ref{prop:negative_types}.
An increase in supplied quality is not itself a claim that effective quality or consumer utility rises.

For users above the median, greater degradation reduces the virtual return to quality at any fixed positive ad load.
For users who locally remain fully advertised, ad intensity stays at one, so the direct effect lowers supplied quality.
For medium types, reducing advertising can offset this direct effect: fewer ads expose less of the supplied capability to degradation and thereby raise the return to quality.
The precise cutoff follows from the baseline functional forms: at an interior optimum, the advertising condition is $q^{int}\Delta=\bar r(1-\alpha^{int})$.
Combining this condition with the quality choice makes the local quality response proportional to $(1-\alpha^{int})-\alpha^{int}$, with a positive multiplier.
Under a light ad load, the optimal reduction in advertising therefore more than offsets the direct loss, so quality rises.
Under a heavy ad load, the adjustment in advertising does not offset the direct loss, so quality falls; at an ad load of one-half, the effects exactly balance to first order.
For consumers who remain ad-free, degradation affects neither their allocation nor supplied quality.

The empirical prediction concerns supplied capability as well as advertising exposure.
Holding the other primitives fixed, greater degradation widens the supplied-quality gap between heavily advertised medium types who locally remain in that segment and users who remain ad-free.
Supplied quality can vary through usage caps, context limits, model routing, or feature access, so access to a newer model alone does not measure this gap.
Testing the prediction would require identifying a degradation change and accounting for concurrent changes in capabilities and costs; observing an upgraded entry plan alone does not establish the model's incentive.

\subsection{Divergent Pricing: Baseline Ad Profitability ($\bar{r}$)}\label{subsec:statics_rbar}

A higher return to impressions improves the revenue available from a given advertised plan, but does not determine how subscription prices change across the menu.
Consumers who pay to avoid advertising can therefore be affected even when their own service is unchanged.
We isolate this response by increasing baseline ad profitability $\bar r$, the quality-independent return to impressions, while holding $\lambda$ and $\delta$ fixed and allowing quality and advertising to adjust.
OpenAI's impression-based advertising option illustrates the billing unit \citep{openai2026adbasics}; separating this return from the quality-driven premium is a modeling assumption.

\begin{figure}[htbp]
    \centering
    \subfloat[Effective Quality $g(q^*, \alpha^*)$]{
        \includegraphics[height=0.31\textwidth]{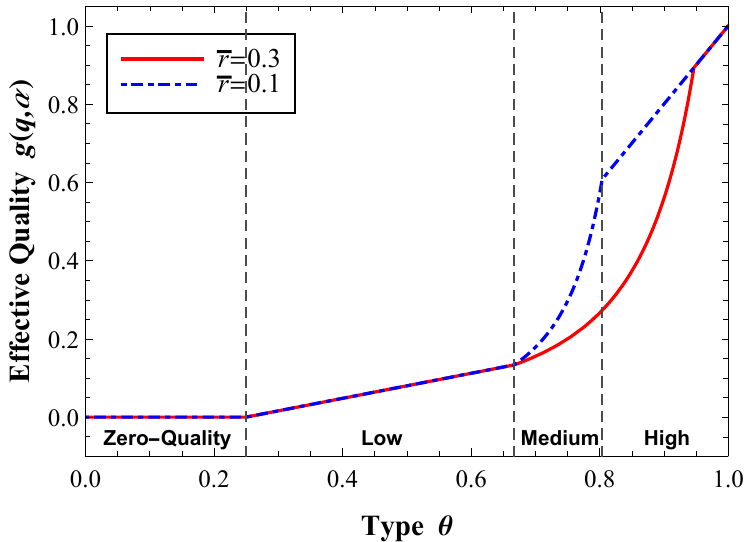}}\quad
    \subfloat[Price $T^*(\theta)$]{
        \includegraphics[height=0.31\textwidth]{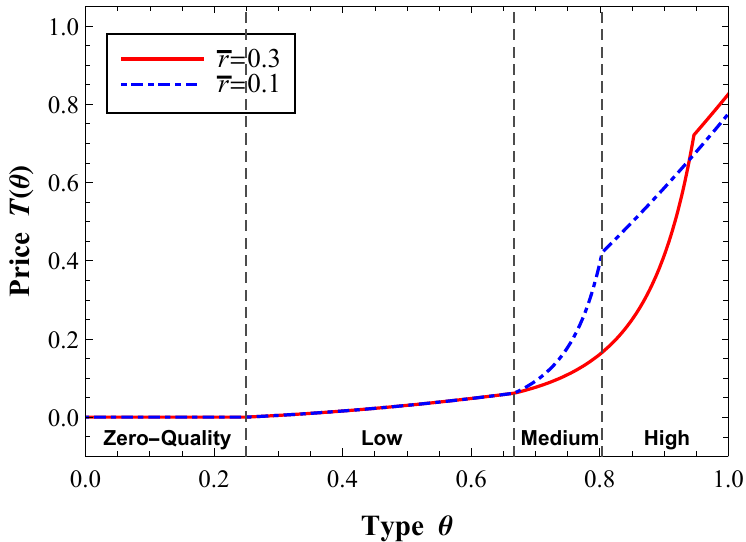}}
    \par\vspace{0.2cm}
    \begin{minipage}{1\textwidth}
        \centering \scriptsize \textit{Notes:} The vertical dashed lines mark, from left to right, $\theta=\underline{\theta}$, $\theta_1$, and $\theta_2$ under $\bar{r}=0.1$; the segment labels also refer to this menu. The curves compare finite changes in $\bar r$. The ad-free price increase concerns users who remain ad-free in both menus; users who cross a segment boundary need not share that price response.
    \end{minipage}
    \caption{The Impact of Ad Profitability ($\bar{r}$) on Platform Allocations and Pricing ($\lambda = 0.2$, $\delta = 0.6$)}
    \label{fig:rbar_impact}
\end{figure}

A higher $\bar r$ makes advertising more profitable, weakly raising ad intensity and weakly lowering supplied and effective quality across the market, as illustrated in Figure~\ref{fig:rbar_impact}(a).
Yet lower effective quality need not reduce every subscription price: the same menu adjustment changes the information rents required to keep higher types on their plans.
Proposition~\ref{prop:rbar_pricing} distinguishes users who locally remain in each segment.

\begin{proposition}[Divergent Pricing]\label{prop:rbar_pricing}
Consider a marginal increase in $\bar r$, holding $\lambda$ and $\delta$ fixed and remaining in the regular region.
For a fixed consumer type away from the segment thresholds:
\begin{enumerate}[$(a)$]
    \item \textbf{Zero-quality and low types ($\alpha^*=1$):} The subscription price is unchanged.
    \item \textbf{Medium types ($\alpha^{int}\in(0,1)$):} The subscription price strictly decreases.
    \item \textbf{High types ($\alpha^*=0$):} The subscription price strictly increases despite unchanged quality and advertising.
\end{enumerate}
\end{proposition}

The price equals gross utility minus the information rent left to the consumer: $T^*(\theta) = \theta \cdot g(q^*(\theta), \alpha^*(\theta)) - U(\theta)$.
Fully advertised plans carry the maximal ad load, and a higher $\bar r$ adds no marginal return to quality, so their allocations and those below them remain unchanged.
Both gross utility and information rent therefore stay fixed, leaving their prices unchanged.
For medium types, a higher $\bar r$ lowers both effective quality and information rents.
The effective-quality loss becomes larger at higher types within this segment, making each user's loss in gross utility exceed the reduction in her information rent; her incentive-compatible price therefore falls.
For users who remain ad-free, gross utility is unchanged, but the deterioration of lower tiers makes downward switching less attractive and reduces the information rent needed to keep them on their assigned tier.
The platform can therefore charge these users more for the same allocation.

The empirical counterpart is a change in the return to impressions, such as an identified CPM shock, holding compute costs, platform capabilities, and market structure fixed \citep{openai2026adbasics}.
For users who remain in their segments, the model predicts lower prices on partially advertised plans and higher prices on unchanged ad-free plans.
Quality and advertising are allowed to adjust; only segment membership must remain locally unchanged.
A change in the incremental advertising return to quality instead maps to $\lambda$, and a change in compute costs maps to $\beta$.
Observed prices and CPMs may reflect several of these forces, as well as competition and menu-adjustment costs, so their co-movement alone does not isolate the predicted cross-tier price divergence.

\subsection{Consumer Utility and Aggregate Welfare}\label{subsec:statics_lambda}

We now compare the optimized hybrid menu with the pure-subscription benchmark to ask which consumers gain after quality, advertising, and payments all adjust.
The hybrid menu has a positive quality-driven ad premium $\lambda$: supplying better service can make interactions more valuable for advertising.
Google's use of queries and AI Overview content in ad serving, Microsoft's use of conversational signals for ad relevance, and Meta's announced use of AI interactions to personalize advertising illustrate the role of interaction content \citep{google2026aioverviewads,microsoft2025copilotads,meta2025recommendations}.
They motivate the channel without establishing the marginal return to supplied quality or calibrating $\lambda$.

\begin{figure}[htbp]
	\centering
	\subfloat[Quality $q^*(\theta)$]{
		\includegraphics[height=0.31\textwidth]{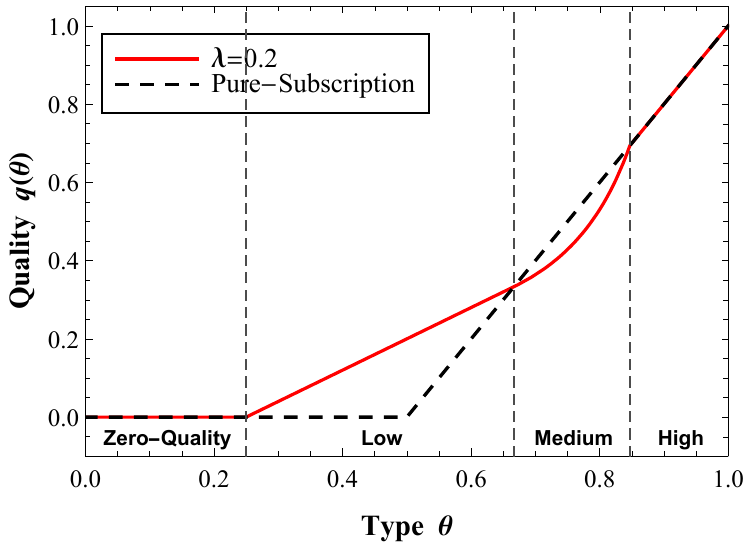}}\quad
	\subfloat[Consumer Utility $U(\theta)$]{
		\includegraphics[height=0.31\textwidth]{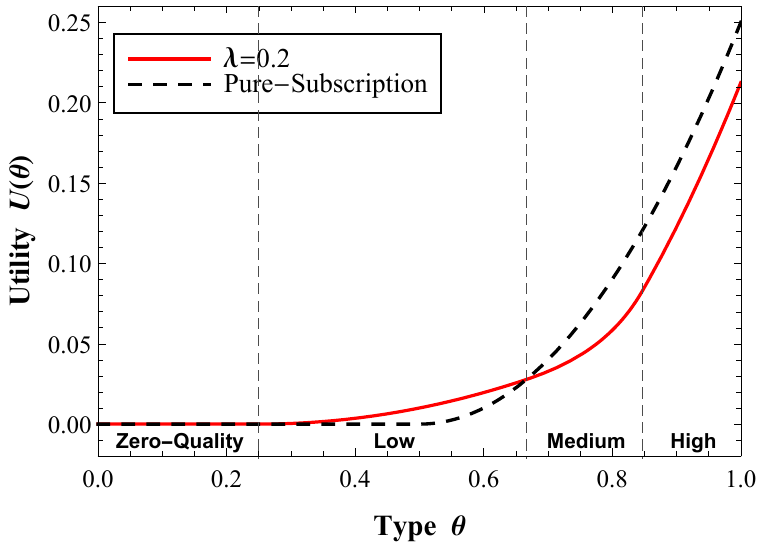}}
    \par\vspace{0.2cm}
    \begin{minipage}{1\textwidth}
        \centering \scriptsize \textit{Notes:} The vertical dashed lines mark, from left to right, $\theta=\underline{\theta}$, $\theta_1$, and $\theta_2$ in the hybrid menu; the segment labels also refer to that menu. Pure subscription imposes $\alpha\equiv0$.
    \end{minipage}
    \caption{Quality and Utility: Hybrid Monetization versus Pure Subscription ($\lambda=0.2$, $\bar{r}=0.15$, $\delta=0.6$)}
    \label{fig:lambda_impact}
\end{figure}

Figure~\ref{fig:lambda_impact} compares supplied quality and consumer utility with their pure-subscription levels.
The ad-free quality curves coincide at the top in panel (a), yet the utility curves remain separated in panel (b): a consumer's unchanged allocation does not guarantee unchanged utility when the rest of the menu changes.
Proposition~\ref{prop:lambda_utility} establishes which consumers gain or lose throughout the regular parameter region.

\begin{proposition}[Utility Polarization]\label{prop:lambda_utility}
Compare the optimal hybrid menu with $\lambda>0$ to the pure-subscription benchmark:
\begin{enumerate}[$(a)$]
    \item \textbf{Positive-quality low types ($\alpha^*=1$):} If $\lambda\le1-\delta$, every type $\theta\in(\max\{0,\underline{\theta}\},\theta_1)$ strictly gains utility, and utility equals the benchmark at $\theta_1$. If $\lambda>1-\delta$, only a subset of these types strictly gains, and utility at $\theta_1$ is strictly below the benchmark.
    \item \textbf{Medium and high types:} Every type in the partial-ad segment ($\alpha^{int}\in(0,1)$) and the ad-free segment ($\alpha^*=0$) strictly loses utility.
\end{enumerate}
Consumers who receive zero quality, and type $\theta=0$, have unchanged zero utility.
\end{proposition}

The utility comparison depends on effective quality along the menu, not supplied quality alone.
By the envelope condition, each type's information rent accumulates the effective quality offered below it.
On the fully advertised segment, this gives $U(\theta)=\bigl([q^*(\theta)]^2-[q^*(0)]^2\bigr)/4$.
When $\lambda\le1-\delta$, the lowest type receives zero quality, so $q^*(0)=0$.
The higher supplied quality of positive-quality low types below $\theta_1$ therefore translates into higher utility; at $\theta_1$, their quality matches the subscription benchmark and the utility gain vanishes.
When $\lambda>1-\delta$, however, even the lowest type receives positive quality, $q^*(0)=\lambda+\delta-1$, while still obtaining zero utility.
The rent formula must therefore subtract the positive term $[q^*(0)]^2/4$.
Because supplied quality matches the benchmark at $\theta_1$, this subtraction makes hybrid utility strictly lower there, leaving only a subset of positive-quality low types better off.

Above $\theta_1$, medium types receive lower effective quality than in the benchmark, so their utility increases more slowly with type and every medium type loses.
The ad-free top retains its benchmark allocation but inherits the lower information rent accumulated below it, allowing the platform to charge more for the same service.
Thus, advertising affects even consumers who see no ads through the quality and pricing of the alternatives available to them.

Strict convexity of quality costs supports the smooth, type-dependent quality choices in the baseline, but Remark~\ref{rmk:convex_necessity} in Appendix~\ref{app:supp_tables} explains why polarization need not require this cost shape.
With linear costs and an independently fixed finite technological ceiling, quality becomes a boundary choice rather than a smoothly varying allocation.
Advertising and information-rent effects can still generate utility polarization in that setting.

\paragraph{Welfare Implications.}

Aggregate consumer surplus adds the utility gains and losses identified in Proposition~\ref{prop:lambda_utility}.
It rises when the integrated gains at the bottom exceed the losses above, falls when the losses exceed the gains, and is unchanged when they balance.
Social welfare is defined here as consumer surplus plus platform profit: subscription payments cancel in this sum, while the value of effective service, advertising revenue, and quality costs remain.
In particular, ad-free consumers receive the same allocation as under pure subscription, so their utility losses are entirely higher payments that transfer surplus to the platform.
The remaining welfare comparison weighs advertising revenue and additional service at the bottom against changes in effective service and production costs on advertised plans.
The proof of Proposition~\ref{prop:lambda_utility} establishes that this balance is strictly positive throughout the regular region, so hybrid monetization raises social welfare even when aggregate consumer surplus falls.
It also establishes that social welfare strictly increases with $\lambda$, holding $\delta$ and $\bar r$ fixed and remaining in that region.

These comparisons take platform operation as given.
The compute-cost projections motivating the paper \citep{reuters_openai_2026} raise a separate viability question: if fixed infrastructure costs made pure subscription unprofitable, its optimized menu might not be an attainable market alternative.
Exit or other adjustments would then require additional analysis; the present polarization result compares consumer utility with the pure-subscription optimum and does not establish the comparison with platform exit.

Product and policy assessments therefore need to consider the distribution of gains across the menu as well as the capability of the entry plan.
An upgraded ad-supported plan can coexist with utility losses elsewhere, and higher social welfare can coexist with lower aggregate consumer surplus.
Table~\ref{tab:cs_summary} in Appendix~\ref{app:supp_tables} summarizes the segment-level results.
The next section asks when limits on advertising or lower quality costs can improve consumer outcomes while preserving the service that advertising helps finance.

\section{Advertising Caps and Compute Efficiency}\label{sec:policy}

Two practical decisions follow from the preceding analysis: how tightly to limit advertising to protect consumers, and how a platform should redesign its menu when quality becomes cheaper to supply.
At a given supplied quality, reducing advertising improves the user experience; the value of a cap also depends on the quality and prices the platform offers in response.
Cheaper compute changes the return to upgrading service and can also change which plans should carry advertising.
Both decisions must account for the alternatives available elsewhere in the menu, because changes to advertised plans affect the surplus left to premium users.

The advertising cap analysis retains $\beta=1$; the compute-efficiency analysis relaxes this normalization and allows the technological ceiling to expand as specified below.

\subsection{Advertising Caps and Consumer Surplus ($\bar{\alpha}$)}\label{subsec:ad_cap}

A platform introducing advertising into a conversational interface must decide how much commercial influence to allow while preserving users' trust in its responses.
OpenAI and Microsoft state that sponsored content should remain distinct from, and should not influence, organic responses \citep{openai2026ads, microsoft2025copilotads}.
Disclosure may not resolve the concern: users can fail to recognize sponsored recommendations embedded in chatbot responses even when disclosure is provided \citep{tang2025ads}.
Policy scrutiny now directly targets advertising in AI assistants.
In 2026, U.S. Senator Edward Markey questioned seven major AI companies about advertising safeguards and introduced the \textit{Youth AI Privacy Act}, which would ban chatbot advertising to minors \citep{markey2026youthai}.
We study the service-quality and pricing consequences of restricting advertising exposure by imposing an intensity cap, $\alpha(\theta)\leq\bar{\alpha}\in[0,1]$.
The cap lets us compare a complete ban with partial limits and unrestricted advertising while allowing the platform to redesign its menu.
Disclosure and separation policies can also affect how much a given ad load degrades user value; Section~\ref{subsec:statics_delta} studies changes in this second margin through $\delta$.

\begin{figure}[htbp]
	\centering
	\subfloat[Non-Monotonic $CS$]{
		\includegraphics[height=0.295\textwidth]{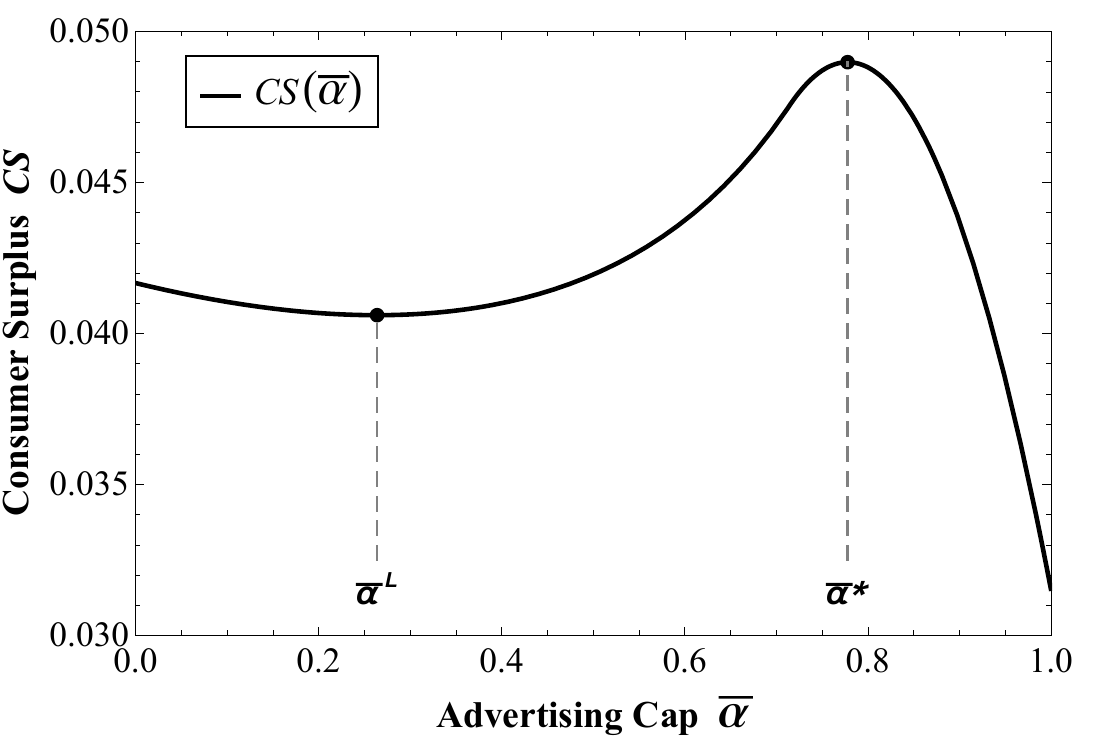}}\quad
	\subfloat[$CS(\bar{\alpha})-CS(1)$]{
		\includegraphics[height=0.295\textwidth]{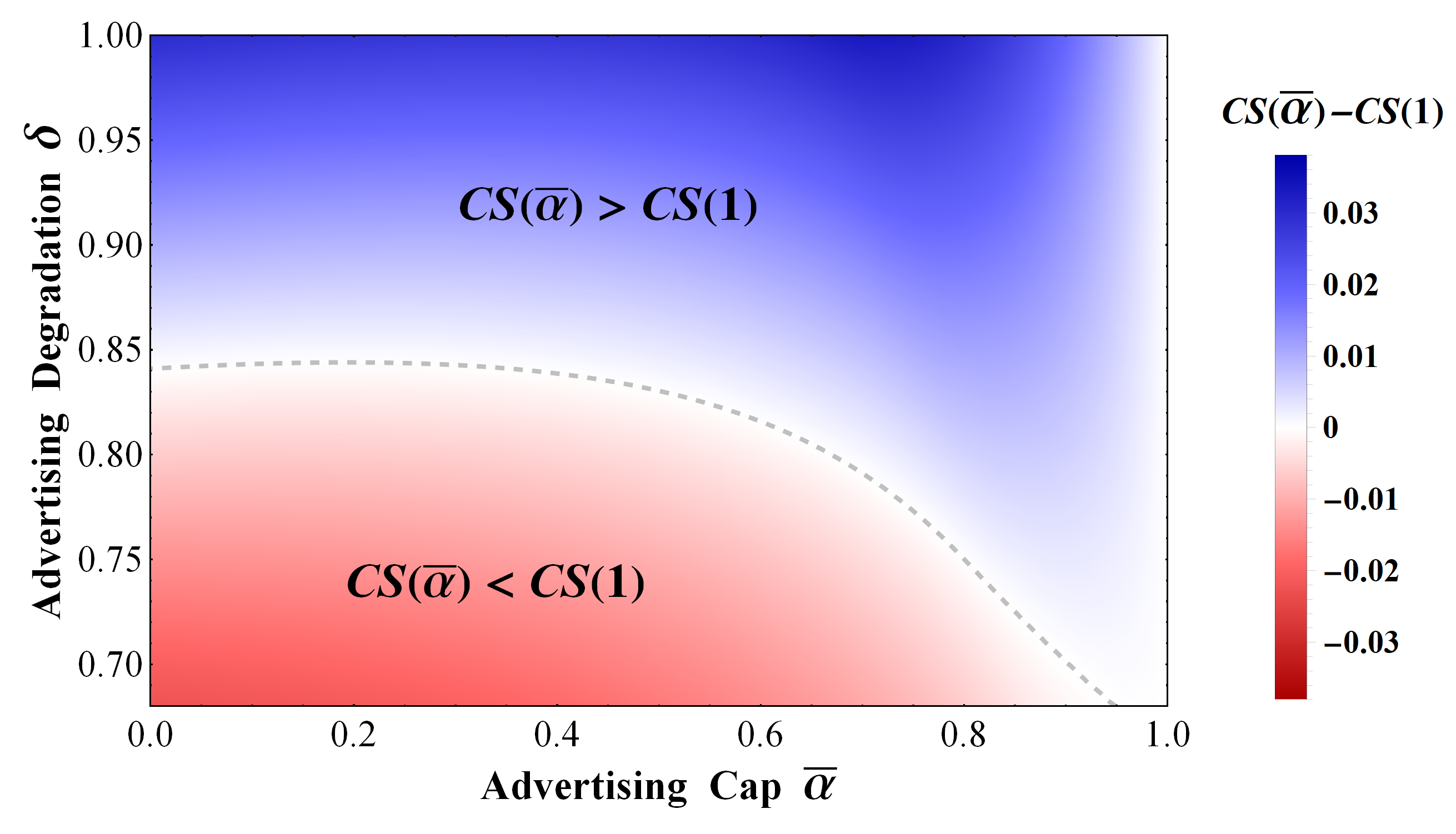}}
    \par\vspace{0.2cm}
    \begin{minipage}{1\textwidth}
        \centering \scriptsize \textit{Notes:} Moving left tightens the advertising cap. In Panel (a), $\bar{\alpha}^*$ marks the illustrated consumer-surplus maximum and $\bar{\alpha}^L$ the low-cap local minimum. In Panel (b), blue indicates $CS(\bar{\alpha})>CS(1)$, and the feasible range $\delta \in (\lambda+\bar{r}, 1)$ follows from market regularity, Assumption~\ref{ass:market_regularity}.
    \end{minipage}
    \caption{Advertising Caps and Consumer Surplus ($\lambda=0.5$, $\bar{r}=0.18$; Panel (a): $\delta=0.9$)}
    \label{fig:cs_cap}
\end{figure}

If quality provision were fixed, reducing advertising would improve the quality consumers experience.
The platform, however, redesigns its menu when a cap restricts the advertising revenue that helps fund quality.
A tighter cap therefore reduces advertising-induced degradation while weakening the return to ad-subsidized quality provision.
Figure~\ref{fig:cs_cap}(a) illustrates how these opposing effects can make consumer surplus ($CS$) non-monotonic in $\bar{\alpha}$ at $(\delta,\lambda,\bar r)=(0.9,0.5,0.18)$.
Holding $\lambda=0.5$ and $\bar r=0.18$, Panel~(b) shows where a cap improves consumer surplus relative to the unregulated benchmark: the blue region satisfies $CS(\bar{\alpha})>CS(1)$, includes sufficiently mild caps throughout the plotted range of $\delta$, and extends to strict caps at higher degradation levels.
This comparison describes a particular parameter slice.
Proposition~\ref{prop:ad_cap_welfare} separates a local condition for improvement from the existence of primitives for which every maximizing cap is interior.

\begin{proposition}[Interior-Cap Optimum]\label{prop:ad_cap_welfare}
Let $CS(\bar{\alpha})$ denote aggregate consumer surplus under the cap $\alpha(\theta)\le\bar{\alpha}$.
\begin{enumerate}[$(a)$]
    \item \textbf{Local improvement:} If $\delta>5/8$ and $\lambda>\hat{\lambda}_{\mathrm{local}}(\delta)$, where $\hat{\lambda}_{\mathrm{local}}(\delta)$ is given in closed form in the proof, then marginally tightening the unregulated benchmark $\bar{\alpha}=1$ strictly raises consumer surplus. Hence $\bar{\alpha}=1$ is not consumer-surplus maximizing.
    \item \textbf{Interior optimality in a nonempty region:} The set of primitives for which every consumer-surplus-maximizing cap lies in $(0,1)$ is nonempty. At $(\delta,\lambda,\bar r)=(0.9,\,0.5,\,0.18)$, direct evaluation of the exact constrained allocation gives $CS(0.78)\approx0.04898>CS(0)\approx0.04167>CS(1)\approx0.03155$. Hence every maximizing cap is strictly interior at this instance, and the same conclusion holds in a neighborhood of these primitives.
\end{enumerate}
\end{proposition}

A tighter cap trades less degradation against weaker quality funding, and changes in lower-tier effective quality alter the information rents passed to users above them.
Proposition~\ref{prop:ad_cap_welfare}(a) gives sufficient conditions for the degradation reduction to dominate locally.
This local gain does not by itself establish an interior optimum; part~(b) establishes that stronger conclusion for a nonempty set of primitives.
The opposite local effect arises whenever a zero-quality segment exists ($\lambda<1-\delta$): marginally tightening the unregulated cap strictly reduces consumer surplus.
Here, a tighter cap weakens ad funding and raises the cutoff for positive quality, so some marginal consumers lose effective service; under the maintained specification, the loss from reduced ad-funded quality dominates locally.
The following numerical example illustrates how these forces can instead favor a moderate cap, with the allocation and rent responses shown in Figure~\ref{fig:cap_micro} in Appendix~\ref{app:supp_tables}.

In the illustrated economy, first tighten the cap from $\bar{\alpha}=1$ toward the consumer-surplus peak $\bar{\alpha}^*$.
Severe ad degradation initially leaves low tiers with little effective quality (Figure~\ref{fig:cap_micro}(a)), making them unattractive alternatives for higher-value users and allowing the platform to leave those users less information rent.
{ A moderate cap reduces advertising-induced degradation, so some lower tiers can offer higher \textit{effective quality} even as their supplied quality $q$ falls.}
As these alternatives become more attractive, the platform must leave higher types more utility to keep them on their intended plans (Figure~\ref{fig:cap_micro}(b)).
Over this range of caps, the resulting gains on balance raise aggregate consumer surplus.

Tightening the cap further can reverse this gain.
In the illustrated economy, consumer surplus falls as the cap moves below $\bar{\alpha}^*$ toward the low-cap trough $\bar{\alpha}^L$ in Figure~\ref{fig:cs_cap}(a).
A tighter cap reduces the advertising return earned on each unit of quality, so the platform cuts quality provision below the median valuation; by the trough, most of these users receive zero quality.
Their lower effective quality also reduces the information rents passed to users above them through the incentive-compatible menu.
For users in the original partial-ad middle whose advertising choices the cap now constrains, both supplied quality and effective quality rise as the cap tightens.
Between the peak and the trough, however, these improvements are insufficient to offset the loss of ad-funded quality at the bottom, and aggregate consumer surplus falls.

In the illustrated economy, further tightening from $\bar{\alpha}^L$ toward a complete ban ($\bar{\alpha}=0$) raises aggregate consumer surplus again.
By the trough, most below-median users have already lost positive-quality service, leaving little ad-subsidized quality at the bottom to sacrifice as the cap tightens further.
The ban eliminates advertising revenue and restores pure-subscription screening.
It removes the remaining ad-subsidized quality below the median, but it also removes advertising-induced degradation and raises supplied quality for users in the original partial-ad middle.
The higher effective quality offered to these middle users increases the information rents their allocations generate for users above them.
In this instance, the resulting consumer-surplus gain outweighs the loss of the remaining quality at the bottom, so the ban delivers more consumer surplus than the trough.
An intermediate cap nevertheless performs better than the ban: as Proposition~\ref{prop:ad_cap_welfare} establishes for this instance and a neighborhood of its primitives, preserving some ad-subsidized quality while limiting its degradation yields a strictly interior consumer-surplus optimum.

A cap restricts the platform's choices and cannot raise its maximized profit, so a consumer-surplus gain need not raise social welfare.
Appendix~\ref{app:social_welfare} identifies a narrow possibility from marginally tightening the unregulated benchmark, requiring sufficiently severe degradation and an intermediate range of quality-driven advertising premiums.

\subsection{Compute Efficiency, Service Quality, and Advertising ($\beta$)}\label{subsec:beta}

Cheaper compute gives a platform a choice about how to redesign its menu: supply more capability, reduce advertising, or combine the two.
These adjustments pull advertising revenue in different directions.
Better service makes advertising more valuable, but it also gives consumers more useful capability to lose through advertising-induced degradation.
The platform must balance these effects when deciding which plans to upgrade and which should carry fewer ads.
We examine the resulting allocations and payoffs, and then ask whether reduced exposure necessarily means less advertising revenue.

Model distillation, more efficient routing, and hardware acceleration can reduce the compute required to deliver a given capability.
These improvements correspond to a lower $\beta$ in $c(q)=\beta q^2/2$.
Following Appendix~\ref{app:beta_extension}, we hold $(\delta,\lambda,\bar r)$ fixed and vary $\beta\in(0,1]$; the maintained technological ceiling expands with efficiency because $\bar q=1/\beta$.
The advertising regions retain their structure: the positive-quality threshold $\underline{\theta}$ and ad-saturation threshold $\theta_1$ remain fixed, while the ad-free threshold becomes $\theta_{2,\beta}=\frac{2\delta+\lambda+\sqrt{\lambda^2+4\delta\beta\bar r}}{4\delta}$.
As $\beta$ falls, this upper boundary moves left, so more consumers receive ad-free service and allocations change within the remaining regions.

\begin{proposition}[Effects of Improved Compute Efficiency]\label{prop:beta_allocations}
Improved compute efficiency ($\beta\downarrow$) has the following effects:
\begin{enumerate}[$(a)$]
    \item \textbf{Allocations:} Optimal quality weakly increases and advertising intensity weakly decreases for every consumer type.
    \item \textbf{Payoffs:} Each consumer type's utility and the platform's profit weakly increase.
    \item \textbf{Segment membership:} The ad-free premium segment expands.
\end{enumerate}
\end{proposition}

Proposition~\ref{prop:beta_allocations} identifies a tier-specific response to cheaper quality.
Positive-quality users on fully advertised plans receive more capability while their ad load remains unchanged.
In the partial-ad middle, the platform both raises quality and reduces advertising: degradation imposes a positive net cost after accounting for quality-linked advertising revenue, and raising quality increases this cost.
Once the net cost of even the first increment of advertising exceeds its baseline advertising return, a plan becomes ad-free.
Thus, cheaper compute can lead the platform to remove advertising from more plans even though better service also makes advertising more valuable.
The resulting increase in effective quality weakly raises every type's utility through the incentive-compatible rent schedule.
Platform profit also weakly rises: the old menu remains incentive compatible and individually rational, fits within the expanded ceiling, and costs less to supply.
The platform therefore improves the consumer experience without having to sacrifice profit to do so.

Less advertising per user and a larger ad-free segment both put downward pressure on advertising revenue.
Does this mean that cheaper compute weakens the advertising business?
Proposition~\ref{prop:beta_ad_revenue} shows why the quality of the remaining ad-supported interactions also matters.

\begin{proposition}[The Volume-to-Value Pivot]\label{prop:beta_ad_revenue}
Improved compute efficiency weakly reduces advertising intensity for every user.
Nevertheless, letting $R_{total}(\beta)$ denote aggregate advertising revenue under compute-cost parameter $\beta$, there exists $\hat{\beta}\in(0,1)$ such that $R_{total}(\hat{\beta})>R_{total}(1)$: aggregate advertising revenue can exceed its level at the baseline $\beta=1$.
\end{proposition}

The platform earns advertising revenue from both the intensity of advertising and the quality of the interactions in which it appears.
Cheaper compute weakly lowers advertising intensity and moves some middle types into the ad-free segment, reducing revenue through the volume channel.
Higher quality raises the quality-driven advertising return at a given ad intensity; for middle types who continue to receive ads, total advertising revenue rises only if this gain outweighs the lower intensity.
The fully ad-saturated, positive-quality low-type segment provides a sharper comparison because its mass and advertising intensity ($\alpha^*=1$) do not depend on $\beta$.
Quality in this segment is proportional to $1/\beta$, so its quality-linked advertising revenue grows at the same rate.
Under the linear premium $\lambda q\alpha$ and expanding ceiling $\bar q=1/\beta$, this segment's advertising revenue becomes arbitrarily large as $\beta$ approaches zero.
It therefore eventually offsets revenue lost as other users receive fewer ads, establishing the comparison with the finite baseline revenue in Proposition~\ref{prop:beta_ad_revenue}.\footnote{This comparison with $\beta=1$ does not imply that advertising revenue rises after every marginal cost reduction.}

{
Appendix~\ref{app:ext_concave_ad_premium} examines whether this gain survives diminishing advertising returns to quality.
Under its stated conditions, the comparison between two fixed cost levels remains valid for concave premiums sufficiently close to the linear benchmark.
A separate numerical example with bounded, diminishing advertising returns to quality shows that reducing $\beta$ from $1$ to $0.5$ raises aggregate advertising revenue by about $3.6\%$ while weakly reducing advertising intensity for every type.
Thus, the increase need not rely on unbounded advertising value, although it is not guaranteed for arbitrary concave premiums.
For a platform assessing its revenue outlook, the implication is to track the quality and advertising value of remaining advertised service alongside ad intensity and the share of ad-free users.
}

The compute-efficiency analysis shows how cheaper quality changes capability and advertising across a continuous menu.
The next section asks a different design question: whether stronger advertising returns lead the platform to offer more distinct positive-quality tiers or to combine existing tiers.
It holds quality costs and the technological ceiling fixed while increasing the advertising return to quality, isolating the effect of monetization on the scope for service differentiation.

\section{An Extension: When Service Tiers Converge}\label{sec:tier_design}

Advertising-funded upgrading can do more than widen access: it can erode the differences on which subscription prices depend.
This extension asks when stronger advertising returns make separate contracts cease to be optimal because consumer groups receive the same quality and advertising exposure.
Unlike the compute-efficiency analysis, we hold quality costs and the technological ceiling fixed while increasing the quality-driven ad premium $\lambda$.
A moderate premium can finance entry into positive quality, whereas stronger returns can push groups to the same allocation.
We count distinct positive-quality contracts, excluding the zero-quality baseline in Remark~\ref{rem:zero_quality_baseline}.
In the three-type framework below, Proposition~\ref{prop:discrete_tiers} establishes a $2$--$3$--$2$--$1$ sequence in the number of positive-quality tiers.

We replace the continuous type distribution with equally sized Low ($L$), Middle ($M$), and High ($H$) groups, with $\theta_L=0$, $\theta_M\in(1/2,1)$, and $\theta_H=1$.
The model retains the baseline cost and revenue functions with $\beta=1$, $\bar q=1$, and $\delta\in(0,1)$, but relaxes Assumption~\ref{ass:market_regularity} to include parameter regions without an ad-free premium segment.
The virtual valuations are $v_L=-2\theta_M<0$, $v_M=2\theta_M-1>0$, and $v_H=1$.
Pure subscription supplies qualities $v_M$ and $v_H$ to the upper groups and zero quality to $L$.
At a full ad load, unconstrained quality is $\tilde q_i(1)=(1-\delta)v_i+\lambda$, which rises with the premium for every group.
The three thresholds below mark its becoming positive for $L$, reaching the ceiling for $M$, and reaching the ceiling for $L$; their ordering places entry before upper-type pooling.
The platform must compare interior candidates with boundary allocations, because the ceiling eventually binds.
Appendix~\ref{app:discrete_tiers} provides the discrete calculation and stationary candidates, with exact allocations in Table~\ref{tab:tier_exact_dynamics}.

\begin{proposition}[Expansion-then-Pooling of Service Tiers]\label{prop:discrete_tiers}
In our three-type framework, define the thresholds for type $L$ to receive positive quality, upper-type pooling, and universal pooling as $\lambda_L:=2\theta_M(1-\delta)$, $\lambda_M:=\bar q-(2\theta_M-1)(1-\delta)$, and $\lambda_U:=\bar q+2\theta_M(1-\delta)$.
Suppose $0<\bar r\le\delta-\lambda_L$, which implies $\lambda_L<\lambda_M<\lambda_U$.
The optimal menu has:
\begin{enumerate}[$(a)$]
    \item \textbf{Ad initiation, $\lambda\in(0,\lambda_L]$:} Two positive-quality tiers, serving $M$ and $H$ separately.
    \item \textbf{Full separation, $\lambda\in(\lambda_L,\lambda_M)$:} Three distinct positive-quality tiers, one for each group.
    \item \textbf{Top pooling, $\lambda\in[\lambda_M,\lambda_U)$:} Two positive-quality tiers, with $M$ and $H$ sharing a plan and $L$ remaining separate.
    \item \textbf{Universal pooling, $\lambda\ge\lambda_U$:} One positive-quality tier shared by all three groups.
\end{enumerate}
\end{proposition}

Initially, improving the low group's free baseline would earn more advertising revenue but also make downward switching more attractive to paying consumers, increasing the information rents needed to retain them.
The negative virtual valuation $v_L$ captures this screening cost, which a small quality-driven premium cannot cover.
Type $L$ therefore remains at the fully advertised zero-quality baseline, while $M$ and $H$ receive positive quality.
The high group remains ad-free under the stated restrictions.
If the middle plan carries an interior ad load, its quality-driven advertising return does not yet offset the virtual cost of degradation, so it supplies less quality than under pure subscription.
Once $\lambda$ exceeds $\lambda_L$, advertising revenue justifies upgrading the low group into a third positive-quality tier.
This tier is free and fully advertised: the normalization $\theta_L=0$ leaves no subscription willingness to pay, while the advertiser-side return supports quality provision.
The expansion upgrades an existing layer and need not add a newly listed plan.
Advertising can subsequently reach the high group while all three allocations remain distinct; the detailed ad-load transitions are reported in Table~\ref{tab:tier_exact_dynamics}.

At $\lambda_M$, type $M$ reaches the quality ceiling already attained by $H$.
At these premiums, the quality-driven advertising return exceeds the virtual cost of degradation even for $H$, so both upper groups receive the full ad load.
Identical quality and advertising require the same price, leaving two positive-quality contracts.
During top pooling, upgrading the low plan makes it a better alternative for upper types, so their unchanged common plan must become cheaper.
At $\lambda_U$, the low group also reaches the ceiling, and every group receives $(q^*,\alpha^*)=(\bar q,1)$.
The normalization $\theta_L=0$ makes this common plan free, leaving advertising as the revenue source.
The fixed ceiling is essential: without it, full-ad qualities remain strictly ordered and this pooling mechanism disappears.

For a platform upgrading an ad-supported offering, the practical question is whether capability differences and differences in ad exposure still support separate subscription prices.
The model concerns distinct positive-quality contracts; attributes and plan-management costs outside the benchmark can still support different marketed plans.
Universal pooling requires the strong advertising return specified by $\lambda_U$, above the highest consumer-side valuation under $\bar q=\theta_H=1$; reliance on advertising revenue alone does not establish this condition.
OpenAI's Free, Go, Plus, and Pro offerings illustrate a granular menu, while Meta AI combines a free core service with paid bundles offering higher usage of compute-intensive AI features \citep{meta2026metaone}.
These examples establish the relevance of choosing differentiated offerings, but do not show that advertising causes them to converge.
Evidence for the mechanism would need to connect stronger quality-driven advertising returns to narrower capability differences, converging ad exposure, and the disappearance of associated price differences, while accounting for compute constraints and scarcity, bundling, competition, and legacy ecosystem effects.

Beyond tier counts, the preceding analysis suggests three additional tests of the quality, screening, and compute-efficiency mechanisms.
First, examine whether better service raises advertising revenue per interaction, holding ad exposure, user intent, and advertiser conditions comparable.
Second, tracking capability allowances, subscription prices, advertising exposure, and users' tier switching together could test the predicted screening responses across the menu.
Third, decomposing advertising-revenue growth into user growth, ad frequency, advertiser demand, and revenue per impression could help distinguish exposure-driven growth from the quality-linked value channel in the compute-efficiency result.

\section{Conclusion}\label{sec:conclusion}

An LLM platform's joint choice of quality, advertising, and prices affects the entire subscription menu, including users who never see an ad.
When partly ad-supported and ad-free plans coexist, advertising can finance useful service for consumers who receive none under pure subscription, but middle plans deliver less effective quality than under pure subscription, allowing the platform to charge more for unchanged premium service.
Some fully ad-supported users gain utility, while partly ad-supported and ad-free users lose.
This \textit{utility polarization} means that broader access can coexist with greater rent extraction from existing subscribers.

Advertising caps should therefore be assessed through their effects on service quality and prices, as well as exposure.
A moderate cap can improve lower-tier service and strengthen premium users' alternatives while preserving revenue that supports quality.
Tighter limits can remove the return that sustains quality at the bottom.
In some settings, a moderate cap yields more consumer surplus than either a ban or unrestricted advertising.
The welfare objective remains distinct: a cap cannot raise maximized platform profit, and gains to consumers need not raise total welfare.
Disclosure and separation rules may instead change how exposure degrades assistance and require a different comparison.

Cheaper compute supports better service with fewer ads and broader ad-free access, leaving consumers and the platform at least as well off.
With sufficiently large efficiency gains, higher advertising value from fully advertised users already receiving useful service can outweigh reduced exposure elsewhere and lift aggregate advertising revenue above its initial level.
This \textit{volume-to-value pivot} makes delivered capability important alongside ad counts.

The findings concern a monopolistic platform that can commit to its menu and consumers who differ in their valuation of quality.
In the tier extension, strong advertising returns can bring plans to the same maximum capability and exposure, eliminating price differences supported by those attributes; other forms of differentiation remain possible.
The e-companion examines alternative user-value and advertising-return specifications, with conclusions specific to each environment.
Its competition diagnostic preserves a local pattern of opposite price responses across partly ad-supported and ad-free users, but does not determine how competitors set and revise posted prices.

Future work could endogenize the feedbacks that this framework holds fixed.
If commercial influence weakens trust and context sharing, the loss of future advertising value may discipline current ad loads.
Linking advertiser demand and auction design \citep{duetting2024mechanism} to consumer screening could clarify when thin advertiser markets weaken ad-funded access.
External influence through Generative Engine Optimization or data poisoning, and revenue tied to completed purchases, raise further questions about whether monetization improves matching or intensifies steering.

\makeatletter
\renewcommand{\@biblabel}[1]{}
\let\OLDthebibliography\thebibliography
\renewcommand{\thebibliography}[1]{%
    \OLDthebibliography{#1}%
    \setlength{\itemsep}{1pt}%
    \setlength{\parskip}{1pt}%
    \setlength{\baselineskip}{15pt}%
}
\makeatother
\bibliographystyle{ormsv080}
\bibliography{2026_LLM_Ads_no_urls}
\newpage
\ECSwitch
\vspace{-20pt}
\small
\begin{APPENDICES}
\section{Model Extensions}\label{app:model_extensions}

These extensions examine which parts of the baseline mechanism depend on its specification.
Additive nuisance changes the lowest allocation because the platform must keep effective quality nonnegative; type-dependent and concave advertising premiums change the returns to serving different consumers.
Direct ad utility yields an allocation equivalence after adjusting the effective advertising premium.
The correlated-degradation and competition exercises are diagnostics with more limited interpretations.
Throughout, we distinguish restrictions that preserve the allocation structure from the smaller neighborhoods in which comparative statics and utility comparisons persist.
\subsection{Mixed Ad Degradation}\label{app:ext_mixed_degradation}

Our baseline analysis focuses on quality-dependent proportional ad degradation ($\delta$), a feature central to generative AI platforms. We now allow ads to generate an additional quality-independent nuisance component ($\delta_n$), such as visual interruptions or other disruptions that do not scale with service quality.
To examine the robustness of our main results, we extend the effective quality function to incorporate both effects:
\begin{equation}\label{eq:mixed_effective_quality}
g(q,\alpha)=q(1-\delta\alpha)-\delta_n\alpha,
\end{equation}
where $\delta\in(0,1)$ captures multiplicative degradation and $\delta_n\ge0$ captures additive nuisance. We maintain $\theta\sim U[0,1]$, $v(\theta)=2\theta-1$, $c(q)=q^2/2$, and $R(q,\alpha)=\lambda q\alpha+\bar r(\alpha-\alpha^2/2)$.
Consistent with Assumption~\ref{ass:effective_quality}, the feasible set in this extension explicitly requires $g(q,\alpha)\ge0$.
We impose $\lambda\in(0,\delta)$ and $\bar r\in(0,\delta-\lambda)$.
To state the exact structural restriction, define
\[
\varepsilon_B:=\lambda(1-\delta)(2-\delta),
\qquad
\varepsilon_H:=
\begin{cases}
+\infty, & \bar r\ge \lambda^2,\\[3pt]
\displaystyle
\sqrt{\bar r}+\frac{\delta\bar r}{\lambda-\sqrt{\bar r}},
& 0<\bar r<\lambda^2,
\end{cases}
\qquad
\varepsilon_S:=\min\{\varepsilon_B,\varepsilon_H\}.
\]
The bound $\varepsilon_B$ keeps the zero-effective-quality boundary and the adjacent full-ad segment distinct, whereas $\varepsilon_H$ preserves the ordering of the full-ad/interior and interior/ad-free thresholds and the interior second-order condition.

Let $\Delta_m(\theta):=\delta(2\theta-1)-\lambda$. Define the interior allocations as
\[
\alpha_m^{int}(\theta)=
\frac{\bar r-(2\theta-1)[\Delta_m(\theta)+\delta_n]}
{\bar r-\Delta_m(\theta)^2},
\qquad
q_m^{int}(\theta)=
\frac{\bar r[(2\theta-1)-\Delta_m(\theta)]+(2\theta-1)\delta_n\Delta_m(\theta)}
{\bar r-\Delta_m(\theta)^2}.
\]
The relevant thresholds are
\[
\underline{\theta}^m=
\frac{1}{2}-\frac{\lambda}{2(1-\delta)}
+\frac{\delta_n}{2(1-\delta)^2},
\]
while $\theta_1^m$ and $\theta_2^m$ are defined by
\[
\bigl[\lambda+(2\theta_1^m-1)(1-\delta)\bigr]
\bigl[\lambda-\delta(2\theta_1^m-1)\bigr]
=
\delta_n(2\theta_1^m-1),
\qquad
(2\theta_2^m-1)\bigl[\Delta_m(\theta_2^m)+\delta_n\bigr]=\bar r.
\]

\begin{theorem}[Optimal Mechanism under Mixed Ad Degradation]\label{thm:mixed_optimal_mechanism}
Under \eqref{eq:mixed_effective_quality} and $\delta_n\in[0,\varepsilon_S)$, the optimal mechanism takes the following four-segment form when $\underline{\theta}^m>0$:
\begin{enumerate}[$(a)$]
    \item \textbf{Bottom types} $\theta\in[0,\underline{\theta}^m]$:
    $q^*(\theta)=\frac{\delta_n}{1-\delta}$, $\alpha^*(\theta)=1$, and $g(q^*(\theta),\alpha^*(\theta))=0$.
    \item \textbf{Low types} $\theta\in(\underline{\theta}^m,\theta_1^m]$:
    $q^*(\theta)=(2\theta-1)(1-\delta)+\lambda$ and $\alpha^*(\theta)=1$.
    \item \textbf{Medium types} $\theta\in(\theta_1^m,\theta_2^m]$:
    $q^*(\theta)=q_m^{int}(\theta)$ and $\alpha^*(\theta)=\alpha_m^{int}(\theta)$.
    \item \textbf{High types} $\theta\ge\theta_2^m$:
    $q^*(\theta)=2\theta-1$ and $\alpha^*(\theta)=0$.
\end{enumerate}
If $\underline{\theta}^m<0$, the zero-effective-quality boundary pool is empty and the remaining three segments apply; if $\underline{\theta}^m=0$, it consists only of $\theta=0$.
\end{theorem}

Additive nuisance changes what the platform must provide at the bottom.
When the lowest segment exists, full advertising requires produced quality $\delta_n/(1-\delta)$ merely to offset the nuisance, leaving these consumers with zero effective quality.
Thus the bottom allocation can use costly capability even though it supplies no net user value.
Under $\delta_n<\varepsilon_S$, both this boundary pool and the adjacent positive-effective-quality segment remain fully ad-supported.
The additional condition $\delta_n<\lambda(1-\delta)$ places the boundary threshold below the median type, $\underline\theta^m<1/2$; it is an interpretive restriction rather than the maximal restriction supporting the four-segment structure.

\begin{proposition}[Local Robustness under Mixed Ad Degradation]
\label{prop:mixed_local_robustness}
Fix $(\delta,\lambda,\bar r)$ satisfying the mixed analogue of Assumption \ref{ass:market_regularity}.
There exists $\bar{\delta}_n\in\bigl(0,\min\{\varepsilon_S,\lambda(1-\delta)/2\}\bigr)$ such that, for every $\delta_n\in[0,\bar{\delta}_n)$:
\begin{enumerate}[$(a)$]
    \item \textit{Quality.} As $\delta$ rises, the positive-effective-quality threshold moves left. Within the positive-effective-quality full-ad segment, quality rises below the median and falls above it; the baseline interior-quality signs persist on compact subsets away from the sign-switching locus.
    \item \textit{Prices.} The divergent interior- and ad-free-price signs with respect to $\bar r$ persist on compact subsets away from thresholds.
    \item \textit{Utility.} There exist nonempty compact low- and high-type intervals on which the utility gain and loss, respectively, relative to pure subscription persist.
\end{enumerate}
\end{proposition}

The theorem characterizes allocations throughout the stated structural range, whereas the proposition preserves the baseline comparative statics only for sufficiently small additive nuisance.
The distinction matters because maintaining the order of the segments does not by itself preserve every derivative or utility comparison.
The local argument uses the strict baseline signs on the specified compact sets, which survive a sufficiently small perturbation.
The proofs are provided in Appendices \ref{app:proof_mixed_theorem} and \ref{app:proof_mixed_robustness}.

\subsection{Correlated Ad Degradation: A Diagnostic}\label{app:ext_correlated}

The baseline model assumes a homogeneous ad-degradation parameter. We now consider a diagnostic specification in which users with higher willingness to pay for AI quality are also more likely to be sensitive to ad-induced distortion, as may be the case for professional or enterprise users who place greater weight on reliability, neutrality, and privacy.
The interpretation differs across the two advertising channels introduced in Section~\ref{sec:instruments}. Under on-interface advertising, response interference can make users who value quality and neutrality more exposed to degradation; the December 2025 controversy over ad-like app suggestions shown to some paying ChatGPT subscribers illustrates this concern \citep{ha2025suggestions}. Under off-interface advertising, privacy sensitivity may instead be distinct from quality valuation, so positive correlation should be understood as a diagnostic benchmark rather than a general restriction.
To capture this idea, suppose that, conditional on $\theta\sim U[0,1]$, the consumer is in a high-degradation state with probability $\theta$ and in a low-degradation state with probability $1-\theta$:
\[
\delta_H=2\delta \quad \text{with probability } \theta,
\qquad
\delta_L=\delta \quad \text{with probability } 1-\theta.
\]
We impose $\delta\in(0,1/2)$, together with $\lambda\in(0,\delta)$ and $\bar r\in(0,\delta-\lambda)$, so that both degradation states remain within the admissible range of the baseline model.

A full treatment of this population would require a two-dimensional screening problem over $(\theta,\delta_k)$, with vertical IC constraints in $\theta$ and cross-state IC constraints between high- and low-sensitivity users.
Here we isolate the degradation channel by comparing two baseline calculations, one for each sensitivity state, while holding the virtual type at $2\theta-1$.
The correlation determines how the states are associated with consumer valuation; these calculations do not solve the resulting joint screening problem.
For state $k\in\{L,H\}$, the diagnostic pointwise virtual surplus is
\[
\mathcal V_k(\theta;q,\alpha)
=
(2\theta-1)q(1-\delta_k\alpha)
+\lambda q\alpha
+\bar r\left(\alpha-\frac{\alpha^2}{2}\right)
-\frac{q^2}{2}.
\]
Thus the low-sensitivity benchmark exactly coincides with the baseline model, while the high-sensitivity benchmark is the same model with $\delta$ replaced by $2\delta$.
This diagnostic has three implications.

\begin{enumerate}[$(a)$]
    \item The $\delta$-driven non-monotonic quality-distortion mechanism continues to hold within each state-contingent benchmark: higher degradation weakly expands the positive-quality region, with strict expansion when its threshold moves within the type support; among positive-quality full-ad users, it raises quality below the median and lowers quality above it, preserves the heavy-ad/light-ad sign reversal in the interior region, and leaves ad-free quality unchanged.
    \item The $\bar r$-driven allocation channel remains intact: higher baseline ad profitability increases interior ad load, lowers interior quality and effective quality, and preserves the rent-extraction force that raises prices for remaining ad-free premium users in the state-contingent benchmark.
    \item The utility-polarization comparison with pure subscription remains valid within each state-contingent benchmark. High-sensitivity users become more likely as $\theta$ increases, but the statewise comparisons do not establish a utility ordering for an optimal mechanism in the joint correlated population.
\end{enumerate}

These implications concern the two statewise calculations, including the indirect effect of lower-plan effective quality on ad-free users' information rents. The supporting calculations are provided in Appendix~\ref{app:proof_correlated_diagnostic}.

\subsection{Positive Ad Utility}\label{app:ext_positive_ad_utility}

The baseline model treats embedded advertising as purely disruptive. We now allow ads to generate direct user value by reducing search costs or improving match quality.
For instance, a high-quality model may embed a sponsored recommendation or booking link that improves the user's match relative to unaided search.
To capture this possibility, we augment user utility as
\begin{equation}\label{eq:positive_ad_utility}
U(\theta)=\theta q(1-\delta\alpha)+\eta q\alpha-T,
\end{equation}
where $\eta\ge0$ captures the direct, type-independent matching value generated by advertising.
The negative disruption term $\theta q\delta\alpha$ remains type-dependent because higher-value users suffer more from degraded analytical output, while the positive matching value $\eta q\alpha$ is type-independent.

Let $\tilde{\lambda}:=\lambda+\eta$. To preserve the same ad-free premium structure as in the baseline model, we impose the following effective analogue of Assumption~\ref{ass:market_regularity}:
\[
\tilde{\lambda}\in(0,\delta),
\qquad
\bar r\in(0,\delta-\tilde{\lambda}).
\]
Equivalently, the direct ad benefit must satisfy $0\le\eta<\delta-\lambda-\bar r$.

\begin{proposition}[Equivalence under Positive Ad Utility]
\label{prop:positive_ad_utility_equivalence}
{
Under \eqref{eq:positive_ad_utility} and the effective regularity condition above, equilibrium thresholds, allocations, and consumer utilities coincide with those of the baseline model evaluated at $\tilde{\lambda}=\lambda+\eta$. Subscription payments satisfy
\[
T_\eta(\theta)
=
T_{\mathrm{baseline}}(\theta;\tilde{\lambda})
+\eta q^*(\theta)\alpha^*(\theta).
\]
The allocation and utility results in Sections~\ref{sec:optimal_mechanism}--\ref{sec:comparative_statics} therefore carry over after this substitution. The within-segment subscription-price responses to $\bar r$ in Proposition~\ref{prop:rbar_pricing} also remain valid.
\par}
\end{proposition}

This result shows that positive ad utility does not overturn the baseline mechanism. Instead, it strengthens the effective advertising incentive: {for a given allocation, the platform can extract the type-independent matching value through subscription payments without changing information rents.} The proof is provided in Appendix~\ref{app:proof_positive_ad_utility}.

\subsection{Type-Dependent Ad Monetization}\label{app:ext_type_dependent}

In practice, the attention and inferred intent of high-valuation users may be more valuable for targeted advertising, B2B lead generation, or downstream conversion.
To capture this, we allow the quality-driven ad premium to increase with the consumer's vertical type: $\Lambda(\theta)=\lambda+\gamma\theta$, where $\lambda>0$ is the baseline monetization rate and $\gamma\ge0$ captures the targeted premium.
The pointwise virtual surplus becomes
\begin{equation}\label{eq:type_dependent_VS}
\mathcal{V}_{\gamma}(\theta;q,\alpha)
=
(2\theta-1)q(1-\delta\alpha)
+
(\lambda+\gamma\theta)q\alpha
+
\bar r\left(\alpha-\frac{\alpha^2}{2}\right)
-
\frac{q^2}{2}.
\end{equation}
Let $\Delta_\gamma(\theta):=(2\theta-1)\delta-(\lambda+\gamma\theta)=(2\delta-\gamma)\theta-(\delta+\lambda)$.
We impose $0\le\gamma<\delta-\lambda-\bar r$. This restriction ensures $\Delta_\gamma(1)>\bar r$ and therefore preserves an endogenous ad-free premium tier.
The equilibrium thresholds are
\[
\underline{\theta}^{\gamma}
=
\frac{1-\delta-\lambda}{2(1-\delta)+\gamma},
\qquad
\theta_1^\gamma
=
\frac{\delta+\lambda}{2\delta-\gamma},
\]
and $\theta_2^\gamma$ is the unique solution to
\[
(2\theta_2^\gamma-1)\Delta_\gamma(\theta_2^\gamma)=\bar r.
\]
Interior allocations are obtained from the baseline formulas by replacing $\Delta(\theta)$ with $\Delta_\gamma(\theta)$.

\begin{proposition}[Robustness under Type-Dependent Ad Monetization]
\label{prop:type_dependent_robustness}
Under \eqref{eq:type_dependent_VS} and $0\le\gamma<\delta-\lambda-\bar r$, the baseline equilibrium structure remains intact, with the following implications:
\begin{enumerate}[$(a)$]
    \item \textit{Quality.} Non-monotonic quality distortion with respect to $\delta$ continues to hold.
    \item \textit{Prices.} Higher baseline ad profitability $\bar r$ lowers interior effective quality and raises prices for users locally remaining ad-free. No global sign is asserted for interior prices under type-dependent ad monetization.
    \item \textit{Utility.} The utility-polarization comparison with pure subscription continues to hold.
\end{enumerate}
\end{proposition}

A higher premium for high-valuation consumers changes the thresholds, but the restriction on $\gamma$ keeps advertising returns low enough at the top for an ad-free premium segment to remain.
Within this region, low types remain ad-supported and intermediate types receive partial ad exposure.
Higher baseline ad profitability still reduces effective quality in the interior, making those plans less attractive alternatives to ad-free service and allowing higher prices above.
The type-dependent premium changes the variation in this effect across interior users, so the same allocation argument does not establish their price response.
The proof is provided in Appendix~\ref{app:proof_type_dependent_robustness}.

\subsection{Concave Quality-Driven Advertising Premium}\label{app:ext_concave_ad_premium}

Our baseline analysis assumes that the quality-driven advertising premium is linear in response quality, taking the form $\lambda q\alpha$, so that higher response quality raises advertising value at a constant marginal rate.
In practice, however, the marginal advertising value of quality may diminish: initial quality improvements can substantially improve intent inference and ad relevance, while further improvements may add less incremental advertising value.
To examine the robustness of our main results, we extend the advertising revenue function to
\begin{equation}\label{eq:concave_ad_revenue}
R_h(q,\alpha)=\lambda h(q)\alpha+\bar r\left(\alpha-\frac{\alpha^2}{2}\right),
\end{equation}
where $h(0)=0$, $h'(q)>0$, $h''(q)<0$ for $q>0$, and $h'(0)<\infty$.
The normalization $h(0)=0$ means that the quality-driven premium vanishes at zero quality; baseline ad impressions at zero quality are still captured by $\bar r(\alpha-\alpha^2/2)$.
All other primitives remain as in the baseline model: $g(q,\alpha)=q(1-\delta\alpha)$, $c_\beta(q)=\beta q^2/2$ with $\beta\in(0,1]$, $\theta\sim U[0,1]$, and $v(\theta)=2\theta-1$.
We first set $\beta=1$ to isolate robustness to concavity and characterize the resulting mechanism; after Proposition~\ref{prop:concave_premium_robustness}, we restore $\beta$ to revisit the compute-efficiency result.

Setting $\beta=1$, the pointwise virtual surplus under \eqref{eq:concave_ad_revenue} is
\[
\mathcal V_h(\theta;q,\alpha)
=
v(\theta)q(1-\delta\alpha)
+\lambda h(q)\alpha
+\bar r\left(\alpha-\frac{\alpha^2}{2}\right)
-\frac{q^2}{2}.
\]
An interior solution satisfies $q=v(\theta)(1-\delta\alpha)+\lambda\alpha h'(q)$ and $\delta v(\theta)q-\lambda h(q)=\bar r(1-\alpha)$.
Thus, relative to the baseline model, marginal quality provision is governed by $h'(q)$, while the boundary between ad-supported and ad-free service is governed by the average premium $h(q)/q$.

Because a general concave function can generate multiple crossings, we impose the following regularity condition to preserve the same market region as in the baseline model.

\begin{assumption}[Regularity under Concave Advertising Premium]\label{ass:concave_premium_regular}
Let $q_1(v)$ denote the unique positive solution to $q_1(v)=v(1-\delta)+\lambda h'(q_1(v))$, whenever such a solution exists.
Define $A_h(v):=\delta v q_1(v)-\lambda h(q_1(v))$ and $B_h(v):=\delta v^2-\lambda h(v)$.
The equations $A_h(v)=0$ and $B_h(v)=\bar r$ admit unique solutions $v_1^h$ and $v_2^h$, respectively.
Moreover, these thresholds satisfy $\underline v_h<v_1^h<v_2^h<1$, where $\underline v_h:=-\lambda h'(0)/(1-\delta)$.
Finally, for every $v\in(0,1]$ and every interior stationary point $(q,\alpha)\in(0,\infty)\times(0,1)$ satisfying both first-order conditions, the Hessian of $\mathcal V_h$ is negative definite; equivalently,
\[
\bar r\bigl[1-\lambda\alpha h''(q)\bigr]
>
\bigl[\lambda h'(q)-\delta v\bigr]^2.
\]
\end{assumption}

Assumption~\ref{ass:concave_premium_regular} is the analogue of the market-regularity restriction in the baseline model.
It ensures a unique transition from full advertising to partial advertising and then to the ad-free premium segment.
The linear benchmark $h(q)=q$ satisfies the same Hessian condition under Assumption~\ref{ass:market_regularity} and recovers the baseline thresholds, although it lies outside the strictly concave class considered here.

Define the lower threshold $\underline{\theta}^h=(1+\underline v_h)/2=1/2-\lambda h'(0)/[2(1-\delta)]$.
Let $\theta_1^h=(1+v_1^h)/2$ and $\theta_2^h=(1+v_2^h)/2$, where $v_1^h$ and $v_2^h$ are implicitly defined by $\delta v_1^h q_1(v_1^h)=\lambda h(q_1(v_1^h))$ and $\delta (v_2^h)^2-\lambda h(v_2^h)=\bar r$, respectively.

\begin{theorem}[Optimal Mechanism under a Concave Advertising Premium]\label{thm:concave_ad_premium}
Under \eqref{eq:concave_ad_revenue} and Assumption~\ref{ass:concave_premium_regular}, the optimal mechanism has up to four segments:
\begin{enumerate}[$(a)$]
    \item \textbf{Zero-quality types} $\theta\in[0,\underline{\theta}^h]$ (empty when $\underline{\theta}^h<0$):
    $q^*(\theta)=0$ and $\alpha^*(\theta)=1$.

    \item \textbf{Low types} $\theta\in[0,1]$ with $\underline{\theta}^h<\theta\le\theta_1^h$:
    $\alpha^*(\theta)=1$, and $q^*(\theta)=q_1(v(\theta))>0$ is uniquely defined by $q^*(\theta)=v(\theta)(1-\delta)+\lambda h'(q^*(\theta))$.

    \item \textbf{Medium types} $\theta\in(\theta_1^h,\theta_2^h)$:
    $q^*(\theta)=q_h^{int}(\theta)>0$ and $\alpha^*(\theta)=\alpha_h^{int}(\theta)\in(0,1)$, where $(q_h^{int}(\theta),\alpha_h^{int}(\theta))$ is the unique solution to
    \[
    q_h^{int}(\theta)
    =
    v(\theta)\bigl(1-\delta\alpha_h^{int}(\theta)\bigr)
    +\lambda\alpha_h^{int}(\theta)h'(q_h^{int}(\theta)),
    \qquad
    \delta v(\theta)q_h^{int}(\theta)-\lambda h(q_h^{int}(\theta))
    =
    \bar r\bigl(1-\alpha_h^{int}(\theta)\bigr).
    \]

    \item \textbf{High types} $\theta\in[\theta_2^h,1]$:
    $q^*(\theta)=v(\theta)=2\theta-1$ and $\alpha^*(\theta)=0$.
\end{enumerate}
If $\underline{\theta}^h<0$, the zero-quality segment is empty; if $\underline{\theta}^h=0$, it consists only of $\theta=0$.
\end{theorem}

Under Assumption~\ref{ass:concave_premium_regular}, the mechanism continues to separate consumers into fully ad-supported low types, partially ad-supported middle types, and ad-free high types, with a zero-quality segment when the lower threshold is positive.
Concavity changes the quality conditions and leaves the thresholds implicit rather than closed form.
As quality rises under a given $h$, both the marginal premium $h'(q)$ and the average premium $h(q)/q$ decline: the first governs quality provision, while the second helps determine whether service remains ad-supported.
This is a comparison across quality levels within the same specification; concavity alone does not order advertising returns relative to the linear benchmark.

\begin{proposition}[Local Robustness under a Concave Advertising Premium]
\label{prop:concave_premium_robustness}
Fix $(\delta,\lambda,\bar r)$ satisfying Assumption~\ref{ass:market_regularity}.
Consider a family of twice continuously differentiable functions $\{h_a\}_{a\ge0}$ such that $h_0(q)=q$, $h_a(0)=0$, $h_a'(q)>0$, $h_a''(q)<0$ for $a>0$, and $h_a\to h_0$ in the $C^2$ norm on the relevant quality domain as $a\downarrow0$.
There exists $\bar a>0$ such that, for every $a\in[0,\bar a)$, the baseline quality and within-segment price signs persist on compact subsets away from sign-switching loci and thresholds, while some nonempty low- and high-type intervals retain their respective utility gains and losses relative to pure subscription.
\end{proposition}

The preceding theorem permits a general concave premium satisfying its regularity conditions, but the proposition makes a local claim around the linear benchmark.
On compact sets away from thresholds and sign changes, the baseline derivatives and utility differences have strict margins; small changes in the premium preserve those margins.
This argument does not establish the same signs for every concave function.
We now restore $\beta\in(0,1]$ and let $R_{\mathrm{total}}^{a}(\beta)$ denote aggregate advertising revenue under premium $h_a$ and compute-cost parameter $\beta$.

\begin{corollary}[Local Robustness of the Revenue Reversal]
\label{cor:concave_beta_revenue}
Suppose that, under $h_0(q)=q$, there exists $\hat\beta\in(0,1)$ such that
\[
R_{\mathrm{total}}^{0}(\hat\beta)>R_{\mathrm{total}}^{0}(1).
\]
For the family $\{h_a\}_{a\ge0}$ in Proposition~\ref{prop:concave_premium_robustness}, with convergence on a common compact interval containing the optimal qualities at $\beta\in\{\hat\beta,1\}$, there exists $\bar a_\beta>0$ such that
\[
R_{\mathrm{total}}^{a}(\hat\beta)>R_{\mathrm{total}}^{a}(1)
\]
for every $a\in[0,\bar a_\beta)$, provided the corresponding concave-premium regularity and uniqueness conditions hold at both compute-cost values.
\end{corollary}

The corollary preserves a strict revenue comparison between two compute-cost values under a sufficiently small departure from the linear premium.
It does not extend the linear model's limiting argument in Proposition~\ref{prop:beta_ad_revenue} to every concave function or imply that revenue rises after every cost reduction.
For a separate illustration, consider $h_a(q)=(1-e^{-aq})/a$ and set $a=0.5$, so $h(q)=2(1-e^{-q/2})$ is strictly concave and bounded.
At $(\delta,\lambda,\bar r)=(0.6,0.2,0.15)$, lowering $\beta$ from $1$ to $0.5$ weakly reduces advertising intensity for every type, while aggregate advertising revenue rises from $0.078968$ to $0.081803$.
This calculation illustrates that the volume-to-value reversal can occur with a nonlinear premium; it is not a result for arbitrary concavity.
The proofs are provided in Appendices~\ref{app:proof_concave_ad_premium}, \ref{app:proof_concave_premium_robustness}, and~\ref{app:proof_concave_beta_revenue}.

\subsection{Reduced-Form Competition Diagnostic}\label{app:ext_competition}

To examine how competitive rent pass-through reshapes our monopoly insights, we introduce a reduced-form competition diagnostic motivated by utility-space competition and conduct-parameter approaches \citep{armstrong2001competitive, rochet2002nonlinear, mahoney2017imperfect}. Rather than solve a full structural duopoly screening game, we model competition as partially passing information rents back to consumers.
Specifically, let $\kappa\in[0,1]$ denote the degree of competitive rent pass-through and suppose the platform maximizes the reduced-form objective $\Pi+\kappa CS$.
Applying the envelope representation of consumer surplus to our one-dimensional screening problem gives the competitive virtual type
\begin{equation}\label{eq:competition_virtual_type}
v_\kappa(\theta)
=
\theta-(1-\kappa)\frac{1-F(\theta)}{f(\theta)}
=
(1-\kappa)(2\theta-1)+\kappa\theta,
\end{equation}
where the second equality uses $\theta\sim U[0,1]$.
Thus, $\kappa=0$ recovers the monopoly virtual type $2\theta-1$, while $\kappa=1$ corresponds to the efficient endpoint $\theta$.

Let $a_\kappa:=2-\kappa$, $b_\kappa:=1-\kappa$, so that $v_\kappa(\theta)=a_\kappa\theta-b_\kappa$, and define $\Delta_\kappa(\theta):=\delta v_\kappa(\theta)-\lambda$. The competitive virtual surplus is
\begin{equation}\label{eq:competition_virtual_surplus}
\mathcal V_\kappa(\theta;q,\alpha)
=
v_\kappa(\theta)q(1-\delta\alpha)
+\lambda q\alpha
+\bar r\left(\alpha-\frac{\alpha^2}{2}\right)
-\frac{q^2}{2}.
\end{equation}
Throughout this extension, we maintain $\lambda\in(0,\delta)$ and $\bar r\in(0,\delta-\lambda)$. Define
\[
\underline{\theta}_\kappa=\max\left\{0,\frac{b_\kappa-\lambda/(1-\delta)}{a_\kappa}\right\},\quad
\theta_{1,\kappa}=\frac{b_\kappa+\lambda/\delta}{a_\kappa},\quad
\theta_{2,\kappa}=\frac{b_\kappa+\frac{\lambda+\sqrt{\lambda^2+4\delta\bar r}}{2\delta}}{a_\kappa}.
\]
The interior allocations are
\[
\alpha_\kappa^{int}(\theta)=
\frac{\bar r-v_\kappa(\theta)\Delta_\kappa(\theta)}
{\bar r-\Delta_\kappa(\theta)^2},
\qquad
q_\kappa^{int}(\theta)=
\frac{\bar r[v_\kappa(\theta)-\Delta_\kappa(\theta)]}
{\bar r-\Delta_\kappa(\theta)^2}.
\]

\begin{theorem}[Optimal Mechanism under Reduced-Form Competition]
\label{thm:competition_optimal_mechanism}
Under \eqref{eq:competition_virtual_surplus}, the optimal mechanism takes the following four-segment form when $\underline{\theta}_\kappa>0$:
\begin{enumerate}[$(a)$]
    \item \textbf{Bottom types} $\theta\in[0,\underline{\theta}_\kappa]$: $q^*(\theta)=0$ and $\alpha^*(\theta)=1$.
    \item \textbf{Low types} $\theta\in(\underline{\theta}_\kappa,\theta_{1,\kappa}]$: $q^*(\theta)=v_\kappa(\theta)(1-\delta)+\lambda$ and $\alpha^*(\theta)=1$.
    \item \textbf{Medium types} $\theta\in(\theta_{1,\kappa},\theta_{2,\kappa}]$: $q^*(\theta)=q_\kappa^{int}(\theta)$ and $\alpha^*(\theta)=\alpha_\kappa^{int}(\theta)$.
    \item \textbf{High types} $\theta\ge\theta_{2,\kappa}$: $q^*(\theta)=v_\kappa(\theta)$ and $\alpha^*(\theta)=0$.
\end{enumerate}
If $\underline{\theta}_\kappa=0$, the zero-quality segment consists only of $\theta=0$ when $\lambda=(1-\kappa)(1-\delta)$; it is empty, and the low-type segment includes $\theta=0$, when $\lambda>(1-\kappa)(1-\delta)$.
\end{theorem}

\begin{proposition}[Robustness under Reduced-Form Competition]
\label{prop:competition_robustness}
For every $\kappa\in[0,1]$, the reduced-form competition model preserves the main mechanisms in Section \ref{sec:comparative_statics} in the following sense. The utility comparison uses the pure-subscription benchmark at the same $\kappa$.
\begin{enumerate}[$(a)$]
    \item \textit{Quality.} The $\delta$-driven non-monotonic quality distortion remains valid after replacing the baseline virtual-zero cutoff $1/2$ with the competitive virtual-zero cutoff $\theta_{0,\kappa}:=(1-\kappa)/(2-\kappa)$;
    \item \textit{Prices.} For a marginal increase in $\bar r$ evaluated away from segment thresholds, prices fall for users locally remaining in the interior-ad segment and rise for users locally remaining ad-free;
    \item \textit{Utility.} The utility-polarization mechanism remains valid in a weaker interval sense: an interval of low types gains from advertising, whereas all medium and high types lose relative to the pure-subscription benchmark.
\end{enumerate}
Moreover, the stronger claim that all full-ad low types gain holds if and only if $\lambda\le(1-\kappa)(1-\delta)$, equivalently $\kappa\le1-\lambda/(1-\delta)$.
\end{proposition}

The weighted objective reduces the penalty attached to information rents, which shifts the virtual type used to choose allocations.
The baseline quality and local price mechanisms survive this change, but the condition for every full-ad low type to gain becomes harder to satisfy as $\kappa$ rises: it requires $\lambda\le(1-\kappa)(1-\delta)$.
Without that condition, the proposition retains gains for an interval of low types, together with losses for medium and high types relative to pure subscription at the same $\kappa$.
These are comparisons within the weighted-objective diagnostic, not predictions of list-price responses in a strategic platform market.
The proofs are provided in Appendices~\ref{app:proof_competition_theorem} and~\ref{app:proof_competition_robustness}.

\clearpage
\section{Assumptions and Supplementary Exhibits}\label{app:discussion_supplementary}

\subsection{Discussion of Assumptions}\label{sec:discussion_assumptions}

\textbf{Functional Forms.}
Our analysis combines general structural results with a tractable closed-form baseline.
The virtual-surplus decomposition (Lemma~\ref{lem:virtual_surplus}) and the virtual advertising profit (Definition~\ref{def:virtual_ad_profit}) are derived under the general model; the explicit segment thresholds, closed-form allocations, and sharp comparative statics use the uniform-quadratic, multiplicative baseline.
The multiplicative effective-quality form is economically substantive: a given ad load destroys more user value when more quality is supplied, linking output interference and privacy friction to the quality choice.
The mixed-degradation extension adds a pure nuisance component while retaining this quality-related loss; it does not establish the same conclusions for purely additive nuisance.
The extensions in Appendix~\ref{app:model_extensions} should therefore be read under their stated restrictions as scope diagnostics rather than as proof of functional-form invariance.

\textbf{Consumer-Side Advertising Effects.}
We assume a single dimension of private information, $\theta$, which governs both the consumer's valuation of AI quality and the utility loss from advertising-induced degradation.
Under the baseline specification, $g(q,\alpha)=q(1-\delta\alpha)$, advertising lowers the effective value of quality, so higher-type consumers suffer a greater absolute loss from a given ad load.
In the LLM context, this captures that users who value high-quality, trustworthy outputs are also more exposed to output interference or privacy friction.
The effects are not uniform: in a 179-participant chatbot experiment, \citet{tang2025ads} find that users struggled to identify advertisements and sometimes preferred responses containing hidden advertising; their benchmark evaluations show only small changes in most performance measures.
The baseline therefore focuses on settings in which advertising reduces effective user value.

Three extensions clarify the scope of this consumer-side specification.
Appendix~\ref{app:ext_mixed_degradation} considers $g(q,\alpha)=q(1-\delta\alpha)-\delta_n\alpha$, which separates the quality-related harm emphasized in LLM settings from a pure nuisance component more common in traditional media.
Appendix~\ref{app:ext_correlated} allows ad sensitivity to covary with valuation by assuming that a type-$\theta$ consumer faces the high-degradation state $\delta_H$ with probability $\theta$ and the low-degradation state $\delta_L$ with probability $1-\theta$, capturing the idea that higher-value users may be more sensitive to advertising-induced degradation.
Appendix~\ref{app:ext_positive_ad_utility} augments utility to $U(\theta)=\theta q(1-\delta\alpha)+\eta q\alpha-T$, allowing advertising to generate matching or informational value.
These extensions delineate when the baseline mechanisms continue to operate under additional consumer-side advertising channels and where genuinely two-dimensional screening remains open.
The ad-load screening and divergent-pricing results are derived under this maintained scalar-type structure.
At a fixed quality and ad load, the term $v(\theta)\bigl(g(q,0)-g(q,\alpha)\bigr)=(2\theta-1)q\delta\alpha$ measures the reduction in virtual subscription value relative to ad-free service; it can be negative below the virtual-zero type.
As $\theta$ rises, a given ad load $\alpha$ becomes more costly in these virtual subscription terms.
The correlated-degradation diagnostic (Appendix~\ref{app:ext_correlated}) gives statewise comparisons while preserving the coupling between valuation and advertising sensitivity.
A genuinely two-dimensional model with independent or negatively correlated preferences for quality and advertising avoidance remains open; the existing diagnostics do not establish the same $\alpha$-schedule or four-segment structure in that setting.

\textbf{Platform-Side Advertising Revenue.}
In the baseline, advertising revenue depends on endogenous response quality, $R(q,\alpha)$, rather than directly on the consumer's type.
Because the platform assigns different quality levels $q^*(\theta)$ across segments, ad monetization becomes \textit{endogenously} type-dependent.
This captures the possibility that richer responses reveal more commercial intent and increase advertising value.
Appendix~\ref{app:ext_type_dependent} also allows a directly type-dependent premium, with $R(q,\alpha;\theta)=(\lambda+\gamma\theta)q\alpha+\bar r(\alpha-\alpha^2/2)$ and $\gamma\ge0$.
Appendix~\ref{app:ext_concave_ad_premium} instead allows diminishing marginal returns to quality, with $R(q,\alpha)=\lambda h(q)\alpha+\bar r(\alpha-\alpha^2/2)$, where $h'(q)>0$ and $h''(q)<0$.
Under their respective restrictions, these extensions identify which baseline mechanisms survive direct type dependence or a concave quality-driven premium.

\textbf{Commitment.}
We assume that the platform can commit to the posted menu of quality, advertising intensity, and prices, as is standard in screening models and broadly consistent with subscription markets where plan attributes and ad policies are disclosed ex ante. This full-commitment assumption is what licenses the revelation-principle reduction used in Section~\ref{sec:optimal_mechanism} and the binding-IR, envelope-transfer construction in Lemma~\ref{lem:virtual_surplus}. The assumption is substantive in the LLM context because sponsored responses and data-driven targeting may be difficult for users to fully monitor. We therefore view the model as a full-commitment benchmark that isolates the allocative logic of hybrid monetization.
If commitment were imperfect, the direction of the welfare effect is not obvious: ex post ad expansion would impose an additional trust constraint that plausibly deepens the losses of partial-ad and ad-free users, but reputational discipline or the unraveling of credible ad-free tiers could instead force the platform to concede additional rents.
We accordingly view the no-commitment mechanism-design problem as open rather than conjecture its resolution.

\textbf{Market Structure.}
The monopoly baseline isolates how a single platform designs its menu; its welfare results depend on the information rents left to consumers, including the high-type utility losses in Proposition~\ref{prop:lambda_utility}.
Appendix~\ref{app:ext_competition} examines this boundary through a reduced-form rent pass-through diagnostic in the spirit of conduct-parameter and pass-through approaches \citep{armstrong2001competitive, mahoney2017imperfect, weyl2013pass}: a parameter $\kappa\in[0,1]$ scales down the monopoly screening wedge, with $\kappa=0$ recovering the monopoly benchmark and $\kappa=1$ the efficient endpoint.
The ordering of the advertising regimes and the local divergent-pricing mechanism survive for every $\kappa$, allowing a zero-quality segment to be empty.
The stronger claim that \textit{all} full-ad low types gain holds if and only if $\kappa\le 1-\lambda/(1-\delta)$ (Proposition~\ref{prop:competition_robustness}).
The diagnostic captures rent pass-through, not strategic interaction: business stealing, multi-homing, and competitive discipline on ad loads remain outside the model.
It therefore preserves a conditional cross-tier comparison---a marginal increase in baseline ad profitability lowers prices for users remaining partially ad-supported and raises prices for users remaining ad-free---without establishing the price levels or advertising policies of competing platforms.

\textbf{Scope and Specificity.}
Ultimately, while the paper is motivated by generative AI, our framework should be interpreted as a broader theory of ad-supported digital platforms with endogenous quality provision, with LLM platforms serving as a leading application.
The LLM setting is particularly suited to this framework because it combines three forces that are typically studied separately.
First, quality is not a fixed content attribute: platforms can adjust model capability, inference depth, context length, and usage limits across user segments, all at increasing computational cost.
Second, advertising can interfere with the core user experience through output interference or privacy friction, rather than merely consuming peripheral attention.
Third, richer interactions can increase advertising value by revealing more precise commercial intent.
We do not claim that hybrid monetization is exclusive to LLMs. Rather, the analysis identifies how the interplay of endogenous quality, ad-induced degradation, and quality-driven ad value shapes pricing, coverage, welfare, and tier design.

\subsection{Supplementary Exhibits}\label{app:supp_tables}

Table \ref{tab:equilibrium_regimes_simplified} summarizes the optimal allocation under Assumption~\ref{ass:market_regularity}.
These bounds ($\lambda < \delta$, $\bar{r} < \delta - \lambda$) guarantee an endogenous premium ad-free segment ($\theta_2 < 1$), so the two regimes differ in whether the market's bottom receives positive quality.
Under a low quality-driven ad premium (Regime 1), a positive-measure interval of the lowest types receives zero quality.
Under a high premium (Regime 2), that interval disappears as advertising finances quality at the bottom; at the boundary between the regimes, only the lowest type still receives zero quality.
Appendix~\ref{app:extended_parameters} considers parameters outside these bounds.

\begin{table}[htbp]
\centering
\caption{Optimal Equilibrium Structure under Market Regularity}
\vspace{0.1cm}
\label{tab:equilibrium_regimes_simplified}
\renewcommand{\arraystretch}{1.5}
\resizebox{0.88\textwidth}{!}{%
\begin{tabular}{cccccc}
\toprule
\makecell{\textbf{Ad Degradation} \\ ($\delta$)} &
\makecell{\textbf{Quality-Driven} \\ \textbf{Ad Premium} \\ ($\lambda$)} &
\makecell{\textbf{Base Ad Rev} \\ ($\bar{r}$)} &
\makecell{\textbf{User Segment} \\ ($\theta$)} &
\makecell{\textbf{Ad Load} \\ ($\alpha^*$)} &
\makecell{\textbf{Quality} \\ ($q^*$)} \\
\midrule

\multirow{2}{*}{$(0, \frac{1}{2}]$} & \multirow{2}{*}{$(0, \delta)$} & \multirow{4}{*}{$(0, \delta-\lambda)$}
& $\left[0, \frac{1}{2}-\frac{\lambda}{2(1-\delta)}\right)$ & 1 & 0 \\
& & & $\left[\frac{1}{2}-\frac{\lambda}{2(1-\delta)}, \frac{1}{2}+\frac{\lambda}{2\delta}\right)$ & 1 & $(2\theta - 1)(1 - \delta) + \lambda$ \\
\cmidrule(lr){1-1} \cmidrule(lr){2-2}

\multirow{2}{*}{$(\frac{1}{2}, 1)$} & \multirow{2}{*}{$(0, 1-\delta)$} &
& $\left[\frac{1}{2}+\frac{\lambda}{2\delta}, \theta_2\right)$ & $\alpha^{int}$ & $q^{int}$ \\
& & & $\left[\theta_2, 1\right]$ & 0 & $2\theta - 1$ \\
\midrule

\multirow{3}{*}{$(\frac{1}{2}, 1)$} & \multirow{3}{*}{$[1-\delta, \delta)$}
& \multirow{3}{*}{$(0, \delta-\lambda)$}
& $\left[0, \frac{1}{2}+\frac{\lambda}{2\delta}\right)$ & 1 & $(2\theta - 1)(1 - \delta) + \lambda$ \\
& & & $\left[\frac{1}{2}+\frac{\lambda}{2\delta}, \theta_2\right)$ & $\alpha^{int}$ & $q^{int}$ \\
& & & $\left[\theta_2, 1\right]$ & 0 & $2\theta - 1$ \\
\bottomrule
\end{tabular}
}
\par\vspace{0.25cm}
\begin{minipage}{1\textwidth}
    \scriptsize \textit{Notes:} The ad-free threshold is $\theta_2 \equiv \frac{2\delta+\lambda+\sqrt{\lambda^2+4\delta\bar{r}}}{4\delta}$. Interior solutions $q^{int}$ and $\alpha^{int}$ are defined in \eqref{eq:interior_allocations}. The upper row-block (four segments) corresponds to Regime~1, with a low quality-driven ad premium ($\lambda<1-\delta$); the lower row-block (three positive-measure segments) corresponds to Regime~2, with a high premium ($\lambda\ge1-\delta$). In Regime~2, only the lowest type receives zero quality at equality, and the zero-quality set is empty under a strict inequality.
\end{minipage}
\end{table}

Table~\ref{tab:cs_summary} summarizes the segment-level comparative statics of Section~\ref{sec:comparative_statics}.

\begin{table}[htbp]
\centering
\caption{Comparative Statics across Market Segments (Propositions \ref{prop:delta_quality}--\ref{prop:lambda_utility})}
\label{tab:cs_summary}
\renewcommand{\arraystretch}{1.6}
\small
\begin{tabular}{lcccc}
\toprule
 & \makecell{\textbf{Zero-quality}\\($\alpha^*=1,\,q^*=0$)} & \makecell{\textbf{Low}\\($\alpha^*=1$)} & \makecell{\textbf{Medium}\\($\alpha^{int}\in(0,1)$)} & \makecell{\textbf{High}\\($\alpha^*=0$)} \\
\midrule
\makecell[l]{Quality $q^*$ as $\delta\uparrow$\\(Prop.~\ref{prop:delta_quality})} & upgraded$^{a}$ & \makecell{$\uparrow$ if $\theta<\tfrac12$\\$\downarrow$ if $\theta>\tfrac12$} & \makecell{$\downarrow$ if $\alpha^{int}>\tfrac12$\\$\uparrow$ if $\alpha^{int}<\tfrac12$} & unchanged \\
\makecell[l]{Price $T^*$ as $\bar r\uparrow$\\(Prop.~\ref{prop:rbar_pricing})} & unchanged & unchanged & $\downarrow$ & $\uparrow$$^{b}$ \\
\makecell[l]{Utility vs.\ pure subscription\\(Prop.~\ref{prop:lambda_utility})} & unchanged & $+$$^{c}$ & $-$ & $-$ \\
\bottomrule
\end{tabular}
\par\vspace{0.2cm}
\begin{minipage}{1\textwidth}
\scriptsize \textit{Notes:} Quality and price arrows describe local responses within the indicated regimes. Quality is unchanged at the median full-ad type and to first order at the interior half-ad-load cutoff. $^{a}$When the positive-quality threshold $\underline{\theta}$ lies inside the type support, it shifts leftward and upgrades the marginal users it crosses; users below the new cutoff remain at zero quality. $^{b}$For a marginal increase in baseline advertising profitability evaluated away from segment thresholds, prices fall for users locally remaining in the interior-ad segment and rise for users locally remaining ad-free. $^{c}$All low types gain when $\lambda\le1-\delta$; a subset gains otherwise.
\end{minipage}
\end{table}

Remark~\ref{rmk:convex_necessity} clarifies how strict convexity shapes the form of quality adjustment underlying the baseline result.

\begin{remark}[Role of Convex Quality Costs]\label{rmk:convex_necessity}
Strict convexity supports the smooth, type-dependent quality adjustments underlying the baseline characterization of Proposition~\ref{prop:lambda_utility}, but it is not necessary for utility polarization itself. If quality were costless and unconstrained, the platform's pointwise virtual surplus $\mathcal{V}(\theta;q,\alpha)$ would be unbounded in $q$ for any type with a positive marginal virtual benefit; with an independently fixed finite technological ceiling, the platform would instead choose maximum feasible quality whenever this marginal benefit is positive. More generally, under linear costs ($c(q)=\mu q$), quality becomes a threshold decision: provide maximum feasible quality if the marginal virtual benefit exceeds $\mu$, and zero otherwise. Linear costs therefore replace the baseline's smooth, type-dependent quality adjustments with boundary choices. Advertising and information-rent effects can nevertheless still generate utility polarization, so strict convexity shapes the form of the baseline allocation rather than serving as a necessary condition for the polarization result.
\end{remark}

Figure~\ref{fig:cap_micro} illustrates the allocation mechanism behind the interior-cap optimum of Proposition~\ref{prop:ad_cap_welfare} at the stated parameters.
Relative to the unregulated menu, the interior cap raises effective quality for many low types, while a complete ban leaves the bottom without positive quality.
For the ad-free top, effective quality coincides under the interior cap and the unregulated menu, yet utility differs because changes in lower plans alter the information rents left to higher types.

\begin{figure}[htbp]
	\centering
	\subfloat[Effective Quality $g(q^*, \alpha^*)$]{
		\includegraphics[height=0.30\textwidth]{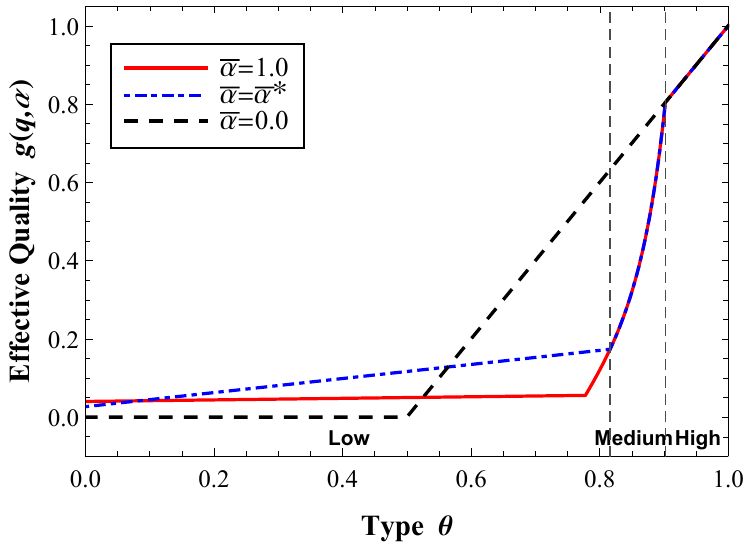}}\quad
	\subfloat[Consumer Utility $U(\theta)$]{
		\includegraphics[height=0.30\textwidth]{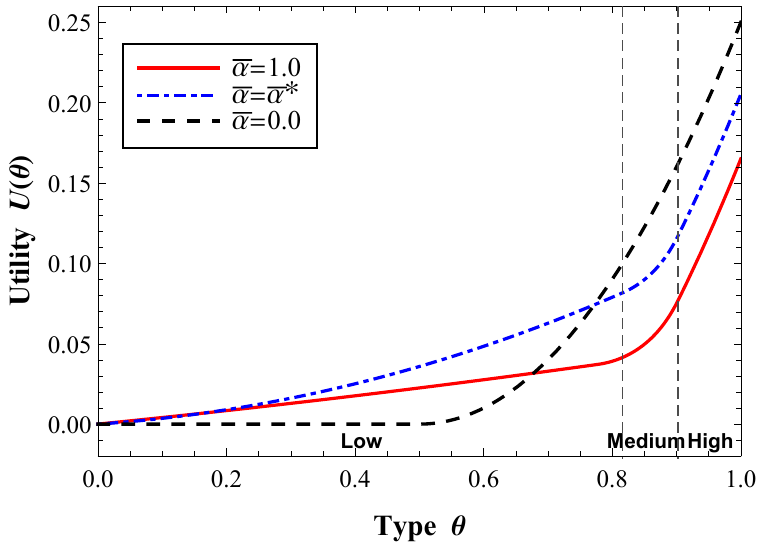}}
    \par\vspace{0.2cm}
    \begin{minipage}{1\textwidth}
        \centering \scriptsize \textit{Notes:} The vertical dashed lines and segment labels refer to the illustrated capped menu at $\bar{\alpha}=\bar{\alpha}^*$. From left to right, the lines mark the boundary between cap-bound and interior advertising allocations, and the boundary between interior advertising and ad-free service. All consumer types receive positive quality in this reference menu. The curves compare different caps, so a given consumer can change segments.
    \end{minipage}
    \caption{Advertising Caps: The Underlying Mechanism ($\lambda=0.5$, $\bar{r}=0.18$, $\delta=0.9$)}
    \label{fig:cap_micro}
\end{figure}

\clearpage

Table~\ref{tab:tier_exact_dynamics} reports the exact tier phase transitions underlying Proposition~\ref{prop:discrete_tiers}.

\begin{table}[htbp]
\centering
\caption{Phase Transitions in Positive-Quality Service Tiers ($\delta = 0.8$, $\bar{r} = 0.2$, $\theta_M = 2/3$, $\bar q=1$)}
\label{tab:tier_exact_dynamics}
\renewcommand{\arraystretch}{1}
\setlength{\tabcolsep}{5pt}
\resizebox{0.95\textwidth}{!}{
\scriptsize
\begin{tabular}{ccccc}
\toprule
\begin{tabular}[c]{@{}c@{}}\textbf{Number of} \\ \textbf{Positive-Quality Tiers}\end{tabular} &
\begin{tabular}[c]{@{}c@{}}\textbf{Quality-Driven Ad Premium} \\ ($\lambda$ Interval)\end{tabular} &
\begin{tabular}[c]{@{}c@{}}\textbf{Type $L$} \\ ($v_L = -4/3$)\end{tabular} &
\begin{tabular}[c]{@{}c@{}}\textbf{Type $M$} \\ ($v_M = 1/3$)\end{tabular} &
\begin{tabular}[c]{@{}c@{}}\textbf{Type $H$} \\ ($v_H = 1$)\end{tabular} \\
\midrule

2 & $\left(0, 4/15\right]$
& \multicolumn{1}{c}{\cellcolor{gray!12}
\begin{tabular}[c]{@{}c@{}}
$\alpha_L^*=1$, $q_L^*=0$ \\
\textit{(Zero-Quality)}
\end{tabular}}
& \begin{tabular}[c]{@{}c@{}}
$\alpha_M^*=\alpha^{int}_M$ \\
$q_M^*=q^{int}_M$
\end{tabular}
& \begin{tabular}[c]{@{}c@{}}
$\alpha_H^*=0$ \\
$q_H^*=v_H$
\end{tabular} \\
\midrule

\multirow{3.5}{*}{3}
& $\left(4/15, 3/5\right]$
& \multirow{3.5}{*}{
\begin{tabular}[c]{@{}c@{}}
$\alpha_L^*=1$ \\
$q_L^*=\tilde{q}_L(1)$
\end{tabular}}
& \multirow{3.5}{*}{
\begin{tabular}[c]{@{}c@{}}
$\alpha_M^*=1$ \\
$q_M^*=\tilde{q}_M(1)$
\end{tabular}}
& $\alpha_H^*=0,\ q_H^*=1$ \\
\cmidrule(lr){2-2} \cmidrule(lr){5-5}

& $\left(3/5, 4/5\right)$
&
&
& $\alpha_H^*=\alpha_H^{int},\ q_H^*=q_H^{int}$ \\
\cmidrule(lr){2-2} \cmidrule(lr){5-5}

& $\left[4/5, 14/15\right)$
&
&
& $\alpha_H^*=1,\ q_H^*=1$ \\
\midrule

2 & $\left[14/15, 19/15\right)$
& \begin{tabular}[c]{@{}c@{}}
$\alpha_L^*=1$ \\
$q_L^*=\tilde{q}_L(1)$
\end{tabular}
& \multicolumn{2}{c}{\cellcolor{blue!10}
\begin{tabular}[c]{@{}c@{}}
$\alpha_{\mathrm{Pool}}^*=1$, $q_{\mathrm{Pool}}^*=\bar{q}$ \\
\textit{(Top Pooling)}
\end{tabular}} \\
\midrule

1 & $\left[19/15, +\infty\right)$
& \multicolumn{3}{c}{\cellcolor{blue!10}
$\alpha_{\mathrm{Pool}}^*=1$, $q_{\mathrm{Pool}}^*=\bar{q}$ \quad \textit{(Universal Pooling)}} \\
\bottomrule
\end{tabular}
}
\par\vspace{0.25cm}
\begin{minipage}{1\textwidth}
    \scriptsize \textit{Notes:} Define $\tilde{q}_i(\alpha)=v_i(1-\delta\alpha)+\lambda\alpha$.
    Interior solutions $q_i^{int}$ and $\alpha_i^{int}$ are defined in \eqref{eq:interior_allocations_discrete}.
    The gray cell denotes the zero-quality baseline in Remark~\ref{rem:zero_quality_baseline}; blue cells denote pooling.
At the tabulated parameters ($\theta_M=2/3$, $\delta=0.8$), the configuration lies on the knife-edge $\delta v_M=\lambda_L=4/15$, so type $M$'s transition to the full-ad corner coincides with the onset of type $L$'s positive quality; hence $\alpha_M^*=1$ throughout the three-tier region shown. For $\theta_M>\delta/[2(2\delta-1)]$ (e.g., $\theta_M=0.7$ at $\delta=0.8$), type $M$ is interior ($\alpha_M^*=\alpha_M^{int}<1$) on $(\lambda_L,\delta v_M)$ before reaching the full-ad corner, while the positive-quality tier count is unchanged.
\end{minipage}
\end{table}

\clearpage

\section{Additional Equilibrium and Welfare Results}\label{app:supplement}
This appendix examines three departures or comparisons beyond the baseline characterization: equilibrium outside the regular parameter region, the local social-welfare effect of advertising caps, and the optimal mechanism under lower compute costs.

\subsection{Equilibrium Structure under Extended Parameter Spaces}\label{app:extended_parameters}

The baseline Market Regularity condition (Assumption~\ref{ass:market_regularity}) imposes $\lambda<\delta$ and $\bar r<\delta-\lambda$ to sustain an endogenous ad-free high-type segment.
Under the normalized baseline ceiling $\bar q=1$, these restrictions also avoid a positive-measure quality-constrained segment.
{We now relax these advertising-return bounds to examine larger quality-driven ad premiums or baseline impression profitability, while maintaining $0<\delta<1$.}
Stronger advertising returns can make advertising profitable even for high types, eliminating the positive-measure ad-free segment.
When the quality-driven premium is sufficiently large, the added advertising return to quality can also push the optimum to the upper bound $\bar q=1$.
Table~\ref{tab:equilibrium_regimes} characterizes the resulting regimes; the discussion below then distinguishes local allocation and price responses from the utility comparison with pure subscription.

\begin{table}[htbp]
\centering
\caption{Optimal Equilibrium Structure outside the Regular Parameter Region}
\vspace{0.1cm}
\label{tab:equilibrium_regimes}
\renewcommand{\arraystretch}{1.3}
\resizebox{0.88\textwidth}{!}{%
\begin{tabular}{cccccc}
\toprule
\makecell{\textbf{Ad Degradation} \\ ($\delta$)} &
\makecell{\textbf{Quality-Driven} \\ \textbf{Ad Premium} \\ ($\lambda$)} &
\makecell{\textbf{Base Ad Rev} \\ ($\bar{r}$)} &
\makecell{\textbf{User Segment} \\ ($\theta$)} &
\makecell{\textbf{Ad Load} \\ ($\alpha^*$)} &
\makecell{\textbf{Quality} \\ ($q^*$)} \\
\midrule

\multirow{3}{*}{{$(0,1)$}} & \multirow{3}{*}{{$[0,\min\{\delta,1-\delta\})$}} & \multirow{3}{*}{{$[\delta-\lambda,+\infty)$}}
& $\left[0, \frac{1}{2}-\frac{\lambda}{2(1-\delta)}\right)$                                      & 1              & 0         \\
& & & $\left[\frac{1}{2}-\frac{\lambda}{2(1-\delta)}, \frac{1}{2}+\frac{\lambda}{2\delta}\right)$ & 1 & $(2\theta - 1)(1 - \delta) + \lambda$ \\
& & & $\left[\frac{1}{2}+\frac{\lambda}{2\delta}, 1\right]$                                        & $\alpha^{int}$ & $q^{int}$ \\
\midrule

\multirow{3}{*}{{$\left(0, \frac{1}{2}\right)$}} & \multirow{3}{*}{$[\delta, 1-\delta)$} & \multirow{3}{*}{$[0, +\infty)$}
& $\left[0, \frac{1}{2}-\frac{\lambda}{2(1-\delta)}\right)$                                      & 1              & 0         \\
& & & $\left[\frac{1}{2}-\frac{\lambda}{2(1-\delta)}, \frac{1}{2}+\frac{1-\lambda}{2(1-\delta)}\right)$                                      & 1              & $(2\theta - 1)(1 - \delta) + \lambda$ \\
& & & $\left[ \frac{1}{2}+\frac{1-\lambda}{2(1-\delta)}, 1\right]$                                      & 1              & $\bar{q}=1$ \\
\midrule

\multirow{2}{*}{{$\left[\frac{1}{2}, 1\right)$}} & \multirow{2}{*}{$[1-\delta, \delta)$} & \multirow{2}{*}{$[\delta-\lambda, +\infty)$}
& $\left[0, \frac{1}{2}+\frac{\lambda}{2\delta}\right)$                                        & 1              & $(2\theta - 1)(1 - \delta) + \lambda$ \\
& & & $\left[\frac{1}{2}+\frac{\lambda}{2\delta}, 1\right]$                                        & $\alpha^{int}$ & $q^{int}$ \\
\midrule

\multirow{2}{*}{{$(0, 1)$}} & \multirow{2}{*}{$[\max\{\delta, 1-\delta\}, +\infty)$} & \multirow{2}{*}{$[0, +\infty)$}
& {$[0,\theta_c)$}                                        & 1              & $(2\theta - 1)(1 - \delta) + \lambda$ \\
& & & {$[\theta_c,1]$}                                        & 1 & $\bar{q}=1$ \\

\bottomrule
\end{tabular}
}
\par\vspace{0.25cm}
\begin{minipage}{1\textwidth}
    \scriptsize \textit{Notes:} Interior solutions $q^{int}$ and $\alpha^{int}$ are defined in \eqref{eq:interior_allocations}.
    {The table includes $\lambda=0$ or $\bar r=0$ as boundary cases and reports an optimal allocation when the optimum is not unique.}
    The four row-blocks, from top to bottom, correspond to Regimes~1--4 referenced in the text below: Regime~1 (interior solutions; $\bar r\ge\delta-\lambda$ with small $\lambda$), Regime~2 ($\lambda\in[\delta,1-\delta)$; quality binds at $\bar q$), Regime~3 ($\lambda\in[1-\delta,\delta)$, $\bar r\ge\delta-\lambda$; interior solutions), and Regime~4 ($\lambda\ge\max\{\delta,1-\delta\}$; quality binds at $\bar q$).
    {In Regime~4, define $\theta_c:=\max\{0,\frac12+\frac{1-\lambda}{2(1-\delta)}\}$. When $\lambda\ge2-\delta$, $\theta_c=0$ and all consumer types receive $(q^*,\alpha^*)=(1,1)$, yielding universal pooling.}
\end{minipage}
\end{table}

\textbf{Quality Responses to Ad Degradation ($\delta$).} Under Assumption~\ref{ass:market_regularity}, Proposition~\ref{prop:delta_quality} leaves users who remain ad-free at their pure-subscription monopoly quality. Outside that region (where $\bar{r} \ge \delta - \lambda$ or $\lambda$ is sufficiently large), the positive-measure ad-free segment disappears, and quality responses depend on both ad intensity and the technological ceiling.
For below-median full-ad users, greater degradation makes their plans less attractive to higher types and lowers the screening cost of quality provision. Provided quality is positive and remains below the ceiling, the platform therefore raises supplied quality, so $\partial q^*/\partial\delta>0$; zero-quality types remain at $q^*=0$ unless a moving threshold brings them into positive quality.
For types $\theta > 1/2$ who remain fully advertised below the ceiling in Regimes 2 and 4, greater degradation lowers the virtual return to quality, giving $\frac{\partial q^*}{\partial \delta} < 0$.
For any type whose upper quality bound remains binding, quality is unchanged ($q^*=1$, yielding $\frac{\partial q^*}{\partial \delta} = 0$).
Within the partial-ad segments of Regimes 1 and 3, the baseline heavy-ad/light-ad sign reversal remains, but there is no positive-measure ad-free endpoint.

\textbf{Price Responses to Baseline Advertising Profitability ($\bar{r}$).} Proposition~\ref{prop:rbar_pricing} establishes that a marginal increase in baseline ad profitability lowers prices for users remaining partially ad-supported and raises prices for users remaining ad-free. The latter segment loses positive measure at $\bar{r} = \delta - \lambda$.
When $\bar{r} \ge \delta - \lambda$ in Regimes 1 and 3, types $\theta > \theta_1$ receive the interior allocation; their ad loads satisfy $\alpha^{int} > 0$ except that the highest type remains ad-free at equality.
{For users who remain in the partial-ad segment, a marginal increase in $\bar r$ lowers prices as their effective quality declines.} The upper-tier price-increase mechanism has no positive-measure ad-free segment on which to operate.
Prices for zero-quality and full-ad low types remain locally unaffected by $\bar r$, because their ad load stays at the upper bound $\alpha^*=1$.

\textbf{Utility Comparisons outside the Regular Region ($\lambda$).} Within the regular region, Proposition~\ref{prop:lambda_utility} shows that some fully ad-supported low types gain relative to pure subscription, while partially ad-supported and ad-free users strictly lose.
Outside that region, positive advertising exposure alone does not determine the utility comparison.
When $\lambda\ge\delta$ (Regimes~2 and~4), the marginal virtual return to advertising is nonnegative for every type, so $\alpha^*(\theta)=1$ and
\[
q^*(\theta)=\min\{1,\max\{0,(1-\delta)(2\theta-1)+\lambda\}\},
\qquad
U^{ad}(\theta)=(1-\delta)\int_0^\theta q^*(t)\,dt.
\]
The pure-subscription benchmark gives $U^{sub}(1)=1/4$.
In Regime~2, $U^{ad}(1)-U^{sub}(1)=(\lambda-\delta)/2$, so the highest type gains when $\lambda>\delta$.
At universal pooling, which occurs when $\lambda\ge2-\delta$, utility becomes $U^{ad}(\theta)=(1-\delta)\theta$ and the highest-type difference is $3/4-\delta$.
Every positive type then gains when $\delta<3/4$; the highest type is indifferent at $\delta=3/4$, while an upper interval loses when $\delta>3/4$.
Thus sufficiently strong advertising returns can generate broad utility gains when degradation is mild, but even universal provision of maximum quality can leave high types worse off when degradation is severe.
Proposition~\ref{prop:lambda_utility}'s utility-polarization result remains specific to its stated regular parameter region.

\subsection{Advertising Caps and Social Welfare}\label{app:social_welfare}

Returning to the baseline specification under Assumption~\ref{ass:market_regularity}, we examine how the advertising-cap mechanism in Section~\ref{subsec:ad_cap} affects social welfare, defined as consumer surplus plus platform profit.
This comparison is more demanding than the consumer-surplus comparison: although a cap can reduce advertising-induced degradation and benefit consumers, a binding cap restricts the platform's feasible choice set and reduces its optimal profit.
A cap therefore raises social welfare only when the consumer-surplus gain exceeds the profit loss.

\begin{corollary}\label{cor:social_welfare}
A marginal tightening of the unregulated advertising level can raise social welfare only under extreme ad-induced degradation.
In particular, such a local social-welfare gain requires
\[
\delta>\frac{1+2\sqrt{2}}{4}\approx0.957.
\]
When this condition holds, the gain occurs for the intermediate interval of the quality-driven ad premium characterized in the proof.
\end{corollary}

Corollary~\ref{cor:social_welfare} identifies a narrow local possibility.
For $\delta\le (1+2\sqrt{2})/4$, marginally tightening the unregulated benchmark cannot raise social welfare through the mechanism analyzed here.
Only at higher degradation, and for the intermediate premium interval specified in the proof, can the gain from improved effective quality and reduced screening distortion exceed the platform's lost profit.
This comparison concerns a sufficiently small tightening from the unregulated benchmark; it does not characterize the globally optimal cap or imply that a larger tightening also improves welfare.

\subsection{Equilibrium under Compute Efficiency}\label{app:beta_extension}

This subsection maintains Assumption~\ref{ass:market_regularity}, relaxes the baseline normalization $\beta=1$, and characterizes the optimal mechanism for $\beta\in(0,1]$.
The cost function is $c(q)=\frac{\beta q^2}{2}$, so a lower $\beta$ reduces the marginal cost of delivering quality.
We hold the other primitives fixed: $\theta\sim U[0,1]$, $g(q,\alpha)=q(1-\delta\alpha)$, and $R(q,\alpha)=\lambda q\alpha+\bar r\left(\alpha-\frac{\alpha^2}{2}\right)$.
We maintain the technological-ceiling convention $c'(\bar q)=1$, which implies $\bar q=1/\beta$; the ceiling therefore expands as compute becomes more efficient, so the allocations characterized below lie within the feasible quality range.
Let $v(\theta):=2\theta-1$ and $\Delta(\theta):=v(\theta)\delta-\lambda$.
The pointwise virtual surplus is
\begin{equation}\label{eq:beta_virtual_surplus}
\mathcal V_\beta(\theta;q,\alpha)
=
v(\theta)q(1-\delta\alpha)
+\lambda q\alpha
+\bar r\left(\alpha-\frac{\alpha^2}{2}\right)
-\frac{\beta q^2}{2}.
\end{equation}

For a fixed ad load $\alpha$, the first-order condition with respect to quality yields
\[
q_\beta^*(\theta,\alpha)
=
\max\left\{0,\frac{v(\theta)(1-\delta\alpha)+\lambda\alpha}{\beta}\right\}.
\]
The marginal virtual surplus with respect to advertising is
\[
\frac{\partial \mathcal V_\beta}{\partial \alpha}
=
-\Delta(\theta)q+\bar r(1-\alpha).
\]
We first define the relevant thresholds.

\begin{definition}[Thresholds under Compute Efficiency]\label{def:beta_thresholds}
Define the following thresholds:
\begin{enumerate}[$(a)$]
    \item \textit{Positive-quality threshold}: $\underline{\theta} := \frac{1}{2}-\frac{\lambda}{2(1-\delta)}$, above which the platform provides strictly positive quality.
    \item \textit{Partial-ad threshold}: $\theta_1 := \frac{1}{2}+\frac{\lambda}{2\delta}$, above which the platform transitions to a partial ad load.
    \item \textit{Ad-free threshold}: $\theta_{2,\beta}\in(\theta_1,1]$ uniquely solving $(2\theta_{2,\beta}-1)\big((2\theta_{2,\beta}-1)\delta-\lambda\big)=\beta\bar r$, or equivalently, $\theta_{2,\beta} = \frac{2\delta+\lambda+\sqrt{\lambda^2+4\delta\beta\bar r}}{4\delta}$.
\end{enumerate}
\end{definition}

The first two thresholds are unchanged relative to the baseline model, while the ad-free threshold shifts with $\beta$. We now characterize the optimal mechanism.

\begin{theorem}[Optimal Mechanism under Compute Efficiency]\label{thm:beta_optimal_mechanism}
The optimal mechanism takes the following form:
\begin{enumerate}[$(a)$]
    \item \textbf{Zero-quality types} ($\theta\in[0,\underline{\theta}]$, empty when $\underline{\theta}<0$):
    $q_\beta^*(\theta)=0$, $\alpha_\beta^*(\theta)=1$, and $T_\beta^*(\theta)=0$.

    \item \textbf{Low types} ($\theta\in[0,1]$ with $\underline{\theta}<\theta\le\theta_1$):
    $\alpha_\beta^*(\theta)=1$ and $q_\beta^*(\theta)=\frac{(2\theta-1)(1-\delta)+\lambda}{\beta}>0$.
    The payment at $\theta=0$ is $T_\beta^*(0)=0$ when $\underline{\theta}<0$.

    \item \textbf{Medium types} ($\theta\in(\theta_1,\theta_{2,\beta})$):
    \[
    \alpha_\beta^{int}(\theta)
    :=
    \frac{\beta\bar r-(2\theta-1)\big((2\theta-1)\delta-\lambda\big)}
    {\beta\bar r-\big((2\theta-1)\delta-\lambda\big)^2},
    \qquad
    q_\beta^{int}(\theta)
    :=
    \frac{\bar r\big[\lambda+(2\theta-1)(1-\delta)\big]}
    {\beta\bar r-\big((2\theta-1)\delta-\lambda\big)^2}.
    \]

    \item \textbf{High types} ($\theta\in[\theta_{2,\beta},1]$):
    $\alpha_\beta^*(\theta)=0$ and $q_\beta^*(\theta)=\frac{2\theta-1}{\beta}$.
\end{enumerate}
Moreover, the effective quality $g(q_\beta^*(\theta),\alpha_\beta^*(\theta))$ remains non-decreasing in $\theta$.
\end{theorem}

Theorem~\ref{thm:beta_optimal_mechanism} preserves the baseline ordering of advertising regimes; subscription payments are recovered from \eqref{eq:transfer}.
A lower $\beta$ raises positive quality by making it cheaper to supply.
For users who remain partially advertised, higher quality increases the net virtual cost of advertising-induced degradation, so the platform reduces ad intensity.
Once even the first increment of advertising is no longer profitable, a plan becomes ad-free; accordingly, the threshold $\theta_{2,\beta}$ falls and the ad-free segment expands.
The condition for a zero-quality segment is unchanged because $\underline{\theta}=1/2-\lambda/[2(1-\delta)]$ does not depend on $\beta$.
That segment has positive measure if $\lambda<1-\delta$, consists only of $\theta=0$ if $\lambda=1-\delta$, and is empty if $\lambda>1-\delta$.
The proof is provided in Appendix~\ref{app:proof_beta_optimal_mechanism}.

\clearpage
\section{Proofs}\label{app:proofs}

We first derive the virtual-surplus representation, solve the pointwise allocation problem, and verify that the resulting menu is incentive compatible. The subsequent proofs establish the quality, price, utility, and welfare comparisons, followed by the compute-efficiency and discrete-tier results. Further proofs cover the model extensions. Each proof maintains the assumptions stated for its result.

\subsection{Proof of Lemma \ref{lem:envelope}}

\begin{proof}{Proof:}
\textbf{Part (a), envelope condition.}
Let $\hat{U}(\theta,\theta'):=\theta g(q(\theta'),\alpha(\theta'))-T(\theta')$ denote the utility of type $\theta$ when reporting $\theta'$.
By incentive compatibility,
\[
U(\theta)=\hat{U}(\theta,\theta)=\max_{\theta'\in[0,1]}\hat{U}(\theta,\theta').
\]
For each report, $\hat U(\theta,\theta')$ is affine in $\theta$, and the maintained allocation bounds keep its slope $g(q(\theta'),\alpha(\theta'))$ uniformly bounded. Thus $U$ is finite and convex, and the bounded slopes make it Lipschitz, hence absolutely continuous and differentiable almost everywhere. The Envelope Theorem \citep{milgrom2002envelope} then gives
\[
U'(\theta)=g(q(\theta),\alpha(\theta)) \quad \text{a.e.}
\]

\textbf{Part (a), monotonicity.}
For any $\theta>\theta'$, IC implies
\begin{align*}
U(\theta) &\ge U(\theta')+(\theta-\theta')g(q(\theta'),\alpha(\theta')),\\
U(\theta') &\ge U(\theta)+(\theta'-\theta)g(q(\theta),\alpha(\theta)).
\end{align*}
Summing and dividing by $\theta-\theta'>0$ gives $g(q(\theta),\alpha(\theta))\ge g(q(\theta'),\alpha(\theta'))$, the monotonicity condition for implementability.

\textbf{Part (b): Reduction to virtual surplus.}

The platform's expected profit is
\[
\Pi = \int_0^1 \left[ T(\theta) + R(q(\theta), \alpha(\theta)) - c(q(\theta)) \right] dF(\theta).
\]
By part (a), $U'=g\ge0$. Feasibility forces $U(0)=0$: since $U(0)=-T(0)$, IR and $T(0)\ge0$ imply $T(0)=0$. Therefore,
\[
T(\theta)
=
\theta g(q(\theta),\alpha(\theta))
-
\int_0^\theta g(q(t),\alpha(t))\,dt.
\]
Substitution and Fubini's theorem give
\begin{align*}
\int_0^1\left(\int_0^\theta g(q(t),\alpha(t))\,dt\right)dF(\theta)
&=
\int_0^1 g(q(t),\alpha(t))\left(\int_t^1 f(\theta)\,d\theta\right)dt \\
&=
\int_0^1 g(q(\theta),\alpha(\theta))[1-F(\theta)]\,d\theta.
\end{align*}
Using $dF(\theta)=f(\theta)d\theta$ then yields
\begin{align*}
\Pi
&=
\int_0^1
\left[
\theta g(q(\theta),\alpha(\theta))
-
\frac{1-F(\theta)}{f(\theta)}g(q(\theta),\alpha(\theta))
+
R(q(\theta),\alpha(\theta))
-
c(q(\theta))
\right] f(\theta)\,d\theta \\
&=
\int_0^1
\left[
\left(\theta-\frac{1-F(\theta)}{f(\theta)}\right)
g(q(\theta),\alpha(\theta))
+
R(q(\theta),\alpha(\theta))
-
c(q(\theta))
\right] dF(\theta).
\end{align*}

Defining $v(\theta)=\theta-[1-F(\theta)]/f(\theta)$ and $\mathcal V(\theta;q,\alpha):=v(\theta)g(q,\alpha)+R(q,\alpha)-c(q)$, the allocation problem becomes
\[
\max_{q(\cdot),\alpha(\cdot)}
\int_0^1 \mathcal V(\theta;q(\theta),\alpha(\theta))\,dF(\theta).
\]
The original allocation bounds continue to apply, together with the monotonicity of $g(q(\theta),\alpha(\theta))$ established in part (a). These constraints complete the virtual-surplus representation. \Halmos
\end{proof}

\subsection{Derivation of Pointwise Optimal Allocations}
\label{app:proof_pointwise}

\begin{proof}{Proof:}
We derive the conditional quality choice and the interior allocation. The proof of Theorem~\ref{thm:optimal_mechanism} below identifies the applicable regimes and verifies implementability. Under the main specification $\theta \sim U[0,1]$, we have $f(\theta)=1$ and $v(\theta)=2\theta-1$.
With $c(q)=q^2/2$ and $g(q,\alpha)=q(1-\delta\alpha)$, pointwise virtual surplus is
\[
\mathcal V(\theta;q,\alpha)
=
(2\theta-1)q(1-\delta\alpha)
+\lambda q\alpha
+\bar r\!\left(\alpha-\frac{\alpha^2}{2}\right)
-\frac{q^2}{2}.
\]

\textbf{First-order conditions.}
At an interior choice, the FOCs with respect to $q$ and $\alpha$ are
\begin{align}
\frac{\partial \mathcal{V}}{\partial q} &= (2\theta - 1)(1 - \delta\alpha) + \lambda\alpha - q = 0, \label{eq:proof_foc_q_spec} \\
\frac{\partial \mathcal{V}}{\partial \alpha} &= -(2\theta - 1)q\delta + \lambda q + \bar{r}(1 - \alpha) = 0. \label{eq:proof_foc_alpha_spec}
\end{align}
Imposing $q\ge0$ in \eqref{eq:proof_foc_q_spec} yields
\[
q^*(\theta,\alpha)
=
\max\{0,(2\theta-1)(1-\delta\alpha)+\lambda\alpha\},
\]
which corresponds to Equation~\eqref{eq:foc_q}.
Define $\Delta(\theta):=(2\theta-1)\delta-\lambda$; then \eqref{eq:proof_foc_alpha_spec} becomes $\bar r(1-\alpha)=q\Delta(\theta)$.

\textbf{Second-order conditions.}
The Hessian is
\[
H=\begin{pmatrix}
-1 & -\Delta(\theta) \\
-\Delta(\theta) & -\bar{r}
\end{pmatrix}.
\]
Negative definiteness requires $|H|=\bar r-\Delta(\theta)^2>0$. Under Assumption~\ref{ass:market_regularity}, this holds on $(\theta_1,\theta_2)$: $\Delta(\theta_1)=0$, while $(2\theta_2-1)\Delta(\theta_2)=\bar r$ implies $\Delta(\theta_2)^2<\bar r$, and $\Delta$ is increasing.

\textbf{Interior solutions.}
For $\alpha\in(0,1)$ and $q>0$, substitute $q=(2\theta-1)-\alpha\Delta(\theta)$ into $\bar r(1-\alpha)=q\Delta(\theta)$ and solve:
\begin{align*}
\alpha^{int}(\theta)
&=
\frac{\bar{r}-(2\theta - 1)\Delta(\theta)}
{\bar{r}-\Delta(\theta)^2},\qquad
q^{int}(\theta)
=
\frac{\bar{r}\bigl((2\theta - 1)-\Delta(\theta)\bigr)}
{\bar{r}-\Delta(\theta)^2}.
\end{align*}
This recovers \eqref{eq:interior_allocations}. \Halmos

\end{proof}

\subsection{Proof of Proposition \ref{prop:negative_types}}

\begin{proof}{Proof:}
For $\theta\le1/2$, $\Delta(\theta):=(2\theta-1)\delta-\lambda<0$. Hence, for $\alpha\in[0,1)$, $\partial\mathcal V/\partial\alpha=\bar r(1-\alpha)-q\Delta(\theta)>0$ because $q\ge0$. Thus $\alpha^*(\theta)=1$.

Substitution into \eqref{eq:proof_foc_q_spec} yields
\[
q^*(\theta)
=
\max\{0,(2\theta-1)(1-\delta)+\lambda\}.
\]
Since $\lambda>0$, the term $(2\theta-1)(1-\delta)+\lambda$ is strictly positive for $\theta$ sufficiently close to $1/2$, so some negative-virtual-type users receive strictly positive quality. \Halmos
\end{proof}

\subsection{Proof of Theorem \ref{thm:optimal_mechanism}}

\begin{proof}{Proof:}
We identify the thresholds in Definition~\ref{def:thresholds} from the quality and advertising boundaries, then verify that effective quality is ordered across types. Let $q^*(\theta,\alpha)$ denote optimal quality conditional on $\alpha$, and set $q^*(\theta):=q^*(\theta,\alpha^*(\theta))$.

\textbf{Threshold characterization.}
On the positive-quality branch of the pointwise solution, quality at $\alpha=1$ satisfies
\[
q^*(\theta,1)=(2\theta-1)(1-\delta)+\lambda.
\]
The condition $q^*(\theta,1)>0$ is equivalent to
\[
(2\theta-1)(1-\delta)+\lambda>0
\;\;\Longleftrightarrow\;\;
\theta>\underline{\theta}
:=
\frac{1}{2}-\frac{\lambda}{2(1-\delta)}.
\]
Thus $\underline\theta<1/2$, with $\underline\theta\le0$ if and only if $\lambda\ge1-\delta$. For feasible types $\theta\in[0,1]$, $(q^*,\alpha^*)=(0,1)$ if and only if $\theta\le\underline\theta$. Hence, the zero-quality segment is $[0,\underline\theta]$ when $\underline\theta\ge0$ and is empty when $\underline\theta<0$.

The transition from full advertising to an interior solution is determined by the marginal virtual surplus at $\alpha=1$:
\[
\left.\frac{\partial \mathcal V}{\partial \alpha}\right|_{\alpha=1}
=
-q^*(\theta,1)\Delta(\theta),
\qquad
\Delta(\theta):=(2\theta-1)\delta-\lambda.
\]
For $q^*(\theta,1)>0$, the sign is governed by $-\Delta(\theta)$, and the transition occurs when
\[
(2\theta-1)\delta-\lambda>0
\;\;\Longleftrightarrow\;\;
\theta>\theta_1
:=
\frac{1}{2}+\frac{\lambda}{2\delta}.
\]
Thus $\theta_1>1/2$ (and $\theta_1\ge1$ if $\lambda\ge\delta$).

The transition to the ad-free regime is determined by the marginal condition at $\alpha=0$, where $q^*(\theta,0)=2\theta-1$:
\[
\left.\frac{\partial \mathcal V}{\partial \alpha}\right|_{\alpha=0}
=
\bar r-(2\theta-1)\Delta(\theta).
\]
Setting this expression to zero gives
\[
(2\theta_2-1)\bigl[(2\theta_2-1)\delta-\lambda\bigr]
=
\bar r,
\]
and hence $\theta_2=[2\delta+\lambda+\sqrt{\lambda^2+4\delta\bar r}]/(4\delta)$.
Thus $\theta_2\ge1$ if $\bar r\ge\delta-\lambda$, and otherwise $\theta_2\in(\theta_1,1)$.

\textbf{Monotonicity and implementability.}
These thresholds induce the stated four segments, with continuity from the matching conditions at $\theta_1$ and $\theta_2$. Let $g^*(\theta):=g(q^*(\theta),\alpha^*(\theta))$ and $\mathcal V^*(\theta):=\max_{q\ge0,\alpha\in[0,1]}\mathcal V(\theta;q,\alpha)$.
For fixed $(q,\alpha)$, $\mathcal V(\theta;q,\alpha)$ is affine in $\theta$, so $\mathcal V^*(\theta)$ is convex.
By the envelope theorem, $d\mathcal V^*(\theta)/d\theta=2g^*(\theta)$ at all differentiability points, implying that $g^*$ is non-decreasing.

Strict monotonicity holds wherever $q^*(\theta)>0$.
In the low and high segments,
\[
\frac{d}{d\theta}
\bigl([(2\theta-1)(1-\delta)+\lambda](1-\delta)\bigr)
=
2(1-\delta)^2>0,
\qquad
\frac{d}{d\theta}(2\theta-1)=2>0.
\]
In the interior segment, take $v_b>v_a$ and denote their optimal allocations and effective qualities by $(q_i,\alpha_i)$ and $g_i:=g(q_i,\alpha_i)$ for $i\in\{a,b\}$. Optimality implies
\begin{align*}
v_bg_b+R(q_b,\alpha_b)-c(q_b)
&\ge v_bg_a+R(q_a,\alpha_a)-c(q_a),\\
v_ag_a+R(q_a,\alpha_a)-c(q_a)
&\ge v_ag_b+R(q_b,\alpha_b)-c(q_b).
\end{align*}
Adding these inequalities yields $(v_b-v_a)(g_b-g_a)\ge0$, and hence $g_b\ge g_a$. If $g_b=g_a$, both inequalities bind. Strict concavity under $\bar r-(\delta v_i-\lambda)^2>0$ then makes the optimizer unique for each type, so $(q_b,\alpha_b)=(q_a,\alpha_a)$. However, the interior $q$-first-order condition
\[
q_i=v_i(1-\delta\alpha_i)+\lambda\alpha_i
\]
cannot hold at the same $(q,\alpha)$ for $v_b>v_a$, because $1-\delta\alpha>0$. Therefore, $g_b>g_a$ throughout the interior segment.
Hence $g^*(\theta)$ is non-decreasing, and strictly increasing whenever $q^*(\theta)>0$.

By Lemma~\ref{lem:envelope}, the allocation is incentive compatible and no ironing is required. \Halmos
\end{proof}
\subsection{Proof of Proposition \ref{prop:delta_quality}}

\begin{proof}{Proof:}
\textbf{Part (a): Zero-quality types.}
Since $\underline\theta=1/2-\lambda/[2(1-\delta)]$, $\partial\underline\theta/\partial\delta=-\lambda/[2(1-\delta)^2]<0$. Thus the threshold shifts left. When it lies inside the type support, types between the new and old cutoffs gain positive quality, while types below the new cutoff remain at zero.

\textbf{Part (b): Low types ($\alpha^* = 1$).}
Here $q^*(\theta)=(2\theta-1)(1-\delta)+\lambda$, so $\partial q^*/\partial\delta=1-2\theta$: quality rises for $\theta<1/2$, is unchanged at $\theta=1/2$, and falls for $\theta>1/2$.

\textbf{Part (c): Medium types ($\alpha^{int} \in (0, 1)$).}
Locally within the interior regime, the multivariate implicit function theorem gives
\[
H \begin{pmatrix} \frac{\partial q^{int}}{\partial \delta} \\ \frac{\partial \alpha^{int}}{\partial \delta} \end{pmatrix} = - \begin{pmatrix} \frac{\partial^2 \mathcal{V}}{\partial q \partial \delta} \\ \frac{\partial^2 \mathcal{V}}{\partial \alpha \partial \delta} \end{pmatrix} = \begin{pmatrix} (2\theta - 1)\alpha^{int} \\ (2\theta - 1)q^{int} \end{pmatrix}.
\]
Using Cramer's rule and the FOC condition $\Delta(\theta)q^{int}=\bar r(1-\alpha^{int})$ from \eqref{eq:proof_foc_alpha_spec}, we obtain
\[
\frac{\partial q^{int}}{\partial \delta}
=
\frac{1}{|H|}
\begin{vmatrix}
(2\theta-1)\alpha^{int} & -\Delta(\theta) \\
(2\theta-1)q^{int} & -\bar r
\end{vmatrix}
=
\frac{\bar r(2\theta-1)}{|H|}
(1-2\alpha^{int}).
\]
Because $|H|>0$ and $2\theta-1>0$ in this segment, the sign is that of $1-2\alpha^{int}$. Thus quality falls under a heavy ad load, rises under a light load, and is locally unchanged at $\alpha^{int}=1/2$.

\textbf{Part (d): High types ($\alpha^* = 0$).}
For users who remain ad-free, $q^*(\theta)=2\theta-1$ is independent of $\delta$, so $\partial q^*/\partial\delta=0$. \Halmos
\end{proof}

\subsection{Proof of Proposition \ref{prop:rbar_pricing}}

\begin{proof}{Proof:}
The optimal subscription price is $T^*(\theta)=\theta g(q^*(\theta),\alpha^*(\theta))-U(\theta)$, where $U(\theta)=\int_0^\theta g(q^*(t),\alpha^*(t))\,dt$ is the information rent. A change in effective quality therefore affects price through both the user's own gross value and the rent accumulated along the menu. We compare these two terms for users who locally remain in each segment.

\textbf{Part (a): Zero-quality and low types ($\theta\le\theta_1$).}
For these fully ad-supported users, $\alpha^*=1$ and
$q^*(\theta)=\max\{0,(2\theta-1)(1-\delta)+\lambda\}$, both of which are independent of $\bar r$.
Consequently, both effective quality and information rent remain unchanged, so $\partial T^*(\theta)/\partial\bar r=0$.

\textbf{Part (b): Medium types ($\theta \in (\theta_1, \theta_2)$).}
Let $\chi(\theta):=\partial g^{int}(\theta)/\partial\bar r$. Writing $s:=2\theta-1$ and $\Delta(\theta)=s\delta-\lambda$, differentiation gives
\[
\chi(\theta)
=
\frac{\Delta(\theta)\bigl(\Delta(\theta)-s\bigr)\bigl[C(\theta)\bar r+\lambda\Delta(\theta)^2\bigr]}{\bigl(\bar r-\Delta(\theta)^2\bigr)^3},
\qquad
C(\theta):=2\delta s(1-\delta)+\lambda(2\delta-1).
\]
On $(\theta_1,\theta_2)$ we have $\Delta(\theta)>0$, $\Delta(\theta)-s<0$, and $\bar r-\Delta(\theta)^2>0$; moreover $C(\theta)$ is increasing in $s$ with minimum value $\lambda>0$ attained at $s=\lambda/\delta$, so the bracket is strictly positive. Hence $\chi(\theta)<0$ strictly on the open interval, with $\chi(\theta_1)=0$ (the factor $\Delta(\theta_1)=0$).
To establish monotonicity, set $x:=\Delta(\theta)=\delta s-\lambda$. Then $s=(x+\lambda)/\delta$, $C(\theta)=\lambda+2(1-\delta)x$, and $\Delta(\theta)-s=-[(1-\delta)x+\lambda]/\delta$, so
\[
-\chi(\theta)
=
\frac{x\bigl[(1-\delta)x+\lambda\bigr]
\bigl\{\bar r[\lambda+2(1-\delta)x]+\lambda x^2\bigr\}}
{\delta(\bar r-x^2)^3}.
\]
On the interior segment, $x>0$ and $\bar r-x^2>0$. Every numerator factor is positive and weakly increasing (the leading factor strictly), whereas the positive denominator decreases. Hence $-\chi$ increases in $x$, and $dx/d\theta=2\delta>0$ implies $\chi'(\theta)<0$. Thus higher $\bar r$ lowers interior effective quality, increasingly so for higher medium types.

Using $T^*(\theta)=\theta g^*(\theta)-U(\theta)$ and $U(\theta)=\int_0^\theta g^*(t)dt$, and noting that allocations below $\theta_1$ are independent of $\bar r$, we obtain
\[
\frac{\partial T^*(\theta)}{\partial \bar r}
=
\theta \chi(\theta)-\int_{\theta_1}^{\theta}\chi(t)dt.
\]
Since $\chi'(t)<0$, $\int_{\theta_1}^{\theta}\chi(t)\,dt>(\theta-\theta_1)\chi(\theta)$, which implies
\[
\frac{\partial T^*(\theta)}{\partial \bar r}
<
\theta \chi(\theta)-(\theta-\theta_1)\chi(\theta)
=
\theta_1 \chi(\theta)
<0.
\]
Therefore, medium-type prices strictly decrease with $\bar r$.

\textbf{Part (c): High types ($\theta\ge\theta_2$).}
The ad-free threshold satisfies
\[
\theta_2=\frac{2\delta+\lambda+\sqrt{\lambda^2+4\delta\bar r}}{4\delta},
\qquad
\frac{\partial\theta_2}{\partial\bar r}
=
\frac{1}{2\sqrt{\lambda^2+4\delta\bar r}}>0.
\]
Thus, a higher $\bar r$ shifts the ad-free threshold rightward. The price comparison below applies to users who remain above that moving threshold.

\textit{Users remaining ad-free.}
Users remaining ad-free keep $(\alpha^*,q^*)=(0,2\theta-1)$, so $\partial g^*(\theta)/\partial\bar r=0$, but their information rent falls because medium-type effective quality falls:
\[
\frac{\partial U(\theta)}{\partial \bar r}
=
\int_{\theta_1}^{\theta_2}
\frac{\partial g^{int}(t)}{\partial \bar r}\,dt
=
\int_{\theta_1}^{\theta_2}\chi(t)\,dt
<0.
\]
The moving-boundary term vanishes because the interior and ad-free allocations deliver the same effective quality at $\theta_2$. Therefore,
\[
\frac{\partial T^*(\theta)}{\partial \bar r}
=
-\frac{\partial U(\theta)}{\partial \bar r}
>0.
\]
Thus, users remaining in the ad-free tier face a strict price increase.

The result is deliberately local: continuity holds at the moving threshold, but no universal sign is asserted for a finite change that moves a user across it.
\Halmos
\end{proof}

\subsection{Proof of Proposition \ref{prop:lambda_utility}}

\begin{proof}{Proof:}
Under pure subscription ($\alpha\equiv0$), $q^{sub}(\theta)=\max\{0,2\theta-1\}$ and $U^{sub}(\theta)=0$ for $\theta\le1/2$, while $U^{sub}(\theta)=\int_{1/2}^{\theta}(2t-1)dt=(\theta-1/2)^2$ otherwise. The comparison holds the primitives fixed and changes the platform's available monetization choices.

\textbf{Utility gains for low types.}
Under the ad-supported mechanism ($\lambda>0$, $\delta\in(0,1)$), the positive-quality threshold is $\underline{\theta}=1/2-\lambda/[2(1-\delta)]<1/2$.
For $\theta\in(\max\{0,\underline\theta\},1/2]$, the full-ad allocation has $q^*>0$, so $U^{ad}(\theta)=\int_{\max\{0,\underline\theta\}}^\theta g^*(t)dt>0=U^{sub}(\theta)$.

For served low types with above-median valuations, $\theta\in(1/2,\theta_1)$, the benchmark utility is positive, $U^{sub}(\theta)=(\theta-1/2)^2$. Writing $u(\theta)=(2\theta-1)(1-\delta)+\lambda$, the ad-supported utility on the full-ad segment is $U^{ad}(\theta)=u(\theta)^2/4$ when $\lambda\le1-\delta$, which we call Regime~1 (integration starts at $\underline{\theta}$ with $u(\underline{\theta})=0$). Hence
\[
U^{ad}(\theta)-U^{sub}(\theta)
=
\tfrac14\bigl[u(\theta)^2-(2\theta-1)^2\bigr]
=
\tfrac14\bigl(-\Delta(\theta)\bigr)\bigl(u(\theta)+(2\theta-1)\bigr),
\]
where $\Delta(\theta)=(2\theta-1)\delta-\lambda$. On $(1/2,\theta_1)$ we have $\Delta(\theta)<0$ and $u(\theta)+(2\theta-1)>0$, so the gain is strictly positive; it vanishes only at $\theta_1$, where $\Delta(\theta_1)=0$, as the boundary calculation below confirms.

\textbf{Utility at the partial-ad threshold.}
At $\theta_1=1/2+\lambda/(2\delta)$, the benchmark gives $U^{sub}(\theta_1)=\lambda^2/(4\delta^2)$.
Under the ad-supported mechanism,
\[
U^{ad}(\theta_1)
=
\int_{\max\{0,\underline{\theta}\}}^{\theta_1}
q(t)(1-\delta)\,dt,
\qquad
q(t)=(2t-1)(1-\delta)+\lambda.
\]
Let $u=(2t-1)(1-\delta)+\lambda$. Then $dt=du/[2(1-\delta)]$, $q(t)(1-\delta)\,dt=u\,du/2$, and $u(\theta_1)=\lambda/\delta$.

\textit{Regime 1: $\lambda\le1-\delta$.}
The zero-quality segment exists ($\underline{\theta}\ge0$), meaning $u(\underline{\theta})=0$. Hence
\[
U^{ad}(\theta_1)
=
\int_0^{\lambda/\delta}\frac{u}{2}\,du
=
\frac{\lambda^2}{4\delta^2}
=
U^{sub}(\theta_1).
\]

\textit{Regime 2: $\lambda>1-\delta$.}
The zero-quality segment is eliminated ($\underline{\theta}<0$), so the integral starts from $\theta=0$, where $u(0)=\lambda+\delta-1>0$. Hence
\[
U^{ad}(\theta_1)
=
\int_{\lambda+\delta-1}^{\lambda/\delta}\frac{u}{2}\,du
=
\frac{\lambda^2}{4\delta^2}
-
\frac{1}{4}(\lambda+\delta-1)^2
<
U^{sub}(\theta_1).
\]
Thus $U^{ad}(\theta_1)\le U^{sub}(\theta_1)$ in both regimes: the earlier utility advantage disappears by the partial-ad threshold. In Regime~2, the strict loss at this boundary, together with the strict gains below the median and continuity of utility, implies that only a subset of positive-quality low types gains. Types receiving zero quality and type $\theta=0$ retain zero utility.

\textbf{Utility losses for medium and high types.}
For the partial-ad medium segment ($\theta\in(\theta_1,\theta_2)$), write $\alpha=\alpha^{int}(\theta)$ for notational simplicity. Substituting the interior allocation $q^{int}(\theta)=(2\theta-1)(1-\delta\alpha)+\lambda\alpha$ gives
\[
\begin{aligned}
g^{sub}(\theta)-g^{int}(\theta)
&=
(2\theta-1)-q^{int}(\theta)(1-\delta\alpha) \\
&=
\alpha\left[(2\theta-1)\delta(2-\delta\alpha)-\lambda(1-\delta\alpha)\right].
\end{aligned}
\]
For interior types, the marginal condition $\Delta(\theta)=(2\theta-1)\delta-\lambda>0$ implies $(2\theta-1)\delta>\lambda$. Since $\alpha\in(0,1)$ and $\delta\in(0,1)$, we have $(2-\delta\alpha)>(1-\delta\alpha)>0$.
This guarantees the bracketed term is strictly positive. Since $\alpha>0$, we establish $g^{int}(\theta)<g^{sub}(\theta)$, so utility increases more slowly with type under the ad-supported mechanism, because its derivative is effective quality.
Since $U(\theta)=U(\theta_1)+\int_{\theta_1}^{\theta}g(t)dt$, combining $U^{ad}(\theta_1)\le U^{sub}(\theta_1)$ with $g^{int}<g^{sub}$ proves $U^{ad}(\theta)<U^{sub}(\theta)$ for all medium types.

For $\theta\ge\theta_2$, the allocation matches the benchmark, so $U^{ad}(\theta)=U^{ad}(\theta_2)+\int_{\theta_2}^{\theta}(2t-1)dt$. Since $U^{ad}(\theta_2)<U^{sub}(\theta_2)$, high types inherit the medium-tier loss. Thus only low types can gain.

\textbf{Aggregate social welfare.} We now establish the supplementary welfare comparisons discussed after the proposition. Subscription payments cancel between consumer surplus and platform profit, so individual utility losses alone do not determine this comparison. Let $x:=2\theta-1\in[-1,1]$. Define the optimized pointwise virtual-surplus values under hybrid monetization and pure subscription as
\[
V^{ad}(x):=\max_{q\ge0,\,\alpha\in[0,1]}
\left\{xq(1-\delta\alpha)+\lambda q\alpha
+\bar r\left(\alpha-\frac{\alpha^2}{2}\right)-\frac{q^2}{2}\right\},
\qquad
V^{sub}(x):=\frac{(x_+)^2}{2},\qquad x_+:=\max\{0,x\},
\]

and let $H(x):=V^{ad}(x)-V^{sub}(x)\ge0$, because $\alpha=0$ is feasible. Since $d\theta=dx/2$, $\Delta\Pi:=\Pi^{ad}-\Pi^{sub}=\frac12\int_{-1}^{1}H(x)dx$.

The envelope theorem gives $H'(x)=g^{ad}(x)-g^{sub}(x)$ almost everywhere. Under the uniform distribution, $CS=\frac14\int_{-1}^{1}(1-x)g(x)\,dx$. Writing $\Delta SW:=\Delta CS+\Delta\Pi$, integration by parts therefore yields
\[
\Delta CS
=-\frac12H(-1)+\frac14\int_{-1}^{1}H(x)\,dx
=\frac12\Delta\Pi-\frac12H(-1),
\qquad
\Delta SW
=\frac32\Delta\Pi-\frac12H(-1).
\]

The identity shows why the nonnegative profit gain alone is insufficient: the welfare comparison also subtracts the virtual-surplus gain at the bottom of the type support. It remains to show $\Delta\Pi>H(-1)/3$. A lower bound using only the negative-virtual-value half of the market will suffice, since $H(x)\ge0$ elsewhere. Set $A:=1-\delta>0$. For every $x\le0$, Proposition~\ref{prop:negative_types} gives $\alpha^*(x)=1$ and $q^*(x)=\max\{0,Ax+\lambda\}$, so
\[
H(x)=\frac{\bar r}{2}+\frac12\bigl[\max\{0,Ax+\lambda\}\bigr]^2,
\qquad x\le0.
\]

If $\lambda\le A$, then $H(-1)=\bar r/2$ and $\Delta\Pi\ge\bar r/4$, so $\Delta SW\ge3\bar r/8-\bar r/4=\bar r/8>0$.

If $\lambda>A$, let $m:=\lambda-A>0$. Then $H(-1)=(\bar r+m^2)/2$ and
\[
\Delta\Pi
\ge
\frac12\int_{-1}^{0}
\left[\frac{\bar r}{2}+\frac{(Ax+\lambda)^2}{2}\right]dx
=
\frac{\bar r}{4}+\frac{\lambda^3-m^3}{12A}.
\]

Using $\lambda=m+A$ gives $\Delta SW\ge[\bar r+A^2+3Am+m^2]/8>0$.

Thus, throughout the regular parameter region, hybrid monetization strictly raises aggregate social welfare relative to pure subscription. For the separate marginal comparison, hold $\delta$ and $\bar r$ fixed and increase $\lambda$ within the regular region. The profit envelope gives $\partial\Pi^{ad}/\partial\lambda=\int_0^1q^*(\theta)\alpha^*(\theta)\,d\theta$. If $\lambda\le A$, then $H(-1)$ is constant in $\lambda$ and the positive-quality advertised segment has positive measure, so $\partial SW^{ad}/\partial\lambda>0$. If $\lambda>A$, then $\partial H(-1)/\partial\lambda=m$ and
\[
\frac{\partial SW^{ad}}{\partial\lambda}
=
\frac32\int_0^1q^*(\theta)\alpha^*(\theta)\,d\theta-\frac m2
\ge
\frac32\left(\frac{\lambda}{2}-\frac A4\right)-\frac{\lambda-A}{2}
=
\frac{\lambda}{4}+\frac A8>0.
\]

Therefore, aggregate social welfare is also strictly increasing in the quality-driven ad premium while the mechanism remains in the regular region. \Halmos
\end{proof}

\subsection{Proof of Proposition \ref{prop:ad_cap_welfare}}

\begin{proof}{Proof:}
Let $CS(\bar{\alpha};\delta,\lambda,\bar r)$ denote aggregate consumer surplus under the advertising cap $\alpha(\theta)\le\bar{\alpha}$. By the envelope condition, $CS(\bar{\alpha};\delta,\lambda,\bar r)=\int_0^1(1-\theta)g^*(\theta;\bar{\alpha})\,d\theta$.
Continuity of the constrained pointwise problem and compactness imply that $CS$ is continuous on $[0,1]$ and admits a maximizer. We first derive the local effect at the unregulated cap, then separately establish a nonempty region in which every maximizing cap is interior.

Because tightening means lowering $\bar\alpha$, a negative left derivative at $\bar\alpha=1$ implies a local consumer-surplus gain. Only positive-quality fully advertised types contribute to this first-order effect. Their constrained effective quality is $g^*(\theta;\bar\alpha)=[(2\theta-1)(1-\delta\bar\alpha)+\lambda\bar\alpha](1-\delta\bar\alpha)$, so the left derivative is
\[
\Phi(\delta,\lambda)
:=
\left.\frac{dCS}{d\bar{\alpha}}\right|_{\bar{\alpha}=1}
=
\int_{\max\{0,\underline{\theta}\}}^{\theta_1}
(1-\theta)\left[-2\delta(1-\delta)(2\theta-1)+\lambda(1-2\delta)\right]d\theta .
\]
If $\lambda<1-\delta$, then $\underline\theta>0$ and the saturated positive-quality region is $[\underline\theta,\theta_1]$, giving
\[
\Phi(\delta,\lambda)
=
\int_{\underline{\theta}}^{\theta_1}
(1-\theta)
\left[
-2\delta(1-\delta)(2\theta-1)+\lambda(1-2\delta)
\right]d\theta
=
\frac{\lambda^3}{24\delta^2(1-\delta)^2}
>0.
\]
Thus marginal tightening reduces consumer surplus in this region.

If $\lambda\ge1-\delta$, then $\underline\theta\le0$ and the saturated region starts at zero, yielding
\begin{align*}
\Phi(\delta,\lambda)
&=
\int_{0}^{\theta_1}
(1-\theta)
\left[
-2\delta(1-\delta)(2\theta-1)+\lambda(1-2\delta)
\right]\,d\theta \\
&=
\frac{\delta+\lambda}{24\delta^2}
\left[
(2\delta+1)\lambda^2-\delta(8\delta+1)\lambda
+10\delta^2(1-\delta)
\right].
\end{align*}
Since $(\delta+\lambda)/(24\delta^2)>0$, we have $\Phi(\delta,\lambda)<0$ if and only if
\[
(2\delta+1)\lambda^2-\delta(8\delta+1)\lambda
+10\delta^2(1-\delta)<0.
\]
Hence, when $\delta>5/8$, marginally tightening the unregulated cap strictly increases consumer surplus whenever
\[
\lambda>\hat{\lambda}_{\mathrm{local}}(\delta)
:=
\frac{
\delta\left[(8\delta+1)-\sqrt{144\delta^2-24\delta-39}\right]
}{
2(2\delta+1)
}.
\]
Under this condition, $\Phi(\delta,\lambda)<0$, so lowering $\bar\alpha$ slightly raises consumer surplus and $\bar\alpha=1$ cannot maximize it.

To establish interior optimality for a nonempty set of primitives, we now compare a feasible interior cap with both endpoints. Under a total ad ban, $\bar{\alpha}=0$, the model collapses to the pure-subscription monopoly, so
\[
CS(0;\delta,\lambda,\bar r)
=
\int_{1/2}^1(1-\theta)(2\theta-1)\,d\theta
=
\frac{1}{24}.
\]
For $(\delta,\lambda,\bar r)=(0.9,0.5,0.18)$, the local condition holds, and direct integration of the globally optimal pointwise allocation yields
\[
CS(0.78)\approx0.04898,
\qquad
CS(0)\approx0.04167,
\qquad
CS(1)\approx0.03155.
\]
The feasible cap $0.78$ strictly dominates both endpoints; it need not be the maximizing cap. Since a maximizer exists, every maximizer at this instance must therefore be interior. Strictness of the comparisons and continuity in the primitives extend the conclusion to a neighborhood of this instance. \Halmos
\end{proof}

\subsection{Proof of Corollary \ref{cor:social_welfare}}

\begin{proof}{Proof:}
Let $SW(\bar{\alpha};\delta,\lambda,\bar r):=CS(\bar{\alpha};\delta,\lambda,\bar r)+\Pi(\bar{\alpha};\delta,\lambda,\bar r)$ denote aggregate social welfare under the cap $\alpha(\theta)\le \bar{\alpha}$. Since the constrained allocation varies continuously with $\bar{\alpha}$ over the compact interval $[0,1]$, $SW(\bar{\alpha};\delta,\lambda,\bar r)$ is continuous and admits a maximizer.

We characterize when marginal tightening from $\bar\alpha=1$ increases welfare. Only fully advertised types contribute to the first-order effect, and zero-quality types have no first-order welfare contribution at this boundary. Write $x:=2\theta-1$. For a positive-quality type at the full ad load, $q^*(x,1)=x(1-\delta)+\lambda$. We first derive its marginal contribution, then integrate over the relevant positive-quality segment.

By the envelope theorem, the marginal profit contribution is $M_\pi(x)=[x(1-\delta)+\lambda](\lambda-x\delta)$; the baseline term $\bar r(1-\alpha)$ vanishes at $\alpha=1$.
The marginal consumer-surplus contribution is obtained from $CS=\int_0^1(1-\theta)g^*(\theta)d\theta$.
Since $\partial g/\partial\alpha|_{\alpha=1}=\lambda(1-2\delta)-2x\delta(1-\delta)$, the corresponding contribution is
$M_{cs}(x)=\frac{1-x}{2}[\lambda(1-2\delta)-2x\delta(1-\delta)]$.
Adding the two terms gives a linear pointwise marginal social-welfare effect, as the quadratic terms cancel:
\[
M_{sw}(x)
=
\left[\lambda^2+\frac{1}{2}\lambda(1-2\delta)\right]
+
x\left[\frac{1}{2}\lambda(1-2\delta)-\delta(1-\delta)\right].
\]

If $0<\lambda\le1-\delta$, the closure of the positive-quality fully advertised segment is $x\in[-\lambda/(1-\delta),\lambda/\delta]$. Using $d\theta=dx/2$ gives
\[
\left.\frac{dSW}{d\bar\alpha}\right|_{\bar\alpha=1}
=\frac12\int_{-\lambda/(1-\delta)}^{\lambda/\delta}M_{sw}(x)\,dx
=\frac{\lambda^3}{8\delta^2(1-\delta)^2}>0.
\]
Thus marginal tightening reduces welfare throughout this branch, including $\lambda=1-\delta$.

If $\lambda>1-\delta$, all fully advertised types receive positive quality and the relevant interval is $x\in[-1,\lambda/\delta]$. Integrating over this interval yields
\[
\left.\frac{dSW}{d\bar{\alpha}}\right|_{\bar{\alpha}=1}
=
-\frac{\delta+\lambda}{8\delta^2}
\left[
2\delta^3-2\delta^2-2\delta\lambda^2+\delta\lambda-\lambda^2
\right].
\]
Therefore, marginal tightening raises welfare if and only if $(2\delta+1)\lambda^2-\delta\lambda+2\delta^2(1-\delta)<0$, equivalently,
\[
\left(2+\frac{1}{\delta}\right)\lambda^2-\lambda+2\delta(1-\delta)<0.
\]
Its solution interval is nonempty exactly when $16\delta^2-8\delta-7>0$, or $\delta>(1+2\sqrt2)/4$.
Thus no $\lambda$ supports a local welfare gain when $\delta\le(1+2\sqrt2)/4$; the possibility is confined to extreme degradation.
For such $\delta$, define the local interval
\[
\underline{\lambda}_{sw}^{\mathrm{loc}}(\delta)
=
\frac{\delta\left[1-\sqrt{16\delta^2-8\delta-7}\right]}{2(2\delta+1)},
\qquad
\overline{\lambda}_{sw}^{\mathrm{loc}}(\delta)
=
\frac{\delta\left[1+\sqrt{16\delta^2-8\delta-7}\right]}{2(2\delta+1)}.
\]
Then for every $\lambda\in(\underline{\lambda}_{sw}^{\mathrm{loc}}(\delta),\overline{\lambda}_{sw}^{\mathrm{loc}}(\delta))$, we have
$\left.dSW/d\bar{\alpha}\right|_{\bar{\alpha}=1}<0$. Hence $\bar{\alpha}=1$ cannot be social-welfare maximizing.
Moreover, under the same parameter restriction, the relevant local interval lies above $1-\delta$, which justifies the integration range $x\in[-1,\lambda/\delta]$ used above.

For every premium in this interval, a sufficiently small tightening from $\bar\alpha=1$ raises welfare. Since the interval exists only for $\delta>(1+2\sqrt2)/4\approx0.957$, this is a boundary-case local possibility; no claim about the globally optimal cap follows without comparison to the total-ad-ban endpoint. \Halmos
\end{proof}

\subsection{Proof of Theorem~\ref{thm:beta_optimal_mechanism}}
\label{app:proof_beta_optimal_mechanism}

\begin{proof}{Proof:}
Let $v(\theta):=2\theta-1$ and $\Delta(\theta):=\delta v(\theta)-\lambda$. Under the generalized cost function $c(q)=\beta q^2/2$, the pointwise virtual surplus is
\[
\mathcal V_\beta(\theta;q,\alpha)
=
v(\theta)q(1-\delta\alpha)
+\lambda q\alpha
+\bar r\left(\alpha-\frac{\alpha^2}{2}\right)
-\frac{\beta q^2}{2}.
\]
For fixed $\alpha$, the $q$-FOC gives $q_\beta^*(\theta,\alpha)=\max\{0,[v(\theta)-\alpha\Delta(\theta)]/\beta\}$.

If $v(\theta)(1-\delta)+\lambda\le0$, then $v(\theta)-\alpha\Delta(\theta)\le0$ for all $\alpha$, so $(q^*,\alpha^*)=(0,1)$. The cutoff is $\underline\theta=1/2-\lambda/[2(1-\delta)]$.

Now consider $\theta>\underline{\theta}$. Substituting the optimal quality choice into the virtual surplus reduces the problem to choosing $\alpha$. For positive-quality types, the derivative of the reduced objective is
\[
\frac{d\mathcal V_\beta}{d\alpha}
=
\bar r(1-\alpha)
-
\frac{\Delta(\theta)\bigl[v(\theta)-\alpha\Delta(\theta)\bigr]}{\beta}.
\]
If $\Delta(\theta)\le0$, equivalently $\theta\le\theta_1:=1/2+\lambda/(2\delta)$, this derivative is nonnegative, so $\alpha^*=1$ and $q_\beta^*=[v(\theta)(1-\delta)+\lambda]/\beta$.

For $\theta>\theta_1$, $\Delta(\theta)>0$, and $\alpha^*=0$ iff $v(\theta)\Delta(\theta)\ge\beta\bar r$. This gives $\theta_{2,\beta}=[2\delta+\lambda+\sqrt{\lambda^2+4\delta\beta\bar r}]/(4\delta)$; for $\theta\ge\theta_{2,\beta}$, $(\alpha_\beta^*,q_\beta^*)=(0,v(\theta)/\beta)$.

On $\theta\in(\theta_1,\theta_{2,\beta})$, $\Delta(\theta)>0$ and $v(\theta)\Delta(\theta)<\beta\bar r$, so the optimum is interior and
\[
\beta q=v(\theta)-\alpha\Delta(\theta),
\qquad
\Delta(\theta)q=\bar r(1-\alpha).
\]
Solving gives
\[
\alpha_\beta^{int}(\theta)
=
\frac{\beta\bar r-v(\theta)\Delta(\theta)}
{\beta\bar r-\Delta(\theta)^2},
\qquad
q_\beta^{int}(\theta)
=
\frac{\bar r\bigl[\lambda+v(\theta)(1-\delta)\bigr]}
{\beta\bar r-\Delta(\theta)^2}.
\]
Here $\alpha_\beta^{int}\in(0,1)$ and the solution joins continuously to $\alpha=1$ at $\theta_1$ and $\alpha=0$ at $\theta_{2,\beta}$.

Feasibility follows from $\bar q=1/\beta$: low types satisfy $q_\beta^*\le1/\beta$ because $v(1-\delta)+\lambda\le\lambda/\delta<1$; medium types satisfy $q_\beta^{int}\le v/\beta\le1/\beta$ because $\alpha_\beta^{int}\ge0$ and $\Delta>0$; and high types satisfy $q_\beta^*=v/\beta\le1/\beta$.

Finally, write $\mathcal V_\beta(\theta;y)=v(\theta)g(y)+H_\beta(y)$, where $H_\beta$ is type-independent. Single crossing makes $g(q_\beta^*(\theta),\alpha_\beta^*(\theta))$ non-decreasing, and the envelope formula supplies IC transfers. \Halmos
\end{proof}

\subsection{Proof of Proposition \ref{prop:beta_allocations}}

\begin{proof}{Proof:}
Improved compute efficiency means lower $\beta$. Let
\[
v(\theta):=2\theta-1,\qquad
\Delta(\theta):=v(\theta)\delta-\lambda,\qquad
D_\beta(\theta):=\beta\bar r-\Delta(\theta)^2.
\]
Theorem~\ref{thm:beta_optimal_mechanism} makes $\underline\theta$ and $\theta_1$ independent of $\beta$, while $\theta_{2,\beta}=(2\delta+\lambda+\sqrt{\lambda^2+4\delta\beta\bar r})/(4\delta)$. In the medium segment, $\alpha_\beta^{int}=(\beta\bar r-v\Delta)/D_\beta$ and $q_\beta^{int}=\bar r[\lambda+v(1-\delta)]/D_\beta$.

\textbf{Part (a): Quality weakly increases and advertising intensity weakly decreases.}
For zero-quality types, the allocation is fixed at $q_\beta^*(\theta)=0$ and $\alpha_\beta^*(\theta)=1$, so the claim is immediate.
For low types, Theorem~\ref{thm:beta_optimal_mechanism} gives $\alpha_\beta^*(\theta)=1$ and $q_\beta^*(\theta)=[v(\theta)(1-\delta)+\lambda]/\beta>0$.
Hence $\partial q_\beta^*(\theta)/\partial\beta=-q_\beta^*(\theta)/\beta<0$, so quality strictly increases as $\beta$ decreases, while ad intensity remains unchanged.

For medium types, differentiating the interior allocation with respect to $\beta$ yields
\[
\frac{\partial q_\beta^{int}(\theta)}{\partial \beta}
=
-\frac{\bar r^2[\lambda+v(\theta)(1-\delta)]}{D_\beta(\theta)^2}<0,
\qquad
\frac{\partial \alpha_\beta^{int}(\theta)}{\partial \beta}
=
\frac{\bar r\Delta(\theta)[\lambda+v(\theta)(1-\delta)]}{D_\beta(\theta)^2}>0.
\]
The inequalities follow because $D_\beta(\theta)>0$, $\Delta(\theta)>0$, and $\lambda+v(\theta)(1-\delta)>0$ in the medium segment.
Thus, as $\beta$ decreases, medium-type quality strictly increases and advertising intensity strictly decreases.
For high types, $\alpha_\beta^*(\theta)=0$ and $q_\beta^*(\theta)=v(\theta)/\beta$, so quality strictly increases as $\beta$ decreases, while ad intensity remains zero.

For endogenous switching, Part (c) shows that lower $\beta$ shifts $\theta_{2,\beta}$ left. The interior and ad-free branches meet continuously at that threshold. On the interior side, $q_\beta^{int}\le v/\beta$, equivalently $\Delta[\beta\bar r-v\Delta]\ge0$, by $\Delta>0$ and $\alpha_\beta^{int}\ge0$. Thus the transition introduces no reversal of the within-segment quality and advertising responses, establishing the market-wide weak inequalities.

\textbf{Part (b): Consumer utility and platform profit weakly increase.}
Because $U_\beta(\theta)=\int_0^\theta g(q_\beta^*(t),\alpha_\beta^*(t))dt$ and $g$ rises with $q$ and falls with $\alpha$, Part (a) implies that every type's utility weakly increases as $\beta$ falls.

For profit, fix $\beta'<\beta$. The mechanism optimal at $\beta$ remains feasible at $\beta'$ because IC and IR do not depend on the cost coefficient and the ceiling expands from $1/\beta$ to $1/\beta'$. Therefore,
\[
\Pi(\beta')
\ge
\Pi(\beta)
+
\frac{\beta-\beta'}{2}
\int_0^1 (q_\beta^*(\theta))^2\,d\theta
\ge
\Pi(\beta).
\]
Thus maximized profit weakly increases.

\textbf{Part (c): The ad-free premium segment expands.}
Since $\partial\theta_{2,\beta}/\partial\beta=\bar r/[2\sqrt{\lambda^2+4\delta\beta\bar r}]>0$, lower $\beta$ shifts the cutoff left and expands the ad-free segment. \Halmos
\end{proof}
\subsection{Proof of Proposition \ref{prop:beta_ad_revenue}}

\begin{proof}{Proof:}
Let $R_{total}(\beta):=\int_0^1 R(q_\beta^*(\theta),\alpha_\beta^*(\theta))d\theta$ denote aggregate advertising revenue under compute-efficiency parameter $\beta$.
By Theorem~\ref{thm:beta_optimal_mechanism}, zero-quality and low types remain fully ad-supported with $\alpha_\beta^*(\theta)=1$, while high types remain ad-free with $\alpha_\beta^*(\theta)=0$.
For medium types, write $v(\theta)=2\theta-1$, $\Delta(\theta)=v(\theta)\delta-\lambda$, and $D_\beta(\theta)=\beta\bar r-\Delta(\theta)^2$.
The interior advertising intensity is $\alpha_\beta^{int}(\theta)=(\beta\bar r-v(\theta)\Delta(\theta))/D_\beta(\theta)$.
Differentiation gives $\partial\alpha_\beta^{int}/\partial\beta=\bar r\Delta[\lambda+v(1-\delta)]/D_\beta^2>0$ because all displayed factors are positive in the medium segment. Thus lower $\beta$ reduces medium-type ad intensity.
In addition, the ad-free threshold $\theta_{2,\beta}=(2\delta+\lambda+\sqrt{\lambda^2+4\delta\beta\bar r})/(4\delta)$ is increasing in $\beta$.
Hence a reduction in $\beta$ shifts the ad-free threshold leftward; any type switching from the medium segment into the high segment moves from $\alpha_\beta^{int}(\theta)\ge0$ to $\alpha^*=0$.
Thus lower $\beta$ weakly reduces ad intensity for every user.

It remains to show that aggregate advertising revenue can nevertheless increase.
Consider the fully ad-supported low-type segment $\theta\in(\max\{0,\underline{\theta}\},\theta_1]$.
This interval has strictly positive measure because $\theta_1=1/2+\lambda/(2\delta)>1/2$ and $\max\{0,\underline{\theta}\}\le 1/2$.
For every type in this segment, $\alpha_\beta^*(\theta)=1$ for all $\beta$, and $q_\beta^*(\theta)=\phi(\theta)/\beta$, where $\phi(\theta):=(2\theta-1)(1-\delta)+\lambda>0$.
Moreover, for $\theta\le\theta_1$, $\phi(\theta)\le \lambda/\delta<1$, so $q_\beta^*(\theta)=\phi(\theta)/\beta$ remains below the technological ceiling $\bar q_\beta=1/\beta$.
The advertising revenue from such a type is therefore $R_L(\theta;\beta)=\lambda \phi(\theta)/\beta+\bar r/2$.

Choose any closed subinterval $J\subset(\max\{0,\underline{\theta}\},\theta_1]$ with positive length.
Since $\phi(\theta)$ is strictly positive and continuous on $J$, there exists $m>0$ such that $\phi(\theta)\ge m$ for all $\theta\in J$.
Therefore,
\[
R_{total}(\beta)
\ge
\int_J R_L(\theta;\beta)d\theta
\ge
|J|\left(\frac{\lambda m}{\beta}+\frac{\bar r}{2}\right).
\]
As $\beta\downarrow0$, the right-hand side diverges to $+\infty$.
Since $R_{total}(1)$ is finite, there exists some $\hat{\beta}\in(0,1)$ such that $R_{total}(\hat{\beta})>R_{total}(1)$.

Thus lower $\beta$ can raise aggregate ad revenue despite reducing ad intensity: on a positive mass of low types, $\alpha=1$ remains fixed while cheaper compute raises quality and the component $\lambda q\alpha$. This is an existence result, not global monotonicity of $R_{total}(\beta)$. \Halmos
\end{proof}

\subsection{Proof of Proposition \ref{prop:discrete_tiers}}\label{app:discrete_tiers}

\paragraph{Discrete-Type Setup.}
The benchmark has three equally sized consumer groups ($f_i=1/3$) and retains the baseline cost and revenue functions with $\beta=1$ and $\bar q=1$.
Normalize their quality valuations to $\theta_H=1$, $\theta_M\in(1/2,1)$,\footnote{The 2-3-2-1 sequence requires the additional restrictions in Proposition~\ref{prop:discrete_tiers}, as well as $\theta_M>1/2$.
Alternative type configurations can change this sequence.
At $\theta_M=1/2$, type $M$ has zero virtual valuation and receives zero quality under pure subscription, but positive quality for every $\lambda>0$.
At $\theta_M=1/4$, the two lower virtual valuations coincide, $v_L=v_M=-1/2$, so bottom pooling can occur without ironing; the strict reversal $v_L>v_M$ occurs only when $\theta_M<1/4$.} and $\theta_L=0$.\footnote{We normalize $\theta_L=0$ to represent consumers who enter at a zero price (e.g., through free versions of ChatGPT and Copilot).
The core 2-3-2-1 evolution in positive-quality tiers remains robust for any sufficiently small $\theta_L>0$.}
For non-top types $i\in\{L,M\}$, the virtual valuation is $v_i=\theta_i-\frac{1-F_i}{f_i}(\theta_{i+1}-\theta_i)$, where $F_i=\sum_{j\le i}f_j$; for the highest type, $v_H=\theta_H=1$.

\begin{proof}{Proof:}
The zero-quality baseline is not counted as a tier. The virtual valuations are $v_L=-2\theta_M<0<v_M=2\theta_M-1<v_H=1$. Define
\[
\lambda_L:=2\theta_M(1-\delta),\qquad
\lambda_M:=\bar q-(2\theta_M-1)(1-\delta),\qquad
\lambda_U:=\bar q+2\theta_M(1-\delta).
\]
The maintained restriction $0<\bar r\le\delta-\lambda_L$ implies $\lambda_L<\delta$.
With $\bar q=1$, we have $\lambda_M-\delta=2(1-\theta_M)(1-\delta)>0$ and $\lambda_U-\lambda_M=(4\theta_M-1)(1-\delta)>0$, yielding $\lambda_L<\lambda_M<\lambda_U$.
The bound on $\bar r$ also keeps type $H$ ad-free in the first regime, as shown below.

Subject to $q_i\le\bar q=1$ and $\alpha_i\in[0,1]$, the platform chooses quality and advertising for each group to maximize the following virtual surplus (Section~\ref{sec:quality_production}):
\begin{equation}\label{eq:virtual_surplus}
    \mathcal{S}_i(v_i;q_i, \alpha_i) = v_i q_i(1 - \delta\alpha_i) + \lambda q_i \alpha_i + \bar{r}\left(\alpha_i - \frac{\alpha_i^2}{2}\right) - \frac{q_i^2}{2}.
\end{equation}

The maintained assumptions already imply $\bar r\le\delta-\lambda_L<\delta=\delta v_H^2$.
The interior first-order conditions for Equation~\eqref{eq:virtual_surplus} give the following stationary allocations for type $i$ before imposing the bounds:
\begin{equation}\label{eq:interior_allocations_discrete}
    \alpha^{int}_i = \frac{\bar{r}-v_i(\delta v_i - \lambda)}{\bar{r}-(\delta v_i - \lambda)^2}, \quad q^{int}_i = \frac{\bar{r}(v_i-\delta v_i + \lambda)}{\bar{r}-(\delta v_i - \lambda)^2}.
\end{equation}

For any stationary allocation with $0<q_i<\bar q$ and $0<\alpha_i<1$, set $d_i:=\delta v_i-\lambda$.
The first-order conditions give $q_i=v_i-\alpha_i d_i$ and $\bar r(1-\alpha_i)=q_i d_i$, so $d_i>0$ and $v_i>0$.
Hence
\[
\bar r-d_i^2
=\frac{d_i[(1-\delta)v_i+\lambda]}{1-\alpha_i}>0.
\]
The second-order condition therefore holds automatically, and strict concavity makes this allocation the unique global maximizer.

These expressions describe interior candidates; the platform must also consider the boundary choices under $\alpha_i\in[0,1]$ and $q_i\in[0,\bar q]$.
We first verify that the typewise virtual-surplus maximizers form an implementable menu. Let $z_i:=(q_i,\alpha_i)$, $x_i:=q_i(1-\delta\alpha_i)$, and
\[
b_i:=\lambda q_i\alpha_i
+\bar r\left(\alpha_i-\frac{\alpha_i^2}{2}\right)
-\frac{q_i^2}{2},
\qquad
\mathcal S_i=v_i x_i+b_i.
\]
Because all types share $[0,\bar q]\times[0,1]$, for $v_j>v_i$ optimality gives
\[
v_jx_j+b_j\ge v_jx_i+b_i,
\qquad
v_ix_i+b_i\ge v_ix_j+b_j.
\]
Adding yields $(v_j-v_i)(x_j-x_i)\ge0$, hence $x_L\le x_M\le x_H$, including at the ceiling.

Set low-type IR and the adjacent downward IC constraints to equality:
\[
T_L=0,
\qquad
T_M=\theta_M(x_M-x_L),
\qquad
T_H=T_M+(x_H-x_M).
\]
Then $U_L=0$, $U_M=\theta_Mx_L$, and $U_H=\theta_Mx_L+(1-\theta_M)x_M$, so IR and, by monotonicity, adjacent upward IC hold. The nonadjacent constraints are
\[
U_H-(x_L-T_L)
=(1-\theta_M)(x_M-x_L)\ge0,
\qquad
U_L-(0\cdot x_H-T_H)
=T_H\ge0.
\]
Thus the menu is globally IC. Moreover,
\[
\frac{T_L+T_M+T_H}{3}
=\frac{v_Lx_L+v_Mx_M+v_Hx_H}{3},
\]
so expected profit equals the sum of typewise virtual surpluses and separate maximization is valid. When adjacent types receive the same quality and ad load, these transfers give them the same price. If all types receive $(\bar q,1)$, all transfers are zero because $T_L=0$. The allocation analysis below shows that type $L$ still has zero quality at $\lambda_L$, types $M,H$ pool at $\lambda_M$, and all types pool at $\lambda_U$.

At full ad load, quality before imposing its bounds is $\tilde q_i(1)=v_i(1-\delta)+\lambda$.
The thresholds solve
\[
\tilde q_L(1)=0,\qquad \tilde q_M(1)=\bar q,\qquad \tilde q_L(1)=\bar q.
\]
respectively marking entry of type $L$, the ceiling for type $M$, and the ceiling for type $L$.

For type $L$, $\lambda-\delta v_L=\lambda+2\delta\theta_M>0$, and base ad revenue also favors advertising. Hence $\alpha_L^*=1$ and
\[
q_L^*=\min\{\bar q,\max\{0,\lambda-\lambda_L\}\}.
\]
Thus $q_L^*$ is zero, interior, and capped over the three respective ranges separated by $\lambda_L$ and $\lambda_U$.

\textit{Ad Initiation.}
For $\lambda\in(0,\lambda_L]$, type $L$ has zero quality and is not counted, while $v_M>0$ and $\lambda<\lambda_M$ give type $M$ positive sub-ceiling quality.
Type $M$'s plan carries advertising when $\bar r>v_M(\delta v_M-\lambda)$ and remains ad-free otherwise.
For type $H$, let $d=\delta-\lambda>0$.
Relative to the ad-free allocation $(\alpha_H,q_H)=(0,1)$, the optimized surplus change from introducing ad load $\alpha\in[0,1]$ is
\[
\Delta_H(\alpha)=\alpha(\bar r-d)+\frac{\alpha^2}{2}(d^2-\bar r).
\]
We have $\Delta_H(0)=0$, and $d\ge\delta-\lambda_L\ge\bar r$ with $d<1$. If $d^2<\bar r$, the quadratic is concave with vertex $\alpha^\star=(d-\bar r)/(d^2-\bar r)\le0$ (since $d\ge\bar r$), so $\Delta_H$ is nonincreasing on $[0,1]$. If $d^2=\bar r$, the quadratic term vanishes and $\Delta_H(\alpha)=\alpha(\bar r-d)<0$ for every $\alpha>0$, because $0<d<1$ implies $\bar r=d^2<d$. If $d^2>\bar r$, it is convex, so its maximum on $[0,1]$ is attained at an endpoint; here $\Delta_H(1)=\tfrac12(d^2-2d+\bar r)$, and since $\phi(d):=d^2-2d+\bar r$ has $\phi'(d)=2(d-1)<0$ on $(\bar r,1)$ with $\phi(\bar r)=\bar r(\bar r-1)\le0$, we have $\phi(d)\le0$ for all $d\in[\bar r,1]$, so $\Delta_H(1)\le0$. In all three cases $\Delta_H(\alpha)\le0$ for all $\alpha\in[0,1]$, with equality only at $\alpha=0$.
Thus type $H$ remains ad-free with $q_H^*=1$, leaving exactly two positive-quality tiers, $M$ and $H$.

\textit{Full Separation.}
For $\lambda\in(\lambda_L,\lambda_M)$, $q_L^*>0$ while $\tilde q_M(1)<\bar q$. The typewise solutions are distinct and their effective qualities are ordered by the IC argument, so no ironing is needed and all three types form distinct positive-quality tiers.
Within this region, changes in type $H$'s ad load do not alter the tier count. The ad-free comparison $\Delta_H(\alpha)$ above is weakly negative whenever $d=\delta-\lambda\ge\bar r$, i.e., $\lambda\le\delta-\bar r$; hence $H$ remains ad-free with $(\alpha_H^*,q_H^*)=(0,1)$ on $(\lambda_L,\delta-\bar r]$. This interval is empty at the allowed equality $\bar r=\delta-\lambda_L$. For $\lambda>\delta-\bar r$ the interior allocation in \eqref{eq:interior_allocations_discrete} becomes admissible ($\alpha_H^{int}>0$), and $\alpha_H^{int}$ reaches $1$ at $\lambda=\delta v_H=\delta$ (where $q_H^{int}=\bar q$), after which $H$ is capped at $(\alpha_H^*,q_H^*)=(1,\bar q)$. 

At the illustrative parameters ($\delta=0.8$, $\bar r=0.2$, $v_H=1$) these sub-transitions occur at $\lambda=3/5$ and $\lambda=4/5$, both interior to $(\lambda_L,\lambda_M)=(4/15,14/15)$, reproducing the $H$ sub-rows of Table~\ref{tab:tier_exact_dynamics}.

\textit{Top Pooling.}
For $\lambda\in[\lambda_M,\lambda_U)$, type $M$ is capped.
Moreover, with type $M$ capped at $q_M=\bar q$, the marginal effect of ad load on surplus is $\bar q(\lambda-\delta v_M)+\bar r(1-\alpha)$, which is nonnegative on $[\lambda_M,\lambda_U)$ because $\lambda-\delta v_M\ge\lambda_M-\delta v_M=\bar q-v_M>0$.
Hence $(\alpha_M^*,q_M^*)=(1,\bar q)$. The preceding high-type characterization gives the same full-ad, capped allocation for type $H$ throughout this range, while $\lambda<\lambda_U$ leaves $q_L^*<\bar q$. Because both quality and advertising coincide for $M,H$, the transfer construction above assigns them the same price. They therefore pool above separated type $L$, yielding two tiers.

\textit{Universal Pooling.}
For $\lambda\ge\lambda_U$, type $L$ also reaches $(\alpha_L^*,q_L^*)=(1,\bar q)$, so all types pool in one tier.

The four ranges therefore yield $2$-$3$-$2$-$1$ tiers; substituting the threshold definitions gives the intervals in Proposition~\ref{prop:discrete_tiers}.
\Halmos
\end{proof}
\subsection{Proof of Theorem \ref{thm:mixed_optimal_mechanism}}
\label{app:proof_mixed_theorem}

\begin{proof}{Proof:}
We first derive the interior allocations and the conditions ordering the advertising thresholds.
We then optimize on the zero-effective-quality boundary and compare the remaining candidates, before verifying that effective quality is nondecreasing.
Under \eqref{eq:mixed_effective_quality}, the pointwise virtual surplus is
\[
\mathcal V_m(\theta;q,\alpha)
=
v(\theta)\bigl[q(1-\delta\alpha)-\delta_n\alpha\bigr]
+\lambda q\alpha
+\bar r\left(\alpha-\frac{\alpha^2}{2}\right)
-\frac{q^2}{2}.
\]
Let $\Delta_m(\theta):=\delta v(\theta)-\lambda$. The $q$-FOC gives $q^*=v(\theta)-\alpha\Delta_m(\theta)$, whose substitution into the $\alpha$-FOC yields the stated interior allocations. Write $v:=v(\theta)$ and $\Delta:=\delta v-\lambda$. Where the effective-quality constraint is slack, profiling out $q$ gives
\[
\widehat{\mathcal V}_m(v,\alpha)
=\frac12(v-\alpha\Delta)^2-v\delta_n\alpha
+\bar r\left(\alpha-\frac{\alpha^2}{2}\right).
\]
Its derivatives at the two advertising boundaries are
\[
D_0(v):=\widehat{\mathcal V}_{m,\alpha}(v,0)
=\bar r-v(\Delta+\delta_n),
\qquad
D_1(v):=\widehat{\mathcal V}_{m,\alpha}(v,1)
=\Delta^2-v(\Delta+\delta_n),
\]
and $\widehat{\mathcal V}_{m,\alpha\alpha}=\Delta^2-\bar r$. The positive roots of $D_1(v)=0$ and $D_0(v)=0$ are, respectively, $v_1^m:=2\theta_1^m-1$ and $v_2^m:=2\theta_2^m-1$. At the first root,
\[
v_1^m\bigl[\Delta(v_1^m)+\delta_n\bigr]
=\Delta(v_1^m)^2,
\qquad
D_0(v_1^m)=\bar r-\Delta(v_1^m)^2.
\]
Thus $v_1^m<v_2^m$ iff $\bar r>\Delta(v_1^m)^2$. If $\bar r\ge\lambda^2$, this holds for every finite $\delta_n$; if $0<\bar r<\lambda^2$, equality is first reached at $\Delta(v_1^m)=-\sqrt{\bar r}$, giving
\[
v_1^m=\frac{\lambda-\sqrt{\bar r}}{\delta},
\qquad
\delta_n
=\sqrt{\bar r}+\frac{\delta\bar r}{\lambda-\sqrt{\bar r}}
=\varepsilon_H.
\]
Hence $\delta_n<\varepsilon_H$ orders the thresholds and ensures $\bar r>\Delta(v)^2$ on $(v_1^m,v_2^m)$. There $D_0(v)>0>D_1(v)$ and strict concavity makes the interior allocation the unique global maximizer.

When $\delta_n>0$, $(q,\alpha)=(0,1)$ is infeasible because $g(0,1)<0$. The bottom pool must satisfy $g=0$, i.e.,
\[
q=\frac{\delta_n\alpha}{1-\delta\alpha}.
\]
On this boundary, virtual surplus is
\[
\Psi(\alpha)
=
\bar r\left(\alpha-\frac{\alpha^2}{2}\right)
+
\frac{\lambda\delta_n\alpha^2}{1-\delta\alpha}
-
\frac{\delta_n^2\alpha^2}{2(1-\delta\alpha)^2}.
\]
Differentiating,
\[
\Psi'(\alpha)
=
\bar r(1-\alpha)
+
\frac{\delta_n\alpha}{(1-\delta\alpha)^3}
\left[
\lambda(2-\delta\alpha)(1-\delta\alpha)-\delta_n
\right].
\]
Because $\lambda(2-\delta\alpha)(1-\delta\alpha)$ decreases to $\lambda(2-\delta)(1-\delta)$, $\delta_n<\varepsilon_S\le\varepsilon_B$ implies $\Psi'\ge0$. Thus $\alpha=1$ maximizes the boundary objective and
\[
\alpha_B^*=1,\qquad
q_B^*=\frac{\delta_n}{1-\delta},\qquad
g(q_B^*,\alpha_B^*)=0.
\]

Under full advertising, $q^F(\theta)=\lambda+(2\theta-1)(1-\delta)$. Solving $g(q^F(\underline\theta^m),1)=0$ gives
\[
\underline{\theta}^m=
\frac{1}{2}-\frac{\lambda}{2(1-\delta)}
+\frac{\delta_n}{2(1-\delta)^2}.
\]
For $v_0^m:=2\underline\theta^m-1$, substitution gives $v_0^m<v_1^m$ iff $\delta_n<\varepsilon_B$; equality collapses the positive-effective-quality full-ad segment. Thus $\delta_n<\varepsilon_S$ gives $v_0^m<v_1^m<v_2^m$, with transitions $D_1(v_1^m)=0$ and $D_0(v_2^m)=0$.

After optimizing over $q$, the full-ad minus ad-free endpoint value is
\[
E(v):=\widehat{\mathcal V}_m(v,1)-\widehat{\mathcal V}_m(v,0)
=\frac{\bar r+\lambda^2}{2}
+[\lambda(1-\delta)-\delta_n]v
-\frac{\delta(2-\delta)}{2}v^2.
\]
For $v\ge0$, both endpoints are feasible, and quadraticity in $\alpha$ gives $E(v)=[D_0(v)+D_1(v)]/2$.
On the nonnegative part of the full-ad region, $v\le v_1^m<v_2^m$ implies $D_0(v)>0$ and $D_1(v)\ge0$, so $E(v)>0$. On the ad-free region, $v\ge v_2^m$ implies $D_0(v)\le0$ and $D_1(v)<0$, so $E(v)<0$.

For $v<0$, the ad-free choice $q=v$ is infeasible. On $[v_0^m,\min\{0,v_1^m\}]$, the optimizer follows $g=0$ until the unconstrained $q$-FOC becomes feasible; $\Psi$ is nondecreasing, and the $\alpha$ derivative joins continuously. Thereafter it is affine and nonnegative at both the transition and $\alpha=1$, where it equals $D_1(v)\ge0$. Hence full advertising is globally optimal for these types.

Below $v_0^m$, the $g=0$ allocation remains optimal because virtual surplus is affine in $v$ with slope $g\ge0$. Thus no separated boundary maximum dominates the candidates, and the two-type revealed-preference argument makes effective quality nondecreasing, so no ironing is required.

Combining the four regions proves the theorem.
If $\underline\theta^m<0$, the boundary pool is absent; if $\underline\theta^m=0$, it consists only of $\theta=0$ and has zero measure. \Halmos
\end{proof}

\subsection{Proof of Proposition \ref{prop:mixed_local_robustness}}
\label{app:proof_mixed_robustness}

\begin{proof}{Proof:}
At $\delta_n=0$, the model coincides with the baseline. For $\delta_n\in[0,\varepsilon_S)$, Theorem \ref{thm:mixed_optimal_mechanism} gives the ordered structure. The explicit $\underline\theta^m$ and, by the implicit function theorem, the smoothly defined $\theta_1^m,\theta_2^m$ vary continuously with $\delta_n$; so do interior allocations on compact subsets. Hence $U^m(\theta;\delta_n)=\int_0^\theta g(q^*(t;\delta_n),\alpha^*(t;\delta_n))dt$ is jointly continuous away from segment boundaries.

We first consider the comparative statics with respect to $\delta$. From the expression for $\underline{\theta}^m$,
\[
\frac{\partial \underline{\theta}^m}{\partial \delta}
=
-\frac{\lambda}{2(1-\delta)^2}
+
\frac{\delta_n}{(1-\delta)^3}
=
\frac{2\delta_n-\lambda(1-\delta)}{2(1-\delta)^3},
\]
which is negative iff $\delta_n<\lambda(1-\delta)/2$; equality makes the threshold stationary and a larger $\delta_n$ reverses it. Thus higher $\delta$ expands positive effective quality. For full-ad types,
\[
q^*(\theta)=(2\theta-1)(1-\delta)+\lambda,
\qquad
\frac{\partial q^*(\theta)}{\partial \delta}=-(2\theta-1),
\]
so quality rises for $\theta<1/2$ and falls for $\theta>1/2$. For medium types, the derivative of $q_m^{int}(\theta)$ with respect to $\delta$ is continuous in $\delta_n$; hence the baseline signs in Proposition \ref{prop:delta_quality} persist on compact subsets away from the sign-switching locus $\alpha^{int}=1/2$. High types remain ad-free, with $q^*(\theta)=2\theta-1$, so their quality is unaffected by $\delta$.

Next consider $\bar r$. At $\delta_n=0$, Proposition \ref{prop:rbar_pricing} gives the divergent-pricing pattern. For small $\delta_n$, the bottom and full-ad low allocations are independent of $\bar r$, while the derivatives of medium- and high-type prices with respect to $\bar r$ are continuous perturbations of their baseline values. Hence, on compact subsets away from thresholds, medium-type prices continue to decrease and ad-free prices continue to increase with $\bar r$.

Finally, let $\Delta U^m(\theta;\delta_n):=U^m(\theta;\delta_n)-U^{sub}(\theta)$.
At $\delta_n=0$, Proposition \ref{prop:lambda_utility} implies that there exist compact nonempty intervals $K_+\subset(0,\tfrac12]$ and $K_-\subset(\tfrac12,1]$ such that $\Delta U^m(\theta;0)>0$ on $K_+$ and $\Delta U^m(\theta;0)<0$ on $K_-$. By compactness, these inequalities hold with a positive margin. Joint continuity of $\Delta U^m$ then implies that, for sufficiently small $\delta_n$, the same signs hold on $K_+$ and $K_-$. Thus some low types still gain and some high types still lose relative to the pure-subscription benchmark.
Choose $\bar\delta_n<\min\{\varepsilon_S,\lambda(1-\delta)/2\}$ to preserve these strict margins. The factor-$1/2$ restriction is needed only for the global leftward threshold shift; the structural range may be larger. \Halmos
\end{proof}

\subsection{Diagnostic Calculations for Correlated Ad Degradation}
\label{app:proof_correlated_diagnostic}

This subsection supports the diagnostic implications in Appendix~\ref{app:ext_correlated} by applying the baseline calculation separately to the two degradation states.
It holds the virtual type at $2\theta-1$ and does not impose the cross-state incentive constraints of the joint correlated population.

For each sensitivity state $k\in\{L,H\}$, let $\delta_L=\delta$ and $\delta_H=2\delta$. In the state-contingent benchmark, the pointwise virtual surplus is
\[
\mathcal V_k(\theta;q,\alpha)
=
(2\theta-1)q(1-\delta_k\alpha)
+\lambda q\alpha
+\bar r\left(\alpha-\frac{\alpha^2}{2}\right)
-\frac{q^2}{2}.
\]
Define $\Delta_k(\theta):=(2\theta-1)\delta_k-\lambda$. The first-order condition with respect to $q$ gives
\[
q_k^*(\theta,\alpha)=(2\theta-1)-\alpha\Delta_k(\theta),
\]
and the interior allocations are
\[
\alpha_k^{int}(\theta)
=
\frac{\bar r-(2\theta-1)\Delta_k(\theta)}
{\bar r-\Delta_k(\theta)^2},
\qquad
q_k^{int}(\theta)
=
\frac{\bar r[(2\theta-1)-\Delta_k(\theta)]}
{\bar r-\Delta_k(\theta)^2}.
\]
Thus the low-sensitivity state is exactly the baseline model, and the high-sensitivity state is the baseline model with $\delta$ replaced by $2\delta$.

In the full-ad region, $q_k^F(\theta)=(2\theta-1)(1-\delta_k)+\lambda$, so
\[
\frac{\partial q_L^F(\theta)}{\partial\delta}=-(2\theta-1),
\qquad
\frac{\partial q_H^F(\theta)}{\partial\delta}=-2(2\theta-1).
\]
Hence full-ad quality rises for $\theta<1/2$ and falls for $\theta>1/2$ in both states. The cutoff $\underline\theta_k=1/2-\lambda/[2(1-\delta_k)]$ shifts leftward. In the interior region, Proposition \ref{prop:delta_quality}'s implicit-function argument gives
\[
\frac{\partial q_k^{int}(\theta)}{\partial\delta_k}
=
\frac{\bar r(2\theta-1)}{|H_k|}
\bigl(1-2\alpha_k^{int}(\theta)\bigr),
\]
where $|H_k|>0$. Therefore, quality decreases under heavy ad loads and increases under light ad loads. Ad-free users receive $q_k^*(\theta)=2\theta-1$ and $\alpha_k^*(\theta)=0$, so their quality is unaffected.

For baseline ad profitability, differentiating the interior formulas yields $\partial q_k^{int}/\partial\bar r<0$ and $\partial\alpha_k^{int}/\partial\bar r>0$ in the interior region. Since $g_k(q,\alpha)=q(1-\delta_k\alpha)$ and $\delta_k<1$, interior effective quality falls with $\bar r$. Thus the allocation channel behind Proposition \ref{prop:rbar_pricing} is preserved in each state-contingent benchmark. Remaining ad-free premium users have unchanged allocations, but the lower effective quality of interior users reduces the information rent passed upward, preserving the premium price-increase force.

Finally, since $\delta_L=\delta$ and $\delta_H=2\delta$ both satisfy the analogue of Assumption \ref{ass:market_regularity}, Proposition \ref{prop:lambda_utility} applies separately to both state-contingent benchmarks. Hence, in each state, the ad-supported mechanism generates gains for a nonempty low-type interval and losses for a nonempty premium interval relative to the pure-subscription benchmark. Because the probability of the high-degradation state is increasing in $\theta$, the correlated specification reinforces the economic interpretation that ad-induced distortion is particularly relevant for high-valuation users.

\subsection{Proof of Proposition \ref{prop:positive_ad_utility_equivalence}}
\label{app:proof_positive_ad_utility}

\begin{proof}{Proof:}
Under \eqref{eq:positive_ad_utility}, define $x(\theta):=q(\theta)(1-\delta\alpha(\theta))$ and $y(\theta):=q(\theta)\alpha(\theta)$. A type-$\theta$ user's utility is $U(\theta)=\theta x(\theta)+\eta y(\theta)-T(\theta)$. Because the term $\eta y(\theta)$ does not depend on the private type $\theta$, the envelope condition is unchanged:
\[
U'(\theta)=x(\theta)=q(\theta)(1-\delta\alpha(\theta)).
\]
With lowest-type IR binding, $T(\theta)=\theta x(\theta)+\eta y(\theta)-\int_0^\theta x(t)dt$.
Substituting this expression into platform profit and integrating the rent term by parts yields the pointwise virtual surplus
\[
\mathcal V_\eta(\theta;q,\alpha)
=
(2\theta-1)q(1-\delta\alpha)
+\eta q\alpha
+R(q,\alpha)
-\frac{q^2}{2}.
\]
Substituting $R(q,\alpha)=\lambda q\alpha+\bar r(\alpha-\alpha^2/2)$ gives
\[
\mathcal V_\eta(\theta;q,\alpha)
=
(2\theta-1)q(1-\delta\alpha)
+(\lambda+\eta)q\alpha
+\bar r\left(\alpha-\frac{\alpha^2}{2}\right)
-\frac{q^2}{2}.
\]
{
Thus the optimal thresholds and allocations coincide with those of the baseline problem evaluated at $\tilde\lambda:=\lambda+\eta$. The unchanged envelope condition and $U(0)=0$ give $U_\eta(\theta)=U_{\mathrm{baseline}}(\theta;\tilde\lambda)$ and hence the stated transfer formula. The allocation and utility results therefore carry over under the effective regularity condition.

For the price responses to $\bar r$, hold $\lambda$, $\eta$, and $\delta$ fixed, remain in the effective regular region, and fix a type away from segment thresholds. Here $\theta_1=\frac12+\frac{\tilde\lambda}{2\delta}$. On the partial-ad segment, let $\chi(\theta):=\partial x(\theta)/\partial\bar r$. Differentiating \eqref{eq:interior_allocations} gives $\partial q(\theta)/\partial\bar r<0$, while the allocation calculations in the proof of Proposition~\ref{prop:rbar_pricing}, evaluated at $\tilde\lambda$, establish $\chi(\theta)<0$ and $\chi'(\theta)<0$. Since $y=(q-x)/\delta$ and allocations below $\theta_1$ are independent of $\bar r$,
\begin{align*}
\frac{\partial T_\eta(\theta)}{\partial\bar r}
&=\left(\theta-\frac{\eta}{\delta}\right)\chi(\theta)
-\int_{\theta_1}^{\theta}\chi(t)\,dt
+\frac{\eta}{\delta}\frac{\partial q(\theta)}{\partial\bar r}\\
&<\left(\theta_1-\frac{\eta}{\delta}\right)\chi(\theta)
+\frac{\eta}{\delta}\frac{\partial q(\theta)}{\partial\bar r}<0.
\end{align*}
The first inequality follows from $\chi'(\theta)<0$; the second uses $\theta_1>\tilde\lambda/\delta\ge\eta/\delta$, which follows from $\tilde\lambda<\delta$. Thus partial-ad subscription prices strictly decrease. For full-ad users, including those with zero quality, allocations and rents are independent of $\bar r$, so subscription prices remain unchanged. For users locally remaining ad-free, $y=0$ and the transfer formula reduces to the baseline price at $\tilde\lambda$, whose derivative with respect to $\bar r$ is strictly positive by Proposition~\ref{prop:rbar_pricing}. \Halmos
\par}
\end{proof}

\subsection{Proof of Proposition \ref{prop:type_dependent_robustness}}
\label{app:proof_type_dependent_robustness}

\begin{proof}{Proof:}
Let $v(\theta)=2\theta-1$ and $\Delta_\gamma(\theta)=v(\theta)\delta-(\lambda+\gamma\theta)$. From \eqref{eq:type_dependent_VS}, the $q$-FOC gives $q^*(\theta,\alpha)=v(\theta)-\alpha\Delta_\gamma(\theta)$. Hence the full-ad quality is $q_\gamma^F(\theta)=v(\theta)(1-\delta)+\lambda+\gamma\theta$, which yields the lower threshold $\underline{\theta}^{\gamma}=(1-\delta-\lambda)/[2(1-\delta)+\gamma]$. The full-ad to interior-ad transition is determined by $\Delta_\gamma(\theta)=0$, so $\theta_1^\gamma=(\delta+\lambda)/(2\delta-\gamma)$. The ad-free threshold solves $(2\theta_2^\gamma-1)\Delta_\gamma(\theta_2^\gamma)=\bar r$. Since $0\le\gamma<\delta-\lambda-\bar r$, we have $\Delta_\gamma(1)=\delta-\lambda-\gamma>\bar r$, so $\theta_2^\gamma<1$ and the ad-free premium segment survives. For interior types, the FOCs yield
\[
\alpha_\gamma^{int}(\theta)
=
\frac{\bar r-v(\theta)\Delta_\gamma(\theta)}
{\bar r-\Delta_\gamma(\theta)^2},
\qquad
q_\gamma^{int}(\theta)
=
\frac{\bar r[v(\theta)-\Delta_\gamma(\theta)]}
{\bar r-\Delta_\gamma(\theta)^2}.
\]
Thus the baseline structure applies with $\Delta$ replaced by $\Delta_\gamma$.

For degradation, $\partial\underline\theta^\gamma/\partial\delta=-(2\lambda+\gamma)/[2(1-\delta)+\gamma]^2<0$, while $\partial q_\gamma^F/\partial\delta=-(2\theta-1)$. In the interior region, Proposition \ref{prop:delta_quality}'s argument gives
\[
\frac{\partial q_\gamma^{int}(\theta)}{\partial\delta}
=
\frac{\bar r(2\theta-1)}{|H_\gamma|}
\bigl(1-2\alpha_\gamma^{int}(\theta)\bigr),
\]
where $|H_\gamma|>0$. Thus interior quality falls under heavy ad loads and rises under light loads, while ad-free quality is unaffected.

For baseline ad profitability, differentiating the interior formulas yields $\partial q_\gamma^{int}/\partial\bar r<0$ and $\partial\alpha_\gamma^{int}/\partial\bar r>0$. Thus higher $\bar r$ increases interior ad load and lowers interior effective quality. Low-type allocations are independent of $\bar r$, while users locally remaining ad-free keep $q^*(\theta)=2\theta-1$ and $\alpha^*(\theta)=0$. The degradation of interior effective quality reduces the information rent passed upward, so prices strictly rise for those remaining ad-free users. Because the type-dependent premium changes how the effective-quality derivative varies across interior types, these allocation signs alone do not determine the interior-price sign; we therefore make no such claim here.

Finally, consider utility polarization. For $\theta\le1/2$, the pure-subscription benchmark gives zero utility, while the ad-supported mechanism provides strictly positive effective quality on the nonempty interval $(\max\{0,\underline{\theta}^{\gamma}\},1/2]$, so low types in this interval strictly gain. At the full-ad/interior boundary $\theta_1^\gamma$, direct integration shows that the ad-supported utility is weakly below the pure-subscription utility; when the zero-quality segment exists, the gap is
\[
-\frac{\gamma(\gamma+2\lambda)^2}
{4(2\delta-\gamma)^2[2(1-\delta)+\gamma]}
\le0,
\]
and the no-zero-quality case is also strictly negative. For all interior types, $\Delta_\gamma(\theta)>0$ implies $(2\theta-1)-q_\gamma^{int}(\theta)(1-\delta\alpha_\gamma^{int}(\theta))>0$, so their effective quality is strictly below the pure-subscription benchmark. Hence medium types strictly lose. High types are ad-free but inherit the negative utility gap accumulated over the interior region, so they also strictly lose. Therefore, utility polarization remains intact. \Halmos
\end{proof}

\subsection{Proof of Theorem \ref{thm:concave_ad_premium}}
\label{app:proof_concave_ad_premium}

\begin{proof}{Proof:}
Maintain $\beta=1$ and let $v=2\theta-1$. Pointwise virtual surplus is
\[
\mathcal V_h(v;q,\alpha)
=
vq(1-\delta\alpha)
+\lambda h(q)\alpha
+\bar r\left(\alpha-\frac{\alpha^2}{2}\right)
-\frac{q^2}{2}.
\]
Consumer utility is unchanged, so the baseline envelope and virtual-surplus arguments apply; the platform maximizes $\mathcal V_h$ over $q\ge0$, $\alpha\in[0,1]$, subject to effective-quality monotonicity.
We first identify the full-ad and ad-free candidates, then use the profiled advertising objective to establish global uniqueness in every segment.
Finally, we verify effective-quality monotonicity so that the pointwise allocations can be implemented by an incentive-compatible menu.

The first derivatives are $\partial \mathcal V_h/\partial q=v(1-\delta\alpha)+\lambda\alpha h'(q)-q$ and $\partial \mathcal V_h/\partial \alpha=\lambda h(q)+\bar r(1-\alpha)-\delta vq$.
The second derivatives are $\partial^2\mathcal V_h/\partial q^2=\lambda\alpha h''(q)-1<0$, $\partial^2\mathcal V_h/\partial\alpha^2=-\bar r<0$, and $\partial^2\mathcal V_h/\partial q\partial\alpha=\lambda h'(q)-\delta v$.

By Assumption~\ref{ass:concave_premium_regular}, for every $v\in(0,1]$, every interior stationary point is a strict local maximizer.
Global uniqueness follows from the profiling argument below.
For $v\le0$, $\partial\mathcal V_h/\partial\alpha=\lambda h(q)+\bar r(1-\alpha)-\delta vq\ge0$, with strict inequality when $\alpha<1$.
Thus full advertising is globally optimal for these types, and it remains to choose quality.

First consider $\alpha=1$.
The pointwise virtual surplus becomes $\mathcal V_h(v;q,1)=vq(1-\delta)+\lambda h(q)+\bar r/2-q^2/2$.
Its derivative with respect to $q$ is $v(1-\delta)+\lambda h'(q)-q$, which is strictly decreasing in $q$ because $h''(q)<0$.
Thus full-ad quality is zero if $v(1-\delta)+\lambda h'(0)\le0$ and otherwise is the unique $q_1(v)=v(1-\delta)+\lambda h'(q_1(v))$. The lower cutoff is $\underline v_h=-\lambda h'(0)/(1-\delta)$, equivalently $\underline\theta^h=1/2-\lambda h'(0)/[2(1-\delta)]$.

For $v\le\underline v_h$, $q=0$ and $h(0)=0$ reduce virtual surplus to $\bar r(\alpha-\alpha^2/2)$, so $(q^*,\alpha^*)=(0,1)$.

For $v>\underline v_h$, full-ad quality is $q_1(v)>0$.
The upper-bound condition for $\alpha=1$ is $\partial \mathcal V_h/\partial\alpha|_{q=q_1(v),\alpha=1}\ge0$, which is equivalent to $\lambda h(q_1(v))-\delta vq_1(v)\ge0$, or $\delta vq_1(v)-\lambda h(q_1(v))\le0$.
Define $A_h(v):=\delta vq_1(v)-\lambda h(q_1(v))$.
By Assumption~\ref{ass:concave_premium_regular}, $A_h(v)=0$ has a unique solution $v_1^h$ satisfying $\underline v_h<v_1^h$.
This identifies $(\alpha,q)=(1,q_1(v))$ as the full-ad candidate for $v\in(\underline v_h,v_1^h]$; global optimality for positive virtual types is established below.

At $\alpha=0$, the pointwise virtual surplus becomes $\mathcal V_h(v;q,0)=vq-q^2/2$, so for positive virtual types the optimal ad-free quality is $q^*=v$.
The lower-bound condition for $\alpha=0$ is $\partial \mathcal V_h/\partial\alpha|_{q=v,\alpha=0}\le0$, which is equivalent to $\lambda h(v)+\bar r-\delta v^2\le0$, or $\delta v^2-\lambda h(v)\ge\bar r$.
Define $B_h(v):=\delta v^2-\lambda h(v)$.
By Assumption~\ref{ass:concave_premium_regular}, the equation $B_h(v)=\bar r$ has a unique solution $v_2^h$ satisfying $v_1^h<v_2^h<1$.
This identifies $(q,\alpha)=(v,0)$ as the ad-free candidate for $v\ge v_2^h$; its global optimality is established below.

For $v\in(v_1^h,v_2^h)$, both boundary conditions fail, so the optimum is interior. To prove existence and global uniqueness, note that $A_h(v_1^h)=0$, $q_1(v_1^h)>0$, and $h(q_1(v_1^h))>0$ imply $v>v_1^h>0$. For each $\alpha$, let $Q_v(\alpha)$ maximize $\mathcal V_h$ over $q\ge0$. Since
\[
\mathcal V_{h,q}(v;0,\alpha)
=v(1-\delta\alpha)+\lambda\alpha h'(0)>0,
\qquad
\mathcal V_{h,qq}
=\lambda\alpha h''(q)-1<0,
\]
the maximizer $Q_v(\alpha)>0$ is unique and continuously differentiable. Define the profiled objective $\Phi_v(\alpha):=\mathcal V_h(v;Q_v(\alpha),\alpha)$. The envelope theorem gives
\[
\Phi_v'(\alpha)
=\lambda h(Q_v(\alpha))
+\bar r(1-\alpha)
-\delta vQ_v(\alpha).
\]
Because $Q_v(0)=v$ and $Q_v(1)=q_1(v)$,
\[
\Phi_v'(0)=\bar r-B_h(v)>0,
\qquad
\Phi_v'(1)=-A_h(v)<0.
\]
The first inequality follows from $B_h(0)=0<\bar r$ and the unique root $v_2^h$. For the second, implicit differentiation gives $q_1'(v)=(1-\delta)/[1-\lambda h''(q_1(v))]>0$. At $v_1^h$, strict concavity and $h(0)=0$ imply $h(q_1(v_1^h))>q_1(v_1^h)h'(q_1(v_1^h))$, so $A_h(v_1^h)=0$ yields $\delta v_1^h>\lambda h'(q_1(v_1^h))$ and therefore
\[
A_h'(v_1^h)
=\delta q_1(v_1^h)
+\bigl[\delta v_1^h-\lambda h'(q_1(v_1^h))\bigr]q_1'(v_1^h)
>0.
\]
Together with uniqueness of the root, this implies $A_h(v)>0$ for $v>v_1^h$. Hence $\Phi_v'$ has at least one zero in $(0,1)$. At any such zero, the corresponding pair $(Q_v(\alpha),\alpha)$ is an interior stationary point, and the implicit-function formula gives
\[
\Phi_v''(\alpha)
=\mathcal V_{h,\alpha\alpha}
-\frac{\mathcal V_{h,q\alpha}^2}{\mathcal V_{h,qq}}
=\frac{\det H_h}{\mathcal V_{h,qq}}
<0,
\]
where $H_h$ is the Hessian. Every zero is therefore a strict downward crossing; two zeros would require an intervening upward crossing. Hence the zero is unique and is the global maximizer, with $\alpha^*\in(0,1)$ and $q^*=Q_v(\alpha^*)>0$.
The interior first-order conditions are $v(1-\delta\alpha^*)+\lambda\alpha^*h'(q^*)-q^*=0$ and $\lambda h(q^*)+\bar r(1-\alpha^*)-\delta vq^*=0$.
Equivalently,
\[
q^*
=
v(1-\delta\alpha^*)+\lambda\alpha^*h'(q^*),
\qquad
\delta vq^*-\lambda h(q^*)
=
\bar r(1-\alpha^*).
\]
These are the medium-segment allocations.

For every $v\in(0,1]$, the same positivity and strict concavity in quality ensure that $Q_v(\alpha)>0$ and $\Phi_v(\alpha)$ are well defined on $[0,1]$.
At every interior zero of $\Phi_v'$, Assumption~\ref{ass:concave_premium_regular} implies $\Phi_v''<0$.
For $0<v<v_1^h$, the threshold signs give $\Phi_v'(0)>0$ and $\Phi_v'(1)>0$.
Any interior zero would be a strict downward crossing and would require a subsequent zero that is not a downward crossing, contradicting the assumption.
Hence $\Phi_v'>0$ throughout $(0,1)$, and full advertising is globally optimal.
For $v_2^h<v\le1$, both endpoint derivatives are negative.
An interior downward crossing would require an earlier zero that is not a downward crossing, again a contradiction.
Thus $\Phi_v'<0$ throughout $(0,1)$, and the ad-free allocation is globally optimal.
The cutoff allocations follow by continuity: full advertising at $v=v_1^h$ and no advertising at $v=v_2^h$.
Together with the direct argument for $v\le0$, this establishes global optimality in every segment.

For implementability, write $s(q,\alpha):=q(1-\delta\alpha)$ and $\mathcal V_h=vs+B(q,\alpha)$, where $B$ is type-independent. For $v'>v$, optimality at allocations with effective qualities $s',s$ gives $v's'+B'\ge v's+B$ and $vs+B\ge vs'+B'$. Adding yields $(v'-v)(s'-s)\ge0$, so $s'\ge s$; the envelope formula then constructs IC transfers.

Translating back gives $\underline\theta^h=(1+\underline v_h)/2$, $\theta_1^h=(1+v_1^h)/2$, and $\theta_2^h=(1+v_2^h)/2$.
The zero-quality segment is empty iff $\underline\theta^h<0$, and consists only of $\theta=0$ when $\underline\theta^h=0$.

When $h(q)=q$, $\underline\theta^h=1/2-\lambda/[2(1-\delta)]$; the full-ad boundary implies $v_1^h=\lambda/\delta$ and $\theta_1^h=1/2+\lambda/(2\delta)$; and the ad-free boundary becomes $(2\theta_2^h-1)[(2\theta_2^h-1)\delta-\lambda]=\bar r$. Thus the baseline thresholds are nested.
\Halmos
\end{proof}

\subsection{Proof of Proposition~\ref{prop:concave_premium_robustness}}
\label{app:proof_concave_premium_robustness}

\begin{proof}{Proof:}
At $a=0$, $h_0(q)=q$ and the model is the baseline. For small $a>0$, Theorem~\ref{thm:concave_ad_premium} preserves the segment structure. The implicit function theorem makes the smoothly defined thresholds continuous in $a$, while interior allocations are continuous on compact subsets; consequently, $U^h(\theta;a)=\int_0^\theta q^*(t;a)(1-\delta\alpha^*(t;a))dt$ is jointly continuous away from boundaries.

For degradation, $\partial\underline\theta^h/\partial\delta=-\lambda h_a'(0)/[2(1-\delta)^2]<0$, so the positive-quality set expands. Full-ad quality satisfies $q^*=v(\theta)(1-\delta)+\lambda h_a'(q^*)$, hence
\[
\frac{\partial q^*(\theta)}{\partial\delta}
=
-\frac{v(\theta)}{1-\lambda h_a''(q^*(\theta))}.
\]
Since $h_a''<0$, the denominator is positive, so quality rises below $1/2$ and falls above it.
For medium types, the derivative of $q_h^{int}(\theta)$ with respect to $\delta$ is continuous in $a$; hence the baseline signs in Proposition~\ref{prop:delta_quality} persist on compact subsets away from the sign-switching locus $\alpha^{int}=1/2$.
High types remain ad-free, with $q^*(\theta)=2\theta-1$, so their quality is unaffected by $\delta$.

For $\bar r$, zero-quality and full-ad allocations remain independent of $\bar r$, while medium- and high-type price derivatives are continuous perturbations of their baseline values. Hence the divergent within-segment signs persist on compact subsets away from thresholds.

Finally, let $\Delta U^h(\theta;a):=U^h(\theta;a)-U^{sub}(\theta)$.
At $a=0$, Proposition~\ref{prop:lambda_utility} implies that there exist compact nonempty intervals $K_+\subset(0,\tfrac12]$ and $K_-\subset(\tfrac12,1]$ such that $\Delta U^h(\theta;0)>0$ on $K_+$ and $\Delta U^h(\theta;0)<0$ on $K_-$.
Compactness gives positive margins, and joint continuity preserves both signs for sufficiently small $a$.

Choose $\bar a$ to preserve threshold ordering, the Hessian condition, and these strict signs. \Halmos
\end{proof}

\subsection{Proof of Corollary~\ref{cor:concave_beta_revenue}}
\label{app:proof_concave_beta_revenue}

\begin{proof}{Proof:}
Fix $\hat\beta\in(0,1)$ satisfying $R_{\mathrm{total}}^{0}(\hat\beta)>R_{\mathrm{total}}^{0}(1)$.
For small $a$, concavity and $C^2$ convergence give a common bound on $h_a'(0)$, so the quadratic cost places optimal qualities at $\beta\in\{\hat\beta,1\}$ in a common compact interval. Uniform convergence and uniqueness imply almost-everywhere convergence of allocations (boundaries have measure zero); dominated convergence yields
\[
R_{\mathrm{total}}^{a}(\beta)\longrightarrow R_{\mathrm{total}}^{0}(\beta),
\qquad \beta\in\{\hat\beta,1\}.
\]
Let $\varepsilon=[R_{\mathrm{total}}^{0}(\hat\beta)-R_{\mathrm{total}}^{0}(1)]/3>0$. For small $a$, both convergence errors are below $\varepsilon$, so $R_{\mathrm{total}}^{a}(\hat\beta)-R_{\mathrm{total}}^{a}(1)>\varepsilon>0$. \Halmos
\end{proof}

\subsection{Proof of Theorem \ref{thm:competition_optimal_mechanism}} \label{app:proof_competition_theorem}

\begin{proof}{Proof:}
Let $x(\theta):=g(q(\theta),\alpha(\theta))$. By incentive compatibility, $U'(\theta)=x(\theta)$ and, under the normalization $U(0)=0$, $CS=\int_0^1U(\theta)d\theta=\int_0^1(1-\theta)x(\theta)d\theta$. Hence the weighted objective $\Pi+\kappa CS$ can be written pointwise as
\[
\mathcal V_\kappa(\theta;q,\alpha)
=
v_\kappa(\theta)q(1-\delta\alpha)
+\lambda q\alpha
+\bar r\left(\alpha-\frac{\alpha^2}{2}\right)
-\frac{q^2}{2},
\]
where $v_\kappa(\theta)=\theta-(1-\kappa)(1-\theta)$. Thus the problem is the baseline with $2\theta-1$ replaced by $v_\kappa(\theta)$.

Let $\Delta_\kappa=\delta v_\kappa-\lambda$. The $q$-FOC gives $q^*=v_\kappa-\alpha\Delta_\kappa$. At $\alpha=1$, the zero-quality boundary $v_\kappa=-\lambda/(1-\delta)$ gives $\underline\theta_\kappa$; $\Delta_\kappa=0$ gives $\theta_{1,\kappa}$; the interior FOCs give the stated allocations; and $v_\kappa\Delta_\kappa=\bar r$ gives $\theta_{2,\kappa}$. Since $\bar r<\delta-\lambda$, the last virtual cutoff is below one, so $\theta_{2,\kappa}<1$ for every $\kappa\in[0,1]$. \Halmos
\end{proof}

\subsection{Proof of Proposition \ref{prop:competition_robustness}}\label{app:proof_competition_robustness}

\begin{proof}{Proof:}
As $\delta$ rises, the lower threshold is weakly decreasing.
Where its underlying cutoff is positive, $\partial\underline\theta_\kappa/\partial\delta=-\lambda/[a_\kappa(1-\delta)^2]<0$; where that cutoff is negative, $\underline\theta_\kappa$ remains zero.
The clipping point is interpreted through these one-sided changes.
Within the positive-quality full-ad segment, quality satisfies $\partial q_\kappa^F/\partial\delta=-v_\kappa(\theta)$, which changes sign at $\theta_{0,\kappa}:=b_\kappa/a_\kappa$.
In the interior region, Proposition \ref{prop:delta_quality}'s argument gives
\[
\frac{\partial q_\kappa^{int}(\theta)}{\partial\delta}
=
\frac{\bar r v_\kappa(\theta)}{|H_\kappa|}
\bigl(1-2\alpha_\kappa^{int}(\theta)\bigr),
\]
where $|H_\kappa|>0$. Thus interior quality falls under heavy ad loads and rises under light loads; ad-free quality is unaffected.

Second, consider $\bar r$. In the interior region, differentiating the closed-form allocations gives $\partial q_\kappa^{int}/\partial\bar r=-(v_\kappa-\Delta_\kappa)\Delta_\kappa^2/(\bar r-\Delta_\kappa^2)^2<0$ and $\partial\alpha_\kappa^{int}/\partial\bar r=\Delta_\kappa(v_\kappa-\Delta_\kappa)/(\bar r-\Delta_\kappa^2)^2>0$. Hence interior effective quality strictly decreases. Let $\chi_\kappa(\theta):=\partial g_\kappa^{int}(\theta)/\partial\bar r$. The factorization used in the proof of Proposition~\ref{prop:rbar_pricing} applies after replacing $2\theta-1$ by $v_\kappa(\theta)$; because $v_\kappa'(\theta)=2-\kappa>0$, it gives $\chi_\kappa(\theta)<0$ and $\chi_\kappa'(\theta)<0$ throughout the interior segment. For a user locally remaining interior,
\[
\frac{\partial T^*(\theta)}{\partial\bar r}
=
\theta\chi_\kappa(\theta)
-\int_{\theta_{1,\kappa}}^\theta\chi_\kappa(t)\,dt
<
\theta_{1,\kappa}\chi_\kappa(\theta)
<0.
\]
Low full-ad allocations are independent of $\bar r$. A locally ad-free user keeps $(q^*,\alpha^*)=(v_\kappa,0)$ but receives less information rent from below, so price rises. No sign is asserted for finite threshold crossings.

Finally, compare the advertising mechanism with the pure-subscription benchmark under the same $\kappa$. The benchmark effective quality is $g_\kappa^{sub}(\theta)=\max\{0,v_\kappa(\theta)\}$. Write $w=v_\kappa(\theta)$, so $dw=a_\kappa d\theta$. In the full-ad region, $g^F(w)=(1-\delta)[w(1-\delta)+\lambda]$. This exceeds $\max\{0,w\}$ on a nonempty interval near the bottom of the served full-ad region, so some low types strictly gain. At the full-ad/interior boundary, direct integration gives $\Delta U_\kappa(\theta_{1,\kappa})\le0$, with equality if and only if the zero-quality segment exists, i.e., $\lambda\le(1-\kappa)(1-\delta)$. In the interior region, $\Delta_\kappa=w\delta-\lambda>0$ implies
\[
w-q_\kappa^{int}(\theta)(1-\delta\alpha_\kappa^{int}(\theta))
=
\alpha_\kappa^{int}(\theta)
\left[
w\delta(2-\delta\alpha_\kappa^{int}(\theta))
-\lambda(1-\delta\alpha_\kappa^{int}(\theta))
\right]
>0.
\]
Thus medium types have lower effective quality and a strictly negative utility gap; high types share the benchmark allocation but inherit that gap. All full-ad low types gain exactly when $\Delta U_\kappa(\theta_{1,\kappa})=0$, i.e., $\lambda\le(1-\kappa)(1-\delta)$ or $\kappa\le1-\lambda/(1-\delta)$. \Halmos
\end{proof}

\end{APPENDICES}

\end{document}